\documentclass[a4paper,fleqn,usenatbib]{mnras}

\usepackage[T1]{fontenc}
\usepackage{ae,aecompl}

\usepackage{graphicx}	
\usepackage{amsmath}	
\usepackage{amssymb}	
\usepackage{gensymb}
\usepackage{deluxetable}	
\let\oldAA\AA
\renewcommand{\AA}{\text{\normalfont\oldAA}}

\title[Early-stage merger LIRGs IC 1623 and NGC 6090]{Star formation 
histories in mergers: The spatially resolved properties of the 
early-stage merger LIRGs IC 1623 and NGC 6090}

\author[C. Cortijo-Ferrero et al.]{C. Cortijo-Ferrero,$^{1}$\thanks{E-mail: clara@iaa.es} R. M. Gonz\'alez Delgado,$^{1}$ E. P\'erez, $^{1}$ R. Cid Fernandes,$^{2}$ 
\newauthor S. F. S\'anchez,$^{3}$ A.L. de Amorim,$^{2}$ P. Di Matteo,$^{4}$ R. Garc\'ia-Benito,$^{1}$
\newauthor E. A. D. Lacerda,$^{2}$ R. L\'opez Fern\'andez,$^{1}$ and C. Tadhunter$^{5}$
\\
$^{1}$Instituto de Astrof\'isica de Andaluc\'ia (CSIC), PO Box 3004 18080 Granada, Spain\\
$^{2}$Departamento de F\'isica, Universidade Federal de Santa Catarina, PO Box 476, 88040-900 Florian\'opolis, SC, Brazil\\
$^{3}$Instituto de Astronom\'ia, Universidad Nacional Auton\'oma de M\'exico, A.P. 70-264, 04510 M\'exico, Mexico\\
$^{4}$Observatoire de Paris, GEPI, Observatoire de Paris, PSL Research University, CNRS, Univ Paris Diderot,\\ 
Sorbonne Paris Cité, Place Jules Janssen, 92195 Meudon, France\\
$^{5}$Department of Physics and Astronomy, University of Sheffield, Sheffield S3 7RH, UK\\
}

\date{Accepted XXX. Received YYY; in original form ZZZ}

\pubyear{2015}

\begin{document}
\label{firstpage}
\pagerange{\pageref{firstpage}--\pageref{lastpage}}
\maketitle

\begin{abstract}
The role of major mergers in galaxy evolution is investigated 
through a detailed characterization of the stellar populations, 
ionized gas properties, and star formation rates (SFR) in 
the early-stage merger LIRGs IC 1623 W and NGC 6090, 
by analysing optical 
Integral Field Spectroscopy (IFS) and high resolution HST imaging.  
The spectra were processed with the {\sc starlight} 
full spectral fitting code, and the emission lines measured in the 
residual spectra. The results are compared with control non-interacting 
spiral galaxies from the CALIFA survey. 
Merger-induced star formation is extended and recent, as revealed 
by the young ages (50--80 Myr) and high contributions to light of young 
stellar populations (50--90$\%$), in agreement with merger simulations 
in the literature. These early-stage mergers have positive 
central gradients of the stellar metallicity, 
with an average $\sim$0.6 Z$_{\odot}$.
Compared to non-interacting spirals, 
they have lower central 
nebular metallicity, and flatter profiles, 
in agreement with the gas inflow scenario.
We find that they are dominated by star 
formation, although shock excitation cannot be discarded in some 
regions, where high velocity dispersion is found (170--200 km s$^{-1}$).
The average SFR in these early-stage mergers 
($\sim$23--32 M$_{\odot}$ yr$^{-1}$) 
is enhanced with respect to main-sequence Sbc galaxies by 
factors of 6--9, slightly above the predictions from classical 
merger simulations, but still possible in about 15$\%$ of major 
galaxy mergers, where U/LIRGs belong.
\end{abstract}

\begin{keywords}
galaxies: interactions -- galaxies: evolution -- galaxies: stellar content -- 
galaxies: ISM -- galaxies: star formation -- techniques: spectroscopic
\end{keywords}



\section{Introduction}\label{1}
Merger-induced star formation was first predicted by \cite{toomre1972}, 
and confirmed by \cite{larson&tinsley1978} in a study  of optical colours 
of Arp galaxies. However, not all merging galaxies have significantly enhanced 
star formation rates (SFR). Recent results show that the enhancement of 
the star formation depends 
on multiple factors, such as the merger geometry and the properties of the progenitor 
galaxies \citep{cox2008,dimatteo2007,dimatteo2008}.  
For example, gas-rich (S+S) merging pairs show significant enhancements in the 
specific star formation rates (sSFR) that are not found in S+E pairs \citep{cao2016}. 
Moreover, a clear 
anti-correlation between the sSFR and the pair separation has been  
established. In particular, close pairs with projected separation 
$\leq$ 20 h$^{-1}$ kpc have sSFR enhancement $\gtrsim$ 2 \citep{ellison2008}, 
while the sSFR enhancement of 
wider pairs is significantly less \citep{patton2013}. 

Besides enhanced SFR, the spatial extent of the star formation 
is other important factor that must 
be taken into account when comparing 
the star formation in mergers with that in non-interacting control galaxies. 
Extended star formation is observed in many galaxy mergers from early-stage merger 
or separated progenitors \citep{wang2004,elmegreen2006} to advanced or 
post-coalescence systems \citep{evans2008,wilson2006}. 
Recent high resolution models determine the SFR efficiency 
in mergers by resolving parsec scale physical 
processes \citep{teyssier2010,hopkins2013,renaud2015}.
These simulations find that extended starbursts 
arise spontaneously 
during the first two pericentre passages, 
due to fragmentation of the gas clouds produced by 
an increase of the ISM supersonic turbulence 
as a consequence of the tidal interaction itseft \citep{renaud2014}.  
In these models, a merger-induced nuclear starburst is also present, 
but it starts later in the merger, during the final passage. 
They conclude that  extended star formation is 
important in the early stages of the merger, while the nuclear 
starbursts will occur in advanced stages.

Mergers of gas rich discs, such as those seen in luminous 
and ultra-luminous infrared galaxies (LIRGs and 
ULIRGs, \citealt{surace1998,surace2000,veilleux2002,kim2013}) are therefore 
extreme examples, where star formation is caught at its peak. 
They are unique laboratories offering insight into the physical 
processes triggered by galaxy mergers.
The observational characterization of star formation 
in mergers at different stages is necessary to test the validity, 
and to put contraints on merger simulations.

To understand the role of major mergers in galaxy evolution we have undertaken 
an integral field spectroscopy study of galaxies in different 
stages across the merger sequence.
In this paper we focus on the detailed study of two LIRGs at an early merger stage: IC 1623 and NGC 6090. Both are classified as type IIIb systems according to the morphological classification of \cite{veilleux2002}. We aim to fully characterize their spatially 
resolved stellar population, and ionized gas properties.
In a future work we plan to extend this study to merging systems 
observed in the Calar Alto Legacy Integral Field Area (CALIFA) 
survey \citep{sanchez2012}, in 
order to cover different merger stages and to statistically 
study the evolution of merger-induced star formation.

The paper is organized as follows: Sect. 2 describes the observations and 
data reduction process. In Sect. 3 we summarize the results from the photometric 
analysis of the star clusters detected in  HST images. 
In Sect. 4 we apply the fossil record method to analize the stellar continuum and 
derive the spatially resolved stellar population properties: stellar mass and 
stellar mass surface density, $\mu_{\star}$; stellar extinction, A$_{V}$; 
luminosity weighted mean age, $\langle \log age \rangle_{L}$; 
mass weighted mean metallicity, $\langle \log Z \rangle_{M}$; and the 
contributions to luminosity and mass of young, intermediate, and 
old stellar populations. Section 5 presents results on the ionized gas 
emission, focusing on the morphology, nebular A$_{V}$, and the ionization 
conditions. We discuss the results in Sect. 6, and 
Sect. 7 presents the conclusions.

Throughout the paper we assume a flat cosmology with
$\Omega_{M}$= 0.272, $\Omega_{\Lambda}$ = 0.728, and 
$H_{0}$ = 70.4 km s$^{-1}$ Mpc$^{-1}$ 
(WMAP, seven years results). 
For IC 1623 redshift (z=0.020067) this results in a 
distance of 86.8 Mpc. At this distance 1$\tt{''}$ corresponds 
to 0.421 kpc. For NGC 6090 (z=0.029304), the distance is 127.7 Mpc, and
1$\tt{''}$ corresponds to 0.619 kpc.\begin{figure*}  
\begin{center}
\includegraphics[width=0.7\textwidth]{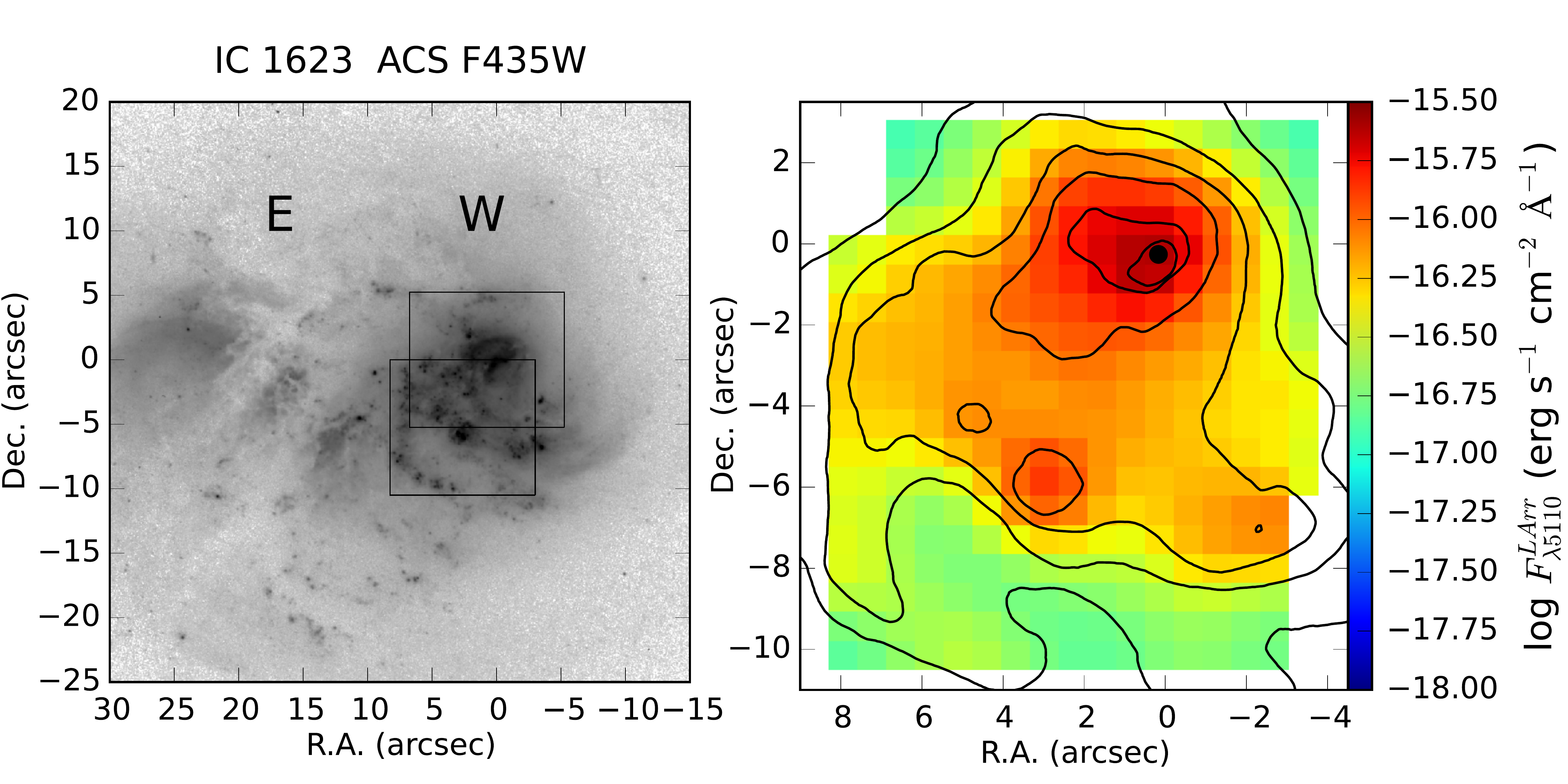} 
\includegraphics[width=0.7\textwidth]{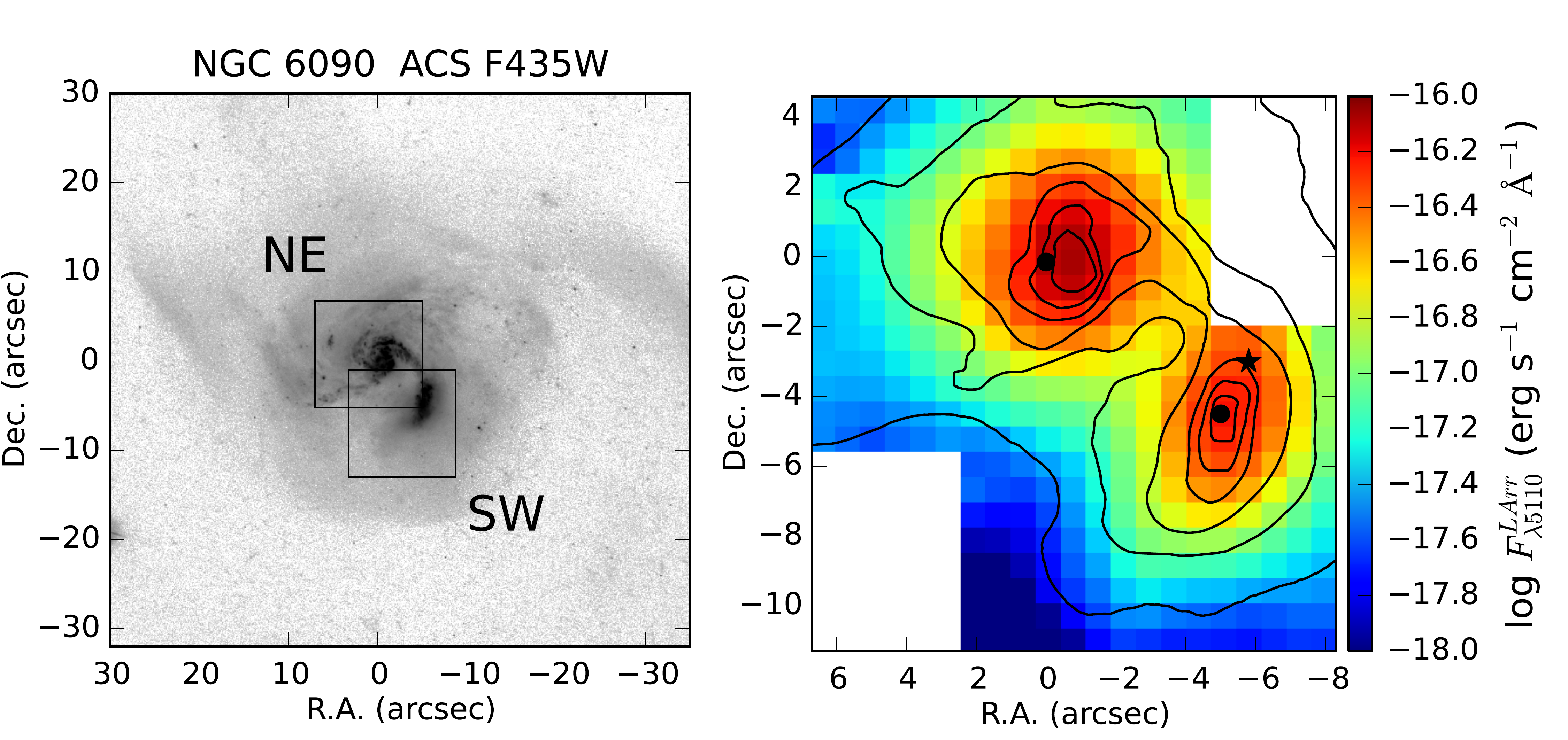} 
\caption{Observations of IC 1623 W (top panels) and NGC 6090 (bottom panels). 
Left: HST ACS F435W images. The black rectangles 
indicate the positions where the IFS observations were obtained. Right: PMAS LArr 
data continuum fluxes at 5110 $\rm \AA$ rest-frame (obtained by averaging 
the continuum from 5050 to 5170 $\rm \AA$) presented on a logarithmic scale. 
Contours are from the HST ACS F435W image smoothed to the PMAS LArr resolution.
One arcsec corresponds to 0.42 kpc for IC 1623 and 0.62 kpc for NGC 6090. 
The black circles indicate the expected position of the 
progenitors nuclei. In the case of NGC 6090 SW, the star symbol 
indicates the position of the brightest knot.}
\label{Fig_1}  
\end{center}
\end{figure*}
\begin{figure*}  
\begin{center}
\includegraphics[width=0.34\textwidth]{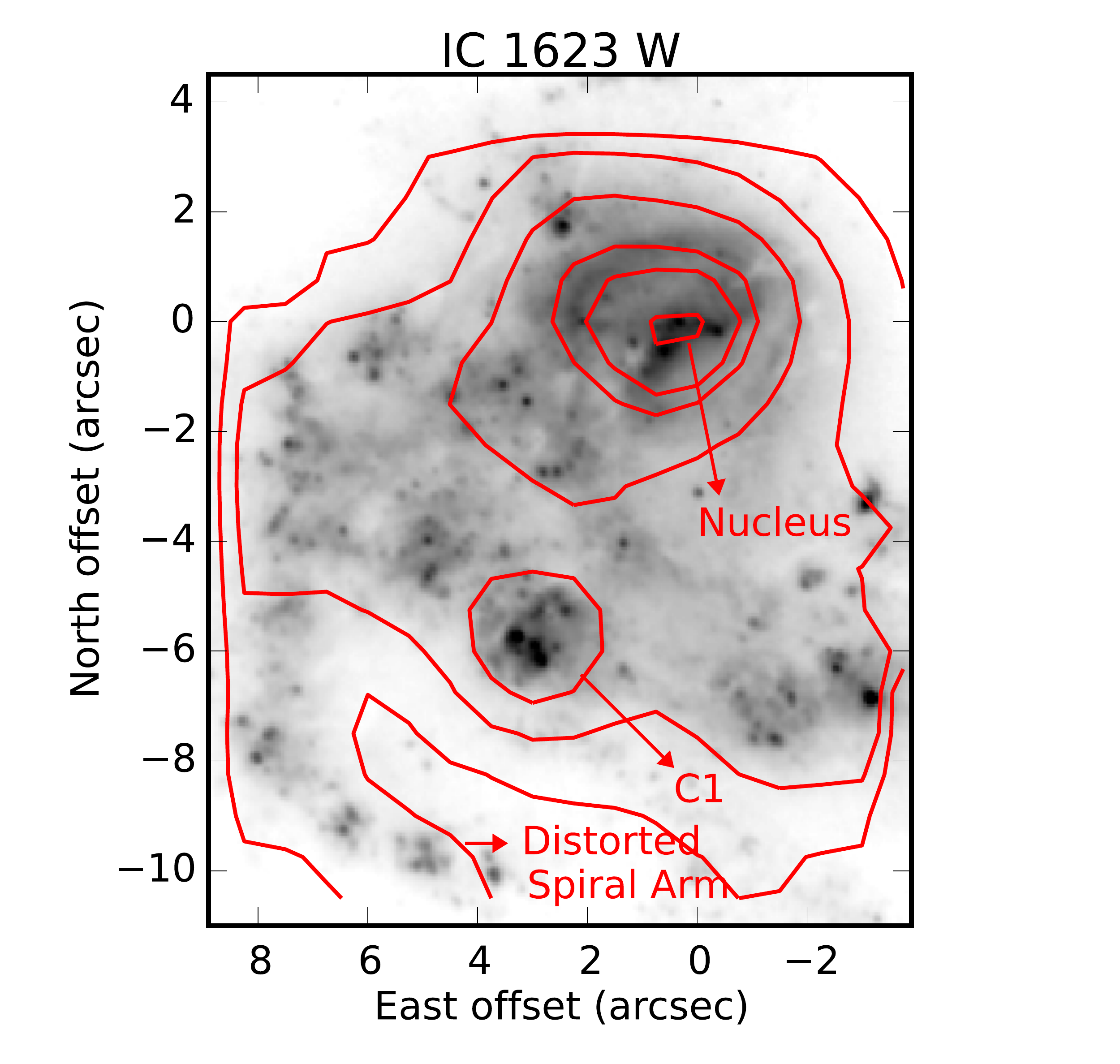} 
\includegraphics[width=0.65\textwidth]{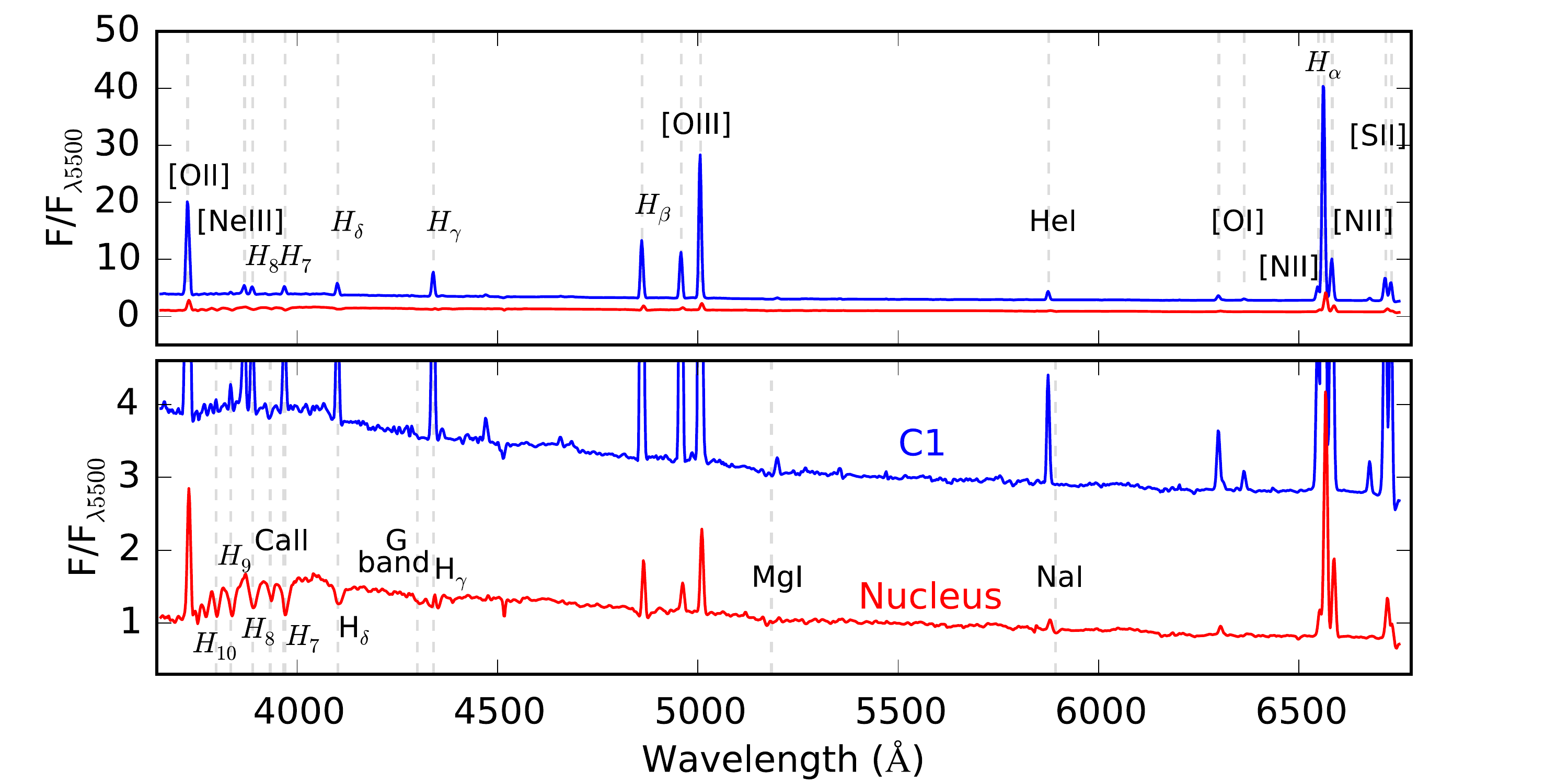} \\
\includegraphics[width=0.34\textwidth]{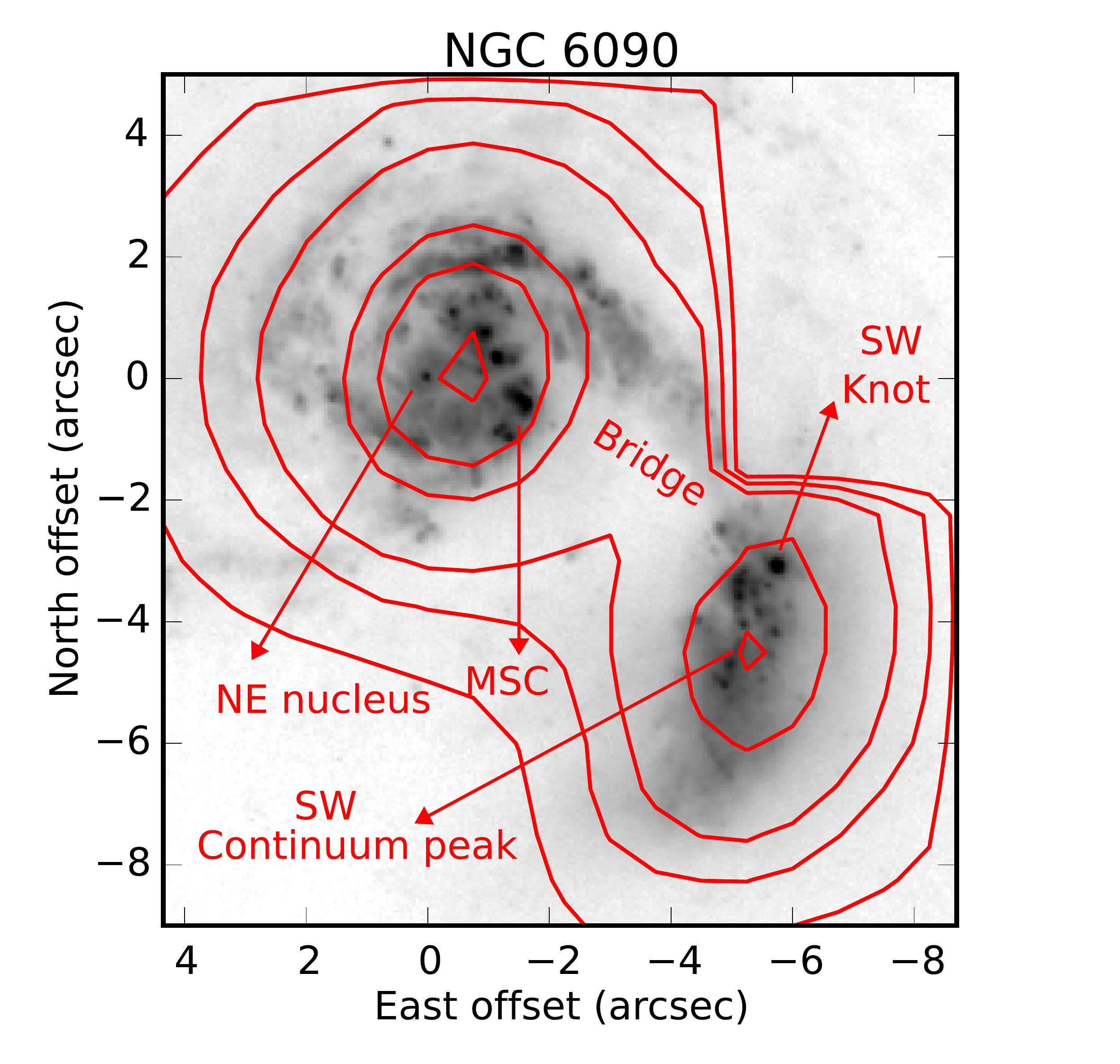} 
\includegraphics[width=0.65\textwidth]{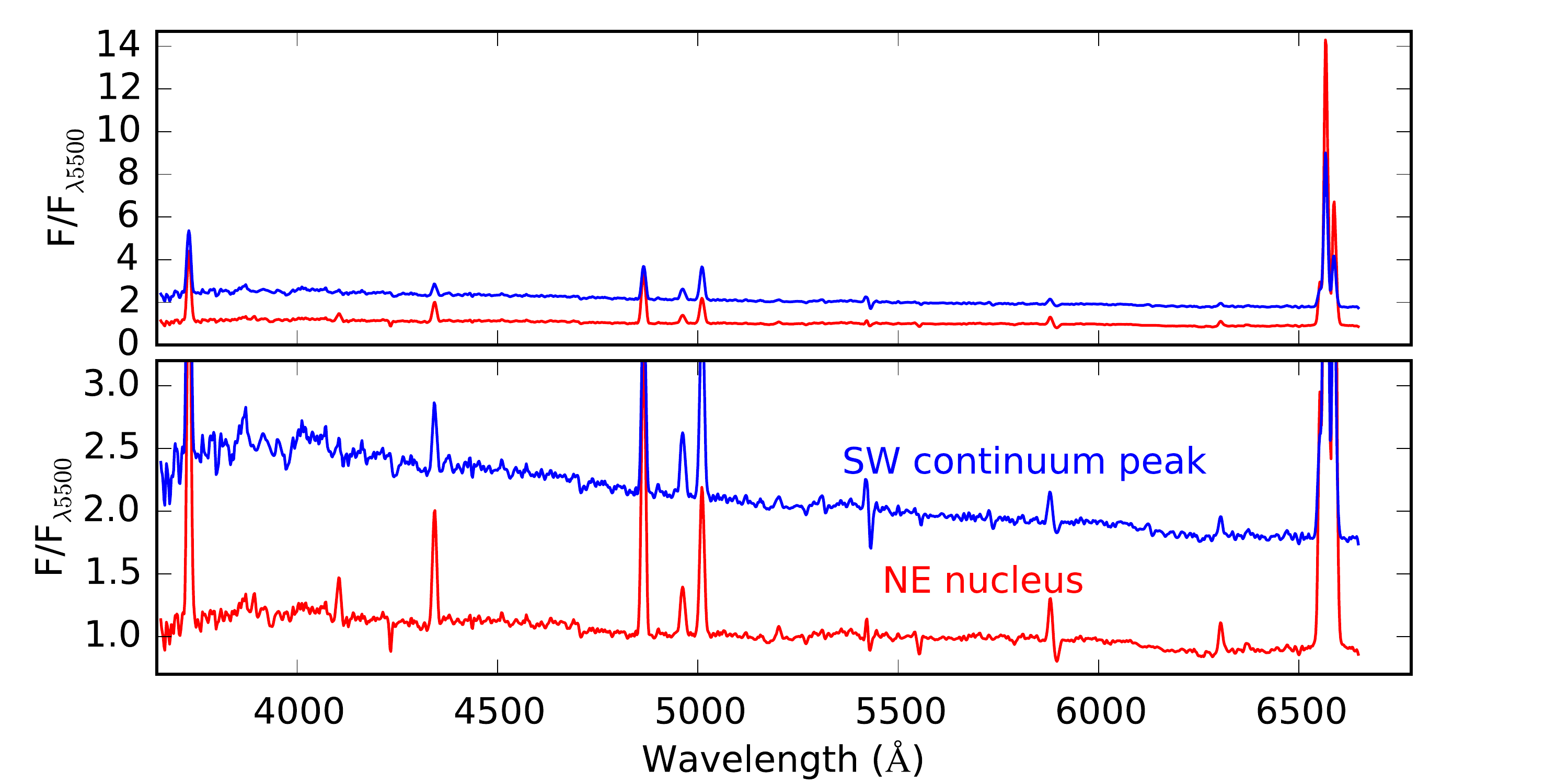} \\  
\caption{Observations of IC 1623 W (top panels) and NGC 6090 (bottom panels). 
Left: HST F435W images presented in greyscale, with the IFS continuum superimposed with red contours. 
The main regions are labelled. Right: Spectra of some of these regions, extracted in 
circular apertures of 2$\tt{''}$ radius. The top and bottom panels are the same 
for each galaxy, but with different y-axis ranges.}
\label{Fig_2}  
\end{center}
\end{figure*}

\subsection{IC 1623}\label{1.1}
IC 1623 (also commonly referred to as VV 114 = Mrk 236; 
L$_{IR}$ = 4.5 $\times$ 10$^{11}$ L$_{\odot}$; e.g., \citealt{sanders2003}) 
is a pair of luminous infrared interacting galaxies at 86.8 Mpc, 
with a projected separation between the two nuclei of 6 kpc. 
The western progenitor, IC 1623 W, is very bright in the UV and optical, 
with many star clusters detected in HST images. Its properties 
are very similar 
to Lyman Break galaxies, and it is considered as 
the nearest Lyman Break Analog \citep{grimes2006}. 
In this paper we are concentrating on the western progenitor, 
which is covered by our IFS data. The top left panel 
of Figure 1 shows the HST F435W image of IC 1623, with
rectangles indicating the positions where 
we have obtained IFS observations. 
In contrast, IC 1623 E is optically obscured but
very infrared-bright. 
About half of the warmer dust traced in the mid-IR 
is associated with the eastern galaxy, where both 
compact (nuclear region) and extended emission is found 
\citep{lefloch2002}. 
Evidence for wide-spread star formation activity and shocks 
across the entire system is found in the UV, optical, 
and mid-IR \citep{knop1994,goldader2002,rich2011}.
Analysing far-UV and X-ray data, 
\cite{grimes2006} found a galactic 
superwind in IC 1623 moving at 300--400 km$\,$s$^{-1}$. 
Assuming a stellar origin for the hard X-ray emission 
(i.e., no contamination by an AGN), they estimate a similar SFR for both 
progenitors: $\sim$28 M$_{\odot}$ yr$^{-1}$ for IC 1623 E 
and $\sim$33 M$_{\odot}$ yr$^{-1}$ for IC 1623 W. 
This is probably an upper limit in the case of 
IC 1623 E, as a highly obscured AGN was identified 
in sub-arcsecond resolution ALMA observations of HCN (4-3) and 
HCO$^{+}$ (4-3) \citep{iono2013}.

\subsection{NGC 6090}\label{1.2}
NGC 6090 (Mrk 496 = UGC 10267; 
L$_{IR}$ = 3.2 $\times$ 10$^{11}$ L$_{\odot}$; e.g., \citealt{sanders2003}) 
is another pair of luminous infrared interacting galaxies
at 127.7 Mpc. The bottom left 
panel of Figure 1 shows the HST F435W image of NGC 6090.
At optical and NIR wavelengths NGC 6090 appears 
as a double nucleus system separated by 3.2 kpc in projection \citep{dinshaw1999}. 
The two galaxies are of roughly equal size, with a pair of 
tidal tails that have a full extent of $\sim$ 50 kpc. The northeastern 
galaxy (NGC 6090 NE) has a clear distorted spiral structure viewed 
face-on, while the  southwestern nucleus (NGC 6090 SW) is seen edge-on, with an 
amorphous morphology and no spiral arms visible.
The most striking feature of NGC 6090 is the abundance of star forming 
knots in the overlapping region, 
on the western side of NGC 6090 NE, and the paucity of similar clusters 
in its eastern spiral arms. 
Moreover, far- and near-UV spectroscopy 
shows that the NE nucleus is also dominated by a young 
starburst \citep{gonzalezdelgado2008}.
NGC 6090 SW also shows a similar 
abundance of blue knots on its the eastern, 
and one extremely luminous knot at the northern end. 
The molecular gas has a single component that
peaks between the two galactic nuclei in the overlap 
region \citep{bryant&scoville1999,wang2004}, 
suggesting that the gas responds to the interaction 
faster than the stars. 
Although the two galaxies show strong H$\alpha$ emission 
in their nuclei, the H$\alpha$ emission appears to be 
brighter in the companion-facing side of each galaxy 
\citep{hattori2004}.
There is considerable evidence 
for starburst activity, but no evidence at optical or 
radio wavelengths for a compact AGN.

\section{Observations and Data reduction}\label{2}
\subsection{IFS Observations}\label{2.1}
The IFS observations we report here were taken using the 
Potsdam Multi-Aperture Spectrophotometer 
(PMAS) spectrograph \citep{roth2005} on the 
3.5-m telescope of the Calar Alto Observatory (CAHA), 
with the Lens Array (LArr) configuration. 
We have used a spatial magnification of 0.75$\tt{''}$/lens, 
covering a 12$\tt{''}$x 12$\tt{''}$ field of view (FoV).  
From the standard stars and  also the CAHA seeing 
monitor observations, we estimate that the seeing was in the range 
FWHM$_{PSF}$ = 1.3--2.0$\tt{''}$ for the observations. 
The V300 grating was used, providing a 3.2 $\AA$/pixel dispersion and 
covering a wavelength range 3700--7100 $\rm \AA$ with a resolution 
of 7.1 $\rm \AA$ FWHM (velocity resolution $\sigma \sim$ 170 km s$^{-1}$). 
For each system two pointings were taken, as shown in Figure 1 
by black rectangles over the HST images, with a total exposure 
time of 3h and 1.75h for NGC 6090 and IC 1623 W, respectively.

\subsection{IFS Data Reduction}\label{2.2}
The PMAS LArr data reduction was performed using the R3D 
package \citep{sanchez2006}, 
in combination with IRAF and our own Python scripts.
The reduction consisted of the standard steps for fibre-based integral-field 
spectroscopy, including bias subtraction, and a flat-field correction that was
based on a master CCD flat provided by the CAHA. 
Calibration frames (continuum and Hg/Ne lamps) were taken before the science 
frames to correct the for the effects of flexure. 
The location of the spectra in the CCD was determined using a 
continuum illuminated exposure. Each spectrum was extracted from 
the science frames by coadding the flux within an aperture of 5 
pixels (3.75$\tt{''}$) around this location along the cross-dispersion 
axis for each pixel in the dispersion axis.
Wavelength calibration was performed using the Hg and Ne lamp exposures.
The final accuracy of the wavelength calibration is $\sim$ 0.38 $\rm \AA$. 
Differences in the fibre-to-fibre transmission throughput were corrected 
by comparing the wavelength-calibrated science frames with 
twilight-sky exposures.
The spectra were then flux calibrated using the 
spectrophotometric standard stars Feige34 and BD+28D4211, observed in the same night. 
We corrected the standard stars datacubes from differential 
atmospheric refraction (DAR, \citealt{filippenko1982}).
The median sky spectrum was obtained from the sky background frames,
and subtracted from the target spectra.
Once we had flux-calibrated
and sky-subtracted all the individual frames for the
same pointing position, they were combined and cosmic 
rays were rejected using IRAF's $\tt{IMCOMBINE}$ task.
Datacubes were created, corrected from DAR, and rotated to the North up, 
East to the left orientation. 
The two exposures at different locations 
were joined together in a mosaic.

Finally, we flux-recalibrated 
our LArr data using broad-band imaging from HST WFC435W 
in the case of IC 1623, and HST WFC435W together with
SDSS g and r filters for NGC 6090. 
From the deviation between the photometry obtained from broad-band 
imaging (in a 4.5 arcsec aperture) and the photometry extracted 
from the IFS in the same aperture, we found, in both systems, an average 
accuracy of the spectrophotometric calibration of 
$\sim$ 15 $\%$ across the wavelength range covered by our data set.

\subsection{HST imaging data}\label{2.3}
In addition to our IFS data, we downloaded from the Hubble 
Legacy Archive (HLA)\footnote {\url{http://hla.stsci.edu/}} 
multiwavelength high resolution images from the Hubble Space 
Telescope (HST) in several broad band 
filters from FUV to NIR: 
STIS F25SRF2, STIS F25QTZ, ACS F435W, ACS F814W, 
NICMOS F110W and NICMOS F160W for IC 1623, and 
ACS F140LP, ACS F330W, ACS F435W, ACS F814W, 
NICMOS F110W and NICMOS F160W for NGC 6090.
 We used these images to perform aperture photometry on the star clusters 
present in these early-stage mergers and roughly determine the 
stellar populations properties.
All these images were retrieved in their pipeline reduced state, 
astrometically corrected, and aligned North up, East left.
Their characteristics are summarized in Appendix A, in 
Tables A.1 and A.2 for IC 1623 and NGC 6090, respectively.

\subsection{Overview of the data}\label{2.4}
In the right panels of Figure 1 we show the PMAS LArr continuum
flux at 5110 $\rm \AA$ rest-frame (obtained by averaging between 
5050 to 5170 $\rm \AA$). 
We have superimposed in contour form the HST F435W image, smoothed to match 
the spatial resolution of the IFS ($\sim$ 1.7 arcsec). 
To better identify the main regions, in the left panels of Fig. 2 we show the 
HST F435W image in greyscale with the IFS continuum superimposed as red contours, 
along with some spectra extracted in circular apertures of $2 \tt{''}$ radius. 

In the case of IC 1623 W, the nuclear region, 
the emission peak C1 (which corresponds to a giant star forming region), 
and the distorted spiral arm are labelled. The C1 spectrum 
is dominated by the emission lines and has a blue continuum. We 
detect broad He II 4686 emission, which is a signature of Wolf Rayet 
stars in this system \citep{crowther2007}.
In the nucleus the line emission is much weaker, and the high order Balmer 
lines are visible in absorption, characteristic of intermediate age 
(100 Myr to 1 Gyr) stellar populations.

In the case of NGC 6090, the NGC 6090 NE nucleus is labelled 
in the bottom left panel of Fig. 2 (and marked with a black dot 
in the bottom right panel of Fig. 1).  
To its west, there is the region 
of multiple star clusters (MSC). 
In NGC 6090 SW we considered the position 
of the peak of the IFS continuum to represent the nucleus
(marked with a black dot in right panel of Figure 1). 
The spectra in the inner 2$\tt{''}$ of both nuclei are dominated by  
line emission, with the SW nucleus having a bluer continuum.

We also note that the brightest point 
source in the HST images (both optical and NIR) of NGC 6090 SW, 
is the knot located to 
the northwest of the progenitor, as previously reported 
by \cite{dinshaw1999}. These authors also found a mismatch between 
the radio emission peak and this knot, and suggested that it may 
not be the nucleus of the SW progenitor but a foreground star instead. 
The IFS has allowed us to confirm that this knot is not a 
foreground star, because the main absorption lines are at 
the redshift of the galaxy. That is the reason why we cannot 
discard it as the true nucleus of NGC 6090 SW, 
and its position will henceforth be shown in the maps with 
a star symbol. We discuss the minor effect this misidentification 
would have on our results in Appendix B.

\section{Star cluster photometry}\label{3} 

Using the HST images from FUV to NIR, 
we performed aperture photometry of the super star clusters 
in IC 1623 and NGC 6090. This was to make a rough estimation 
of their stellar population properties.
Clusters were detected with 
IRAF's  DAOFIND task, and photometry performed using the phot 
task in APPHOT. 

According to \cite{alvensleben2004}, a long-wavelength 
baseline from U-band through to the NIR is necessary to 
determine the properties of star clusters in a 
reliable way, where the availability 
of U-band photometry is crucial. 
In fact, we find that the best colour-colour 
diagrams to break the age-extinction degeneracy are 
FUV-F435W vs. F435W-F814W and 
F435W-F814W vs. F814W-NIC160W. 
A total of 228 and 156 clusters were detected
in IC 1623 and NGC 6090, respectively, with a 
signal-to-noise ratios SNR $>$ 5 in FUV, F435W, and 
F814W filters.
Analogously, 225 and 117 clusters were detected 
in IC 1623 and NGC 6090 with a
SNR $>$ 5 in F435W, F814W, and F160W filters.

All the clusters with positive detections it at least 
one of the diagrams are shown in Figure 3.
In the case of IC 1623, the obscured population of 
clusters in IC 1623 E is shown in red, 
while that of IC 1623 W is shown in blue. 

The star clusters properties (ages, masses, and the amount 
of dust obscuration) 
are estimated by comparing their colours and magnitudes 
with solar metallicity Charlot \& Bruzual (2007, unpublished) 
simple stellar population (SSP) models.
Model colours have been computed using STSDAS.SYNPHOT software. 

Details of the methodology, together with the colour-colour, 
colour-magnitude diagrams are presented in the Appendix A. 
Here we summarize the main results.

\begin{figure}  
\begin{center}
\includegraphics[width=0.45\textwidth]{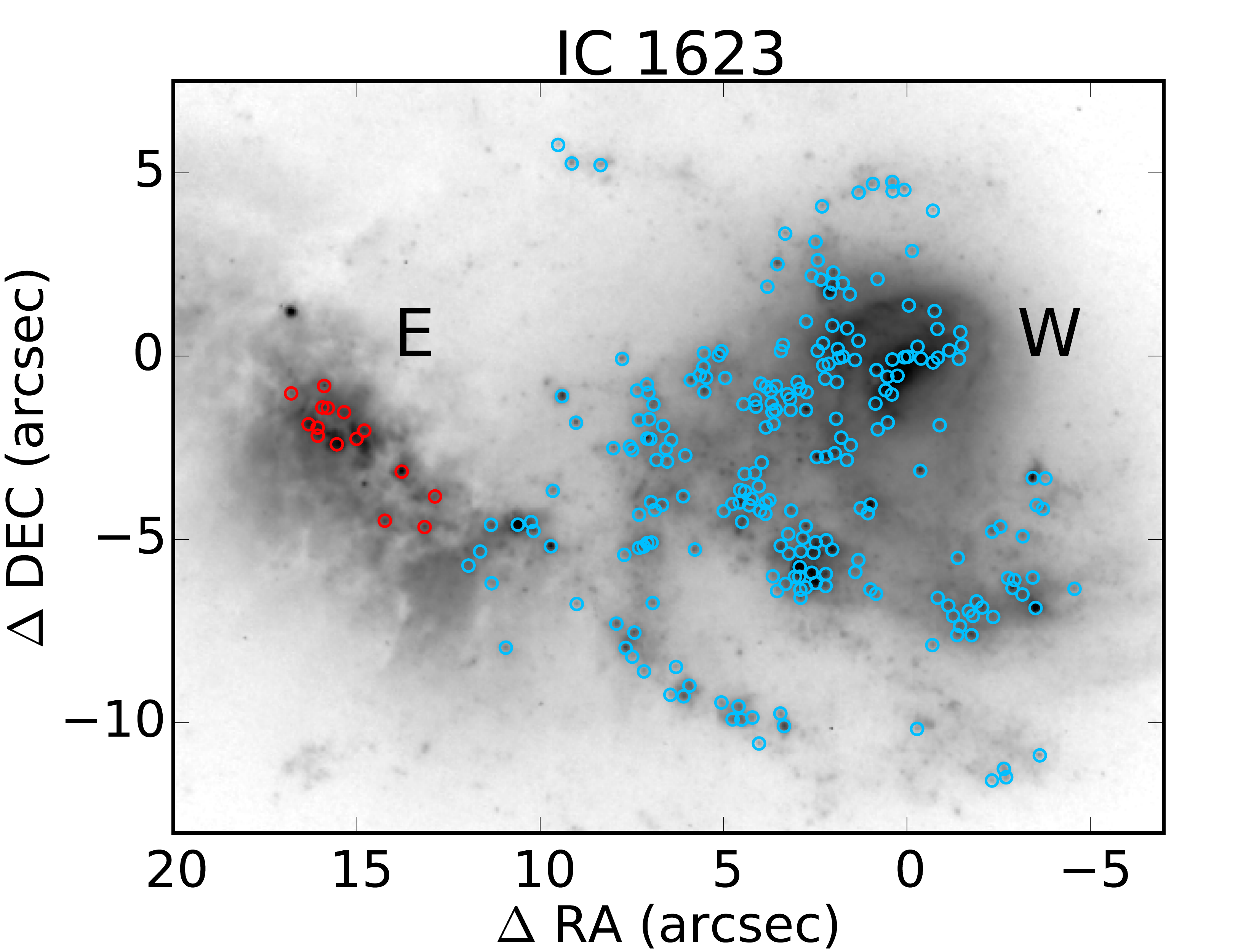}
\includegraphics[width=0.45\textwidth]{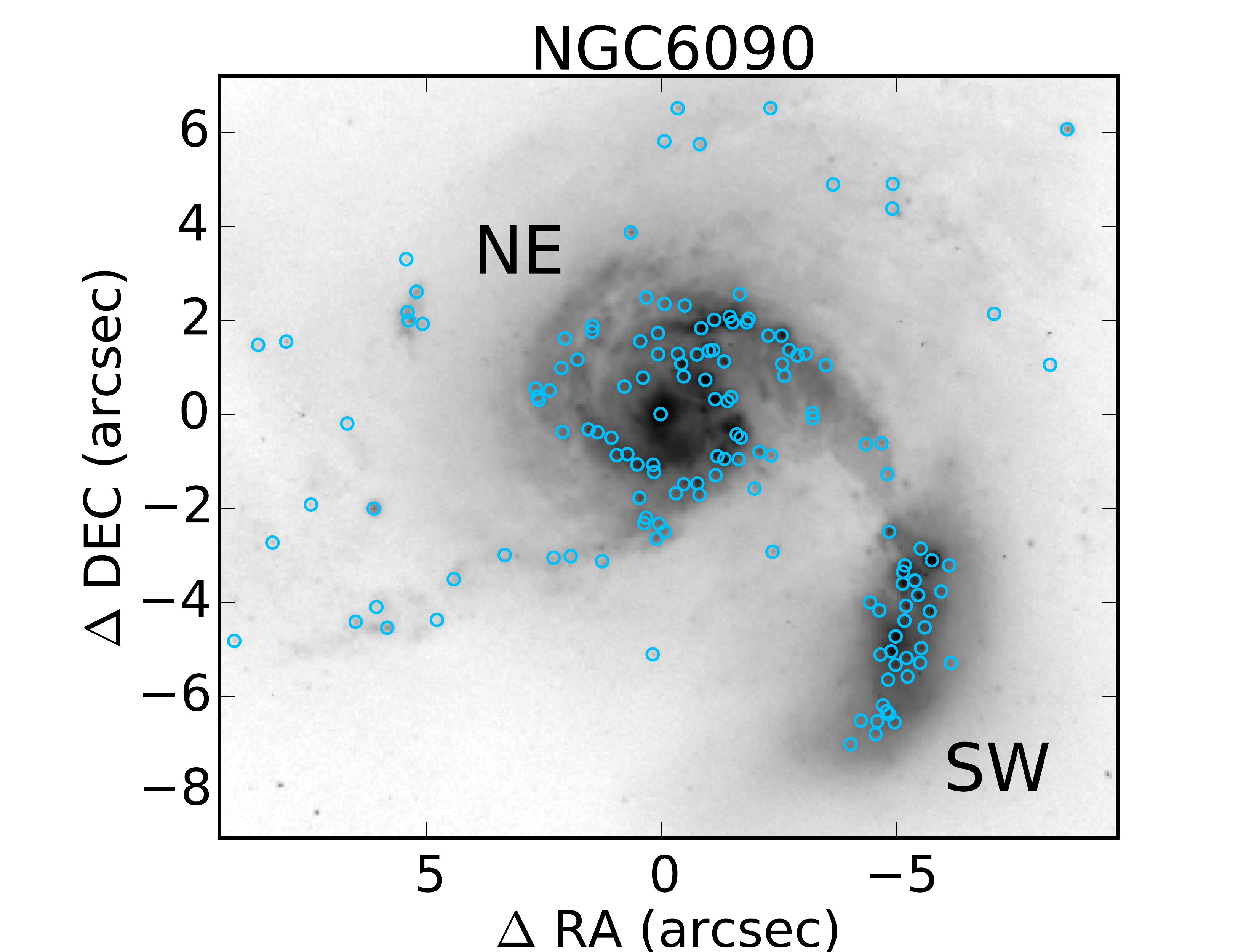}
\caption{F814W images of IC 1623 (top panel) and NGC 6090 (bottom panel). 
North is up, East to left. The detected clusters 
are marked with blue circles. In the case of IC 1623, 
we used red colour for the 
clusters in IC 1623 E, and blue for the ones in IC 1623 W, 
because they clearly represent two different populations 
in the colour-colour and colour-magnitude diagrams.}
\label{Fig_3}  
\end{center}
\end{figure}
 
\textbf{IC 1623}. Most IC 1623 W clusters are very young ($<$ 200 Myr) 
and affected by low to almost no foreground extinction ($<$ 1 mag). 
In contrast, the clusters in IC 1623 E are affected by 2--6 mag 
of extinction.
For the only IC 1623 E cluster emitting in FUV, we derive an age below 10 Myr, 
and A$_{V}$ = 2--3 mag. This suggests that, although affected
by large amounts of dust extinction, some clusters in IC 1623 E are as 
young as or younger than those in IC 1623 W.
Moreover, the  clusters in IC 1623 W have masses between 
10$^{5}$ and 10$^{7}$ M$_{\odot}$, while those in IC 1623 E
they are more massive $\sim$ 
10$^{7}$--$10^{8}$ M$_{\odot}$ \footnote{Probably knots with 
masses higher than 10$^{7}$ M$_{\odot}$ are clumps of several clusters 
rather than individual unique clusters.}. 
The two clusters with masses as high as 
10$^{9}$ M$_{\odot}$ are spatially coincident 
with the two nuclei in IC 1623 E reported by \citep{knop1994}.
A rough estimate of the total mass 
in clusters in IC 1623 W is 
$\sim$ 9 $\times$ 10$^{8}$ M$_{\odot}$. 

\textbf{NGC 6090}. Most of the clusters in NGC 6090 are also 
young ($<$ 300 Myr) and affected by $<$ 1 mag  extinction, 
but some of them could have A$_{V}$ up to 2 mag.
In general, NGC 6090 clusters 
have masses between 10$^{5}$ and 10$^{7}$ M$_{\odot}$. 
We find that the two ``clusters'' with higher masses (around 
10$^{7}$--10$^{8}$ M$_{\odot}$) are NGC 6090 NE nucleus and 
NGC 6090 SW nucleus.
The total stellar mass in clusters in NGC 6090 is 
$\sim2\times 10^9$ M$_{\odot}$.

As noted in Section \ref{A.5}, star cluster 
photometry allow us to trace the star formation which 
occurred in the last $<$ 500 Myr -- 1 Gyr (at best, for no extinction), 
but not in previous epochs. 
Although spectral synthesis reveals that 
intermediate-age stellar populations are not important 
contributors in these galaxies (Fig. 8, Section \ref{4.7}), 
the star cluster detection limits are relevant to be known 
when studying more advanced merger stages, where a significant 
intermediate-age population could be present. 

\section{Stellar populations}\label{4}
In this section we use the stellar continuum shape 
to characterize the spatially resolved stellar population 
properties and constrain the star formation histories
of IC 1623 and NGC 6090.

In order to quantify the effects of mergers 
on the  properties of the stellar populations, 
the results of our LIRGs are compared to control samples of CALIFA 
spirals from \cite{gonzalezdelgado2015}, which have been calibrated with 
version 1.5 of the reduction pipeline, 
explained in detail in the Data Release 2 article \citep{garcia-benito2015}.
The comparisons are made with Sbc/Sc spirals because the 
progenitor galaxies of LIRGs are gas-rich spirals without 
significant bulges \citep{sanders1996,kim2013}, and because they have 
stellar masses similar to IC 1623 and NGC 6090. 
We have defined two possible control samples. 
On the one hand, the average 
of all the Sbc galaxies in CALIFA, because they approximately
cover the same mass range as our LIRGs, with 
$\langle$$\log$ M$^{Sbc}_{\star}$(M$_{\odot}$)$\rangle$ = 10.70 $\pm$ 0.34. 
On the other hand, using later Sc galaxies, but averaging only 
those in a mass range compatible with our LIRGs, 
$\langle$$\log$ M$^{Sc}_{\star}$(M$_{\odot}$)$\rangle$ = 10.45--10.85. 
There are 14 Sc galaxies in the CALIFA sample meeting 
this requirement, and $\sim$ 70 Sbc galaxies.

\subsection{Methodology}\label{4.1}
Our method to extract the stellar population information 
from the IFS data cubes is based on the full spectral 
synthesis approach.
This technique has been extensively tested and applied 
\citep{cidfernandes2010,gonzalezdelgado2010,
cidfernandes2013,perez2013,cidfernandes2014,gonzalezdelgado2014a}, 
and it has been proven to reduce 
the age-metallicity degeneracy \citep{sanchez-blazquez2011}.
It is also notable that, despite the diversity 
of spectral synthesis methods, substantial changes in the 
results are more likely to come from revisions in the input data 
and from updates in the base models, 
the single most important ingredient in any spectral synthesis 
analysis \citep{cidfernandes2013}. 
The general consensus in the field is that uncertainties 
in the results for individual objects average out for large
statistical samples \citep{panter2007}. With IFS, 
each galaxy is a statistical sample per se, 
and even if the results for single spaxels (or zones) 
are uncertain, the overall trends should be robust.

We have analyzed the stellar population properties 
with the {\sc starlight} code 
\citep{cidfernandes2005}, which fits an observed spectrum (O$_{\lambda}$) 
in terms of a model (M$_{\lambda}$) built by a non-parametric linear 
combination of N$_{\star}$ SSPs from a 
base spanning different ages ($t$) and metallicities ($Z$).  
Dust effects are modeled as a foreground screen with a \cite{calzetti2000} 
reddening law with R$_{V}$ = 4.5, assuming the same reddening 
for all the SSP components. 
Kinematical effects are also accounted for by assuming a Gaussian line-of-sight 
velocity distribution. The fits were carried out in the rest-frame 
3650--6950 $\rm \AA$ interval. Windows around the  
[OI], [OII], [OIII], [NII], HeI, and Balmer series
from H$\alpha$ to H$\epsilon$ emission lines were masked in all fits. 
Because of its interstellar absorption component, 
the NaI D doublet was also masked. 
The process is similar to that previously applied 
in CALIFA data \citep{cidfernandes2013,cidfernandes2014}.

The results reported in this paper rely
on the GM base  (\citealt{cidfernandes2013}, CF2013 hereafter)
 of SSP spectra, that 
combines the $\textsc{granada}$ models of 
\cite{gonzalezdelgado2005} for $t < 63$ Myr with those of 
\cite{vazdekis2010}. They are based on the Salpeter IMF.
It contains 156 SSPs, comprising four metallicities, 
$Z = 0.2$, 0.4, 1, and 1.6 Z$_{\odot}$, and 39 ages between
$t = 10^{6}$ and $1.4 \times 10^{10}$ yr.
For comparison purposes, we have also performed the spectral
synthesis using the CB model base,
which is built from an update of the \cite{bruzual&charlot2003} models, 
replacing STELIB by a combination of the MILES \citep{sanchez-blazquez2006}
and $\textsc{granada}$ \citep{martins2005} spectral libraries (the same ones used in base GM). 
The IMF is that of \cite{chabrier2003} and it comprises 160 SSPs of
four metallicities, $Z = 0.2$, 0.4, 1, and 2.5 Z$_{\odot}$, 
and 40 ages between
$t = 10^{6}$ and $1.4 \times10^{10}$ yr.

We discarded from the analysis all the spaxels with 
SNR $<$ 5 in the continuum, measured in the rest-framed window
between 5590 and 5680 $\rm \AA$. Also, the index 

\begin{equation} 
\overline{\Delta} = \frac{1}{N_{\lambda}^{eff}} \sum_{\lambda} \frac{|O_{\lambda}-M_{\lambda}|}{M_{\lambda}}
\end{equation} 

\noindent proposed by CF2013 has been used to quantify the quality of the spectral fits (see their Section 5.1).
Again, O$_{\lambda}$ and M$_{\lambda}$ are the observed and model spectra,
respectively, and the sum is carried over the N$_{\lambda}^{eff}$
wavelengths actually used in the fit, i.e., discarding masked, 
flagged, and clipped pixels. 
$\overline{\Delta}$ = 10$\%$ is a reasonable 
quality-control limit. 
Typical median values of $\sim$ 3$\%$ and $\sim$ 5$\%$ are found 
in IC 1623 W and NGC 6090, respectively. 
From now on, only the spaxels with $\overline{\Delta}$ below the control 
limit are shown in the maps, and considered in the analysis. 

\begin{figure*}  
\begin{center}
\includegraphics[width=0.7\textwidth]{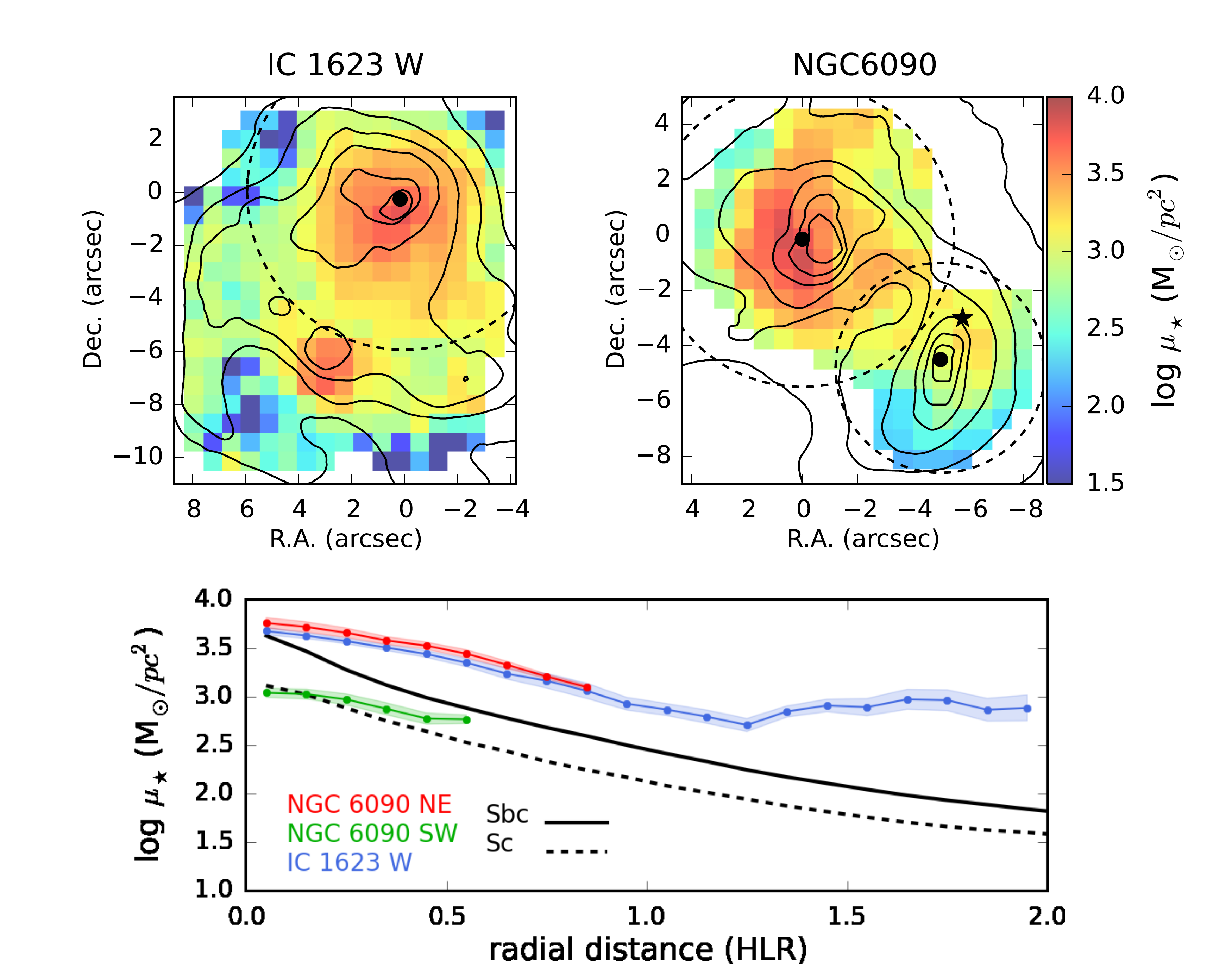}
\caption{Observed stellar mass surface density for IC 1623 (upper left) 
and NGC 6090 (upper right). 
The colour scale is the same for both maps. 
The black dashed lines indicate the position of 1 half-light radius (HLR) in IC 1623 W 
and NGC 6090 NE, and 0.5 HLR in NGC 6090 SW. 
The solid contours correspond to the smoothed HST F435W images. 
Lower panel: mean radial profiles of the stellar mass surface density 
as a function of the radial 
distance in HLR, in red for NGC 6090 NE, green for NGC 6090 SW, and
blue for IC 1623 W. The uncertainties, calculated 
as the standard error of the mean, are shaded in light red, green, 
and blue, respectively. 
For comparison, the black lines are the  average
profiles for Sbc (solid) and Sc (dashed) 
spiral galaxies in CALIFA (GD2015), with masses consistent with our LIRGs.}
\label{Fig_4}  
\end{center}
\end{figure*}

\begin{figure*}
\begin{center}
\includegraphics[width=0.7\textwidth]{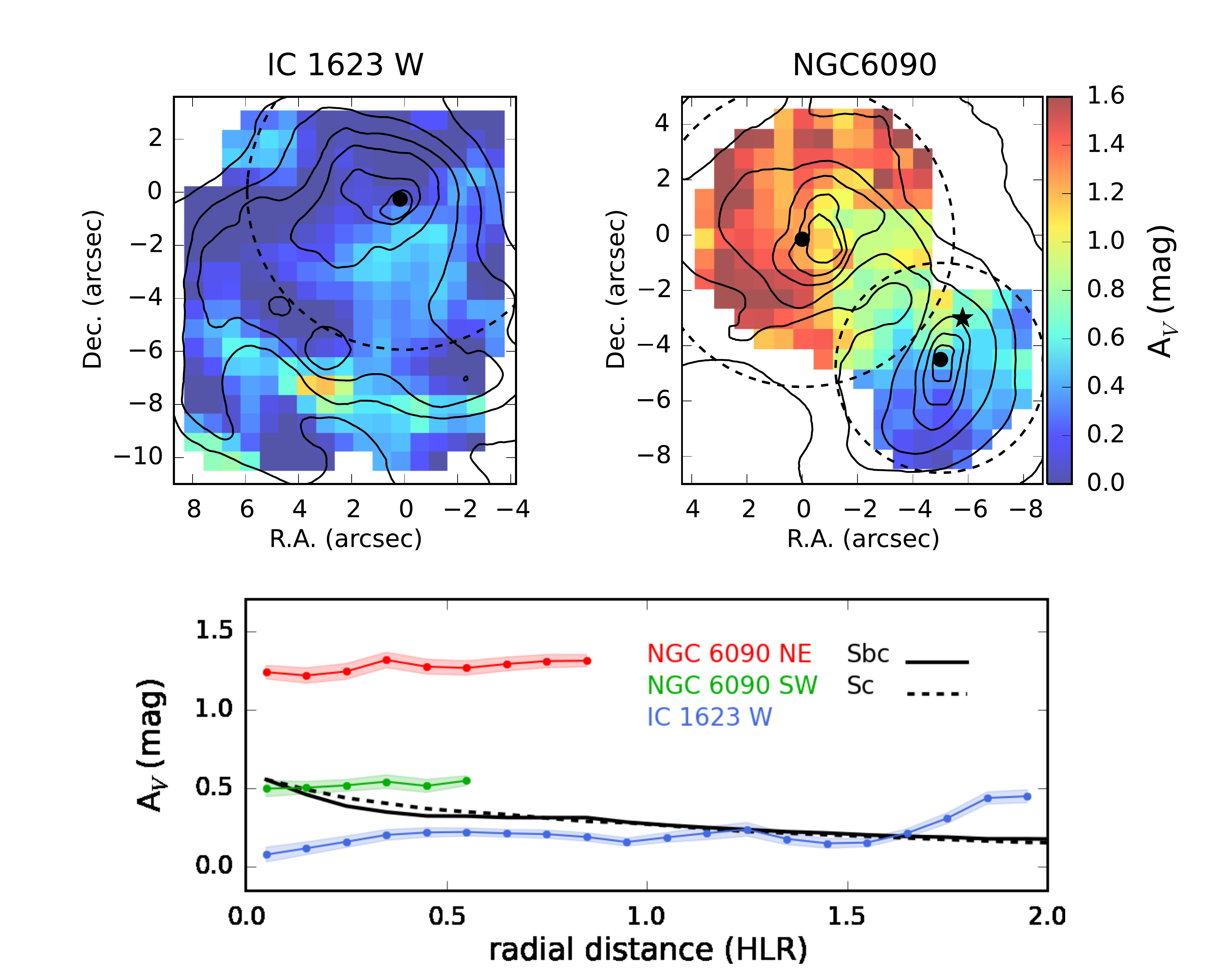}
\caption{As Figure 4 but for stellar dust extinction.}
\label{Fig_5}  
\end{center}
\end{figure*}

\begin{figure*}
\begin{center}
\includegraphics[width=0.7\textwidth]{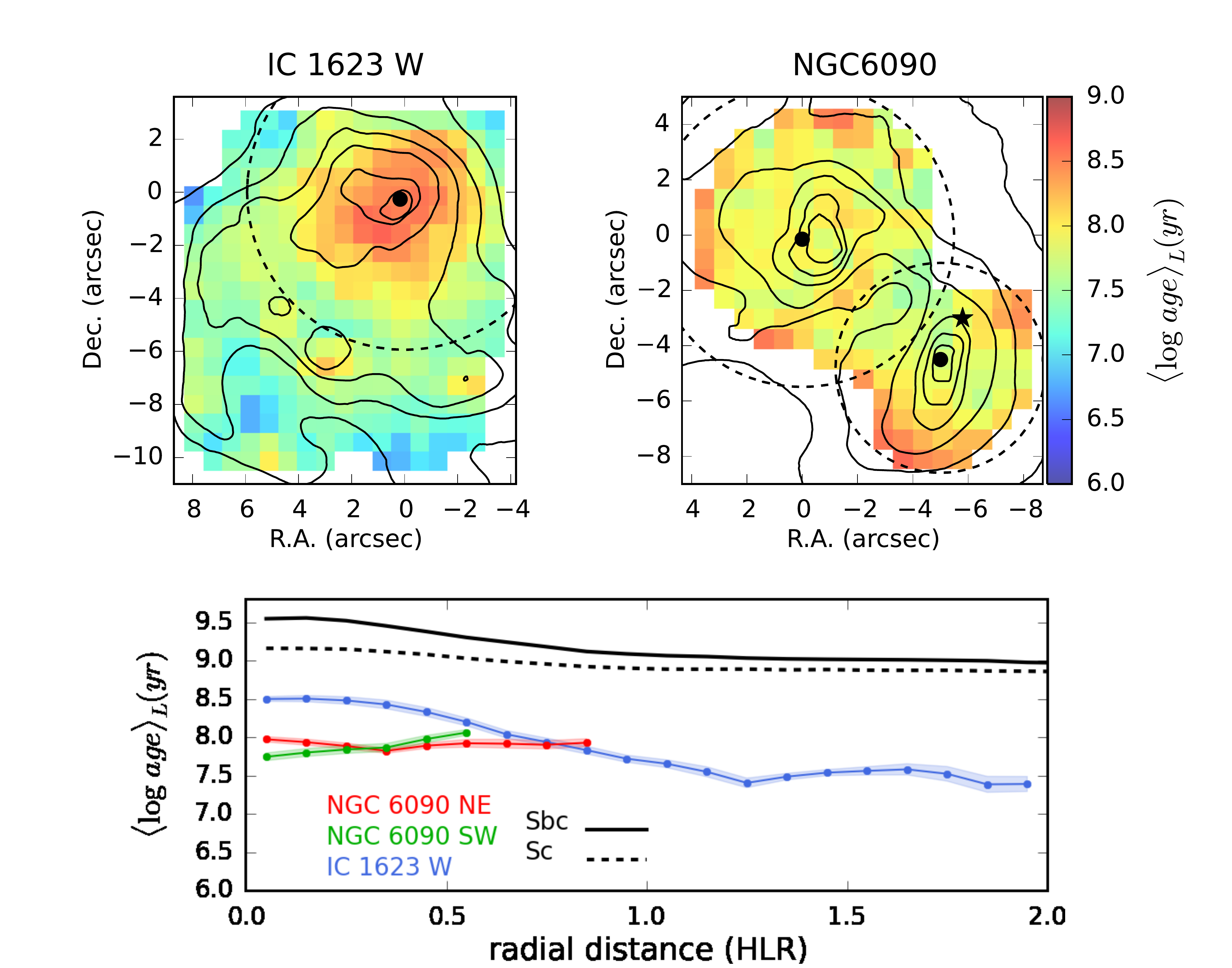}
\caption{As Figure 4 but for mean light weighted stellar ages.}
\label{Fig_6}  
\end{center}
\end{figure*}

\begin{figure*} 
\begin{center}
\includegraphics[width=0.7\textwidth]{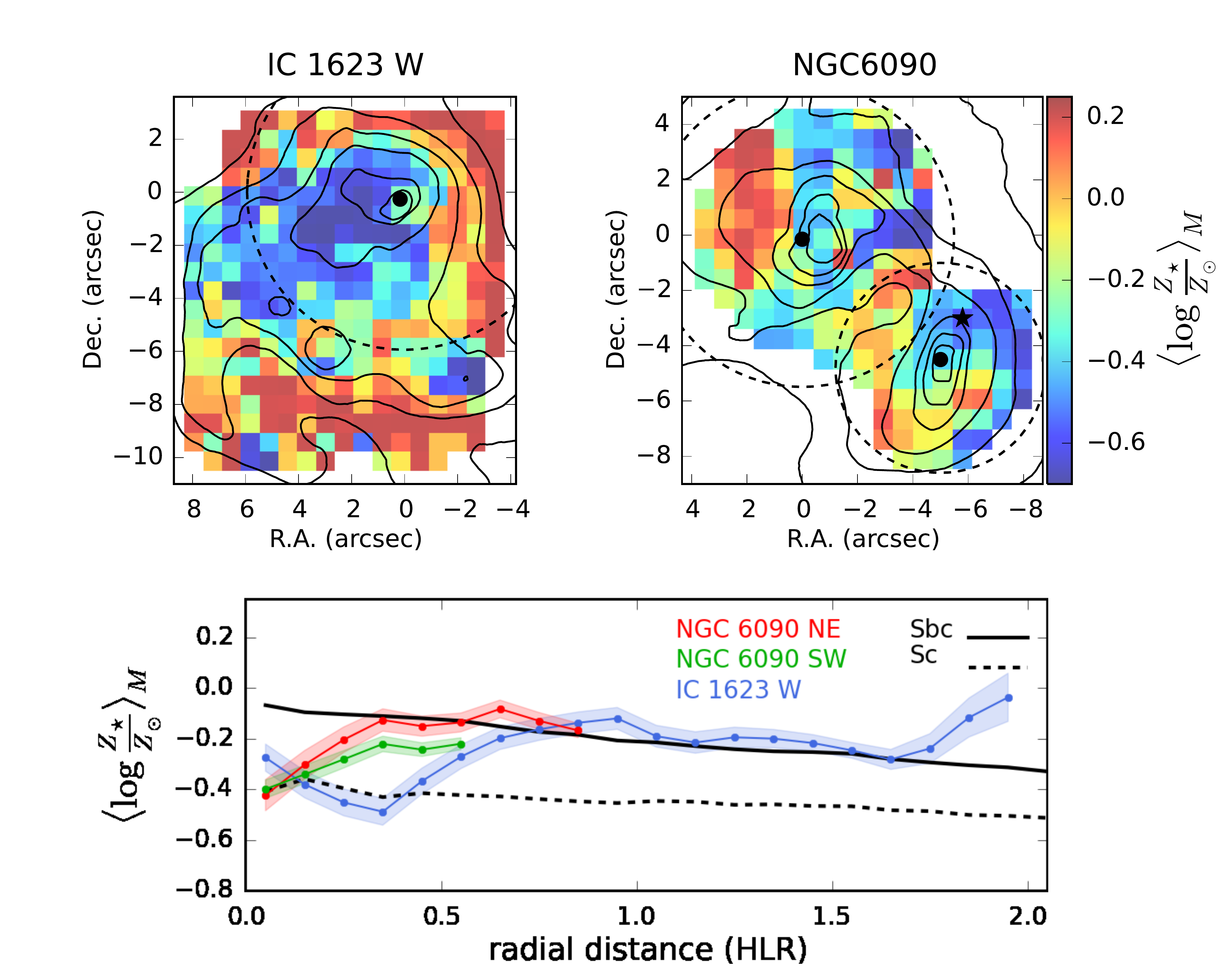}
\caption{As Figure 4 but for mean mass weighted metallicity in solar units.}
\label{Fig_7}  
\end{center}
\end{figure*}

\subsection{Stellar mass}\label{4.2}
An important {\sc starlight} output is the stellar 
mass, which is calculated taking into account the spatial 
variations of the mass to light ratio.
The total stellar mass obtained from the sum of all
spaxels is M = 3.9$\times$10$^{10}$ M$_{\odot}$
and 6.8$\times$10$^{10}$ M$_{\odot}$ for IC 1623 W and
NGC 6090 respectively (for Salpeter IMF).
This is the mass locked in stars today. 
Also counting the mass returned by stars to the interstellar 
medium  M' = 5.5$\times$10$^{10}$M$_{\odot}$ and 
9.6$\times$10$^{10}$M$_{\odot}$ were involved in 
star formation for IC 1623 W and NGC 6090, respectively.

Since in LIRGs the optical light is dominated by young stellar 
populations and the extinction may be important in some regions, 
we could be underestimating the mass of the  old stars.  
The old stellar populations emit the bulk 
of their light in the 1--3 $\mu$m near-IR window, 
and assuming that they are the principal contributors 
to the IR light, we have determined an alternative estimate of
the stellar mass using MIR data from Spitzer at 3.6 
and 4.5 $\mu$m, following the approach proposed 
by \cite{jarrett2013}. 
The global mass to light ratio 
at 3.6$\mu$m ($\zeta_{3.6}^{d}$) can be expressed as:
\begin{equation}
\log \ \zeta_{3.6}^{d} (M_{\odot}/L_{\odot}) = -0.22 + 3.42 ([3.6] - [4.5])
\end{equation}
where [3.6] and [4.5] are the magnitudes  in the  
3.6 and 4.5 $\mu$m bands. This expression was calculated
under several assumptions that we have to take into account. 
On the one hand, it considers 
the SSPs of \cite{bruzual&charlot2003} with a Chabrier IMF. 
On the other hand, it considers that the average mass-to-light 
ratio (at 3.6 $\mu$m) for strictly 
old stars is log $\zeta_{3.6} = -0.23 \pm 0.1$ dex, as
determined by \cite{meidt2012,meidt2014}, and valid in the 
range of metallicities covered by our models. 
With this procedure we found that the stellar mass derived 
from the MIR is 2.8$\times$10$^{10}$M$_{\odot}$ and 
4.2$\times$10$^{10}$M$_{\odot}$ for IC 1623 W and NGC 6090, 
respectively (Chabrier IMF), in good agreement with our 
stellar masses derived from the spectral synthesis, if MIR 
masses are scaled by a factor $\sim$ 1.8 as expected due 
to the change in the IMF.

This agreement between the mass estimations derived from 
optical spectra and MIR imaging confirms that for IC 1623 W and 
NGC 6090 we are not underestimating 
the stellar masses with spectral synthesis, and also indicates that 
the extinction estimates are approximately correct.

\subsection{Stellar mass surface density}\label{4.3}
In Figure 4 we show the stellar mass surface 
density ($\mu_{\star}$) maps 
for IC 1623 W (upper left panel) and NGC 6090 (upper right). 
The black dashed circles indicate the position of 1 half-light 
radius (HLR)\footnote{The HLR is defined 
as the axis length of the circular aperture which 
contains half of the total light of the galaxy at the 
rest-frame wavelength. It was obtained from the continuum 
flux image at 5635 $\rm \AA$ using the PyCASSO package.} 
in IC 1623 W and NGC 6090 NE, and 0.5 HLR in NGC 6090 SW. 
1 HLR $\sim$ 2.5 kpc, 3.4 kpc and 5.0 kpc for IC 1623 W, 
NGC 6090 NE and NGC 6090 SW, respectively, on average, while 
for the Sbc and Sc control galaxies, 1 HLR $\sim$ 5.0 kpc, 
and 4.1 kpc, respectively. 

The 2D maps of the stellar population properties were
azimuthally averaged to study their radial 
variations using PyCASSO (CF2013). 
Radial apertures 0.1 HLR in width were used to extract 
the profiles. 
Expressing radial 
distances in units of HLR allows the profiles of the early-stage mergers
to be compared on a common metric with the control samples 
of Sbc and Sc galaxies. 
The $\mu_{\star}$ profiles are shown in the lower panel, 
in red for NGC 6090 NE, 
NGC 6090 SW in green, and IC 1623 W in blue. 
The uncertainties are shaded in red, 
green, and blue, respectively. They represent the standard 
error of the mean, calculated as the standard deviation divided 
by the square root of the number of points in each distance bin, 
with 0.2 HLR of width. 
As it is normalized by the square root of the number of points/spaxels 
in each HLR bin, the differences in the uncertainties between the inner 
and outer regions are visually not so strong, as the larger the radius, 
the larger is N. In terms of dispersion (=standard deviation) we find 
that with respect to IC 1623 W centre (N=8), the dispersion is a factor 
2.3 larger at 0.5 HLR (N=41), 2.4 larger at 1 HLR (N=47), 
and 1.5 larger at 2 HLR (N=18). In the case of NGC 6090 NE, the dispersion 
is a factor of 2.3 larger at 0.5 HLR (N=42) than at the centre (N=8), 
and for NGC 6090 SW is a factor 2.5 larger at 0.5 HLR (N=74) 
than at the centre (N=12). 
The black lines are the profiles from Sbc (solid) and Sc (dashed) 
spiral galaxies from CALIFA (\citealt{gonzalezdelgado2015}, 
GD2015 hereafter).

The highest values of the mass surface density are reached 
in \textbf{IC 1623 W} nucleus and in the cluster 
concentration C1, with a logarithmic 
value of around 3.7 for $\mu_\star$ in units of $M_\odot\,$pc$^{ -2}$. 
From the radial profile we find a negative 
gradient with distance, going from 
log $\mu_{\star}$ = 3.7 at the nucleus 
to 2.8 at 1.2 HLR, then 
increasing again up to 
log $\mu_{\star}$ = 3.0 at 2 HLR.

In \textbf{NGC 6090}  the 
stellar mass surface density in the NE progenitor  is significantly 
higher than in the SW.
The region with the 
highest value, 
log $\mu_{\star}$ $\sim$ 3.8, 
is located to the south east of the NGC 6090 NE nucleus.
In NGC 6090 SW, log $\mu_{\star}$ is higher towards the north. 
From the radial profile, we also found
a negative gradient of the stellar mass density 
with distance:
log $\mu_{\star}$ goes from 3.8 to 3.1 at 0.9 HLR for 
NGC 6090 NE, and from 3.0 to 2.8 at 0.6 HLR, for NGC 6090 SW.

The central density in IC 1623 W and NGC 6090 NE
is similar to the central density of Sbc galaxies, while 
NGC 6090 SW is more similar to Sc. 
We have compared the inner gradients of the stellar 
mass surface density (from 0 to 1 HLR, as defined in equation 
6 of GD2015) in early-stage mergers and in 
the control spirals. We find
$\Delta_{in}$ log $\mu_{\star}$ = $-0.66$, $-0.79$, $-0.99$, $-1.15$ for 
NGC 6090 NE (in 0.9 HLR), IC 1623 W, Sc, and Sbc, respectively. 
In the case of IC 1623 W we can also compare the 
outer gradient (from 1 to 2 HLR, defined as in equation 
7 of GD2015), 
$\Delta_{out}$ log $\mu_{\star}$ = 0, $-0.54$, $-0.64$ for 
IC 1623 W, Sc, and Sbc galaxies, respectively.
We find that the early-stage merger radial profiles of $\mu_\star$  are flatter 
than in spirals, but not so significantly if we consider the dispersion 
of the control samples, between 0.24 and 0.28 dex (GD2015).
Moreover, we note that given the recent star formation 
in IC 1623 W, its
HLR is $\sim$ half the HLR of the control samples, 
and the apparent flattening of the mass density profile 
may be also due to this, instead of representing a
real evolutionary change, at least in the inner 1 HLR. 
This does not affect NGC 6090 NE $\&$ SW, as their HLRs are
similar to Sbc/Sc galaxies.

\subsection{Stellar dust extinction}\label{4.4}
{\sc starlight}-based A$_{V}$ maps are shown 
in Figure 5. 
The scale is the same for both systems
to facilitate the comparison.

\textbf{IC 1623 W} (upper left) has little extinction, 
with A$_{V}$ below 0.4 mag nearly everywhere.
From the radial profile, we found a slight but nonuniform 
increase of the average extinction from 0.06 mag in 
the nucleus to 0.2 mag at 1.5 HLR. 
The regions most affected by extinction are located 
near areas where HII regions accumulate, 
such as C1, where the extinction reaches values up to 1.2 mag.
As an explanation we suggest that the enhancement of A$_{V}$
could be due to the accumulation of dust swept by the winds and 
supernovae of the star forming regions, at least in 
some locations. Another possibility is that the interaction 
itself has led to the accumulation of dust in these regions. 

In \textbf{NGC 6090 NE} (upper right), 
A$_{V}$ ranges from 1.2 to 1.6 mag, while in 
NGC 6090 SW it runs from from 0.6 mag in the North to 
about 0.2 mag in the South. 
To the West of NGC 6090 NE and in the bridge 
between the two galaxies the dust extinction has an 
intermediate value of 0.8--0.9 mag. 
There is approximately 1 mag difference in
the stellar dust extinction between the two progenitors. 
When averaging (lower panel), 
we do not find significant gradients 
of the dust distribution across the main bodies of the 
individual galaxies, reflected in the nearly flat 
radial profiles.
We note that in these two early-stage merger systems 
one of the progenitors is significantly more obscured 
than the other. In these cases it happens that 
the eastern progenitor is more obscured, 
although the difference is higher for IC 1623 
than for NGC 6090.

The radial profiles of A$_{V}$ in IC 1623 and NGC 6090, 
are flat. This is why the $\mu_{\star}$ profiles shown 
in the previous section 
resemble the surface brightness profiles (not shown). 
In contrast, typical spiral galaxies show a negative radial 
gradient independently of the morphology (GD2015).  

\subsection{Ages}\label{4.5}
The simplest way to quantify the star formation 
history (SFH) of a system is 
to compress the age distribution encoded in the light 
population vectors to their first moments. 
For this purpose we will use the following definition 
for the mean stellar age:
\begin{equation}
\langle \log \ age \rangle_{L}=\sum_{t,Z} x_{t,Z} \log t 
\end{equation}
where $x_{t,Z}$ is the fraction of light at the
normalization wavelength (5635 $\rm \AA$) due to the base 
population with age $t$ and metallicity $Z$. 

The mean stellar age maps are shown in Figure 6. 
The scale is the same  for both maps
to facilitate the comparison and goes 
from $\langle \log \ age \rangle_{L}$ = 6.0 to 9.0.

In \textbf{IC 1623 W} the nuclear regions 
are clearly older than the outer parts. From the radial 
profile we find a negative trend of the mean age 
with distance: from $\sim$ 300 Myr in the nucleus to
$\sim$ 30 Myr 1.2 HLR away. From 1.2 to 2 HLR
the age remains approximately constant at 15--40 Myr, 
indicating the IC 1623 W disc is still forming stars.

In \textbf{NGC 6090} the older spaxels are located in the outskirts 
for both progenitors, but especially for the SW galaxy, 
where they reach ages up to $\sim$ 300 Myr.
The youngest spaxels ($\sim$ 50 Myr) are 
located in the bridge. 
When averaging we found that both progenitors
have the same mean ages, $
\langle \log \ age \rangle_L \sim 50$--100 Myr.

In both early-stage merger systems the 
age profiles are 
significantly flatter than in Sb to Sc galaxies 
(GD2015). 
Even for IC 1623 W, where we find a negative 
gradient of -280 Myr in the inner 1 HLR, it 
is still much lower than the -2.4 Gyr(-620 Myr) gradient 
in the inner 1 HLR of Sbc(Sc) galaxies.

\begin{figure*}  
\begin{center}
\includegraphics[width=0.56\textwidth]{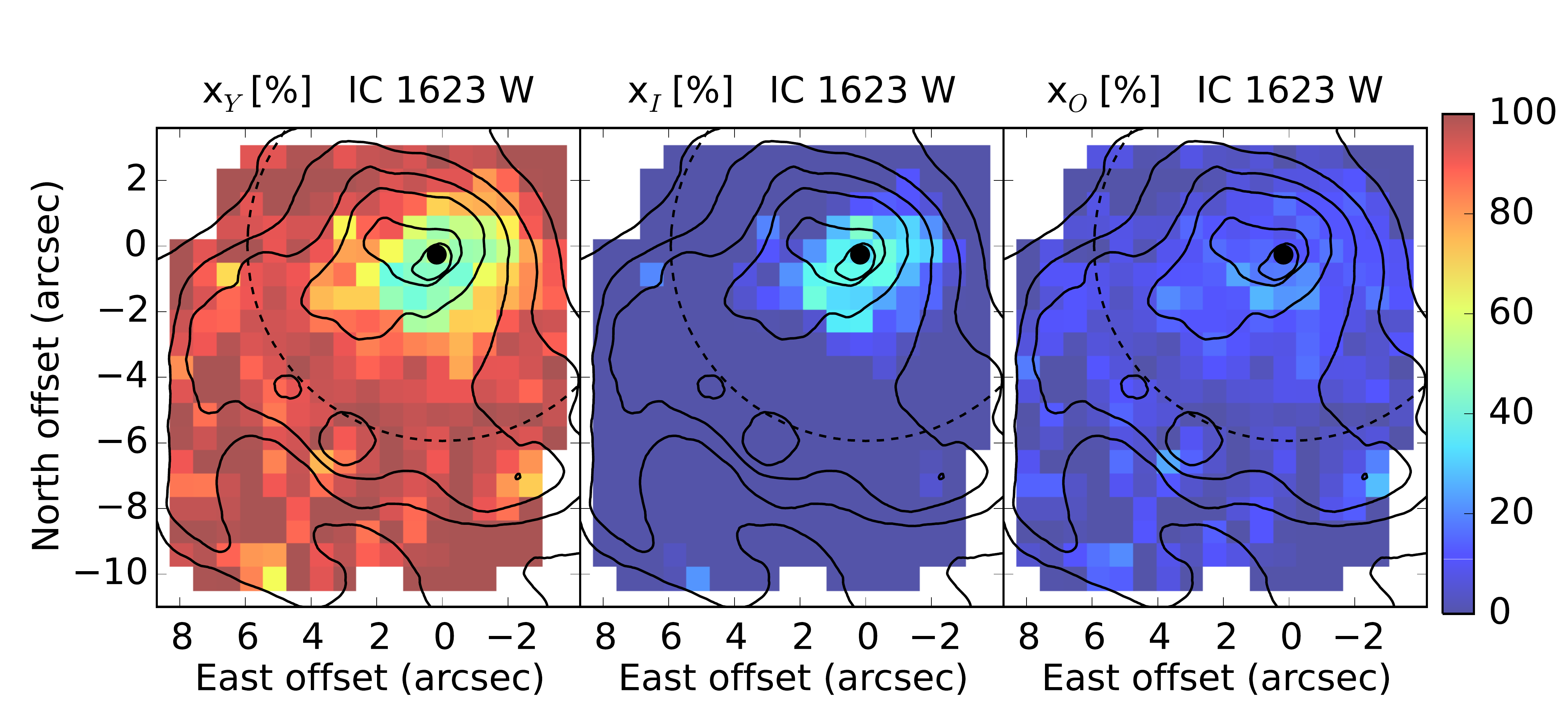} 
\includegraphics[width=0.43\textwidth]{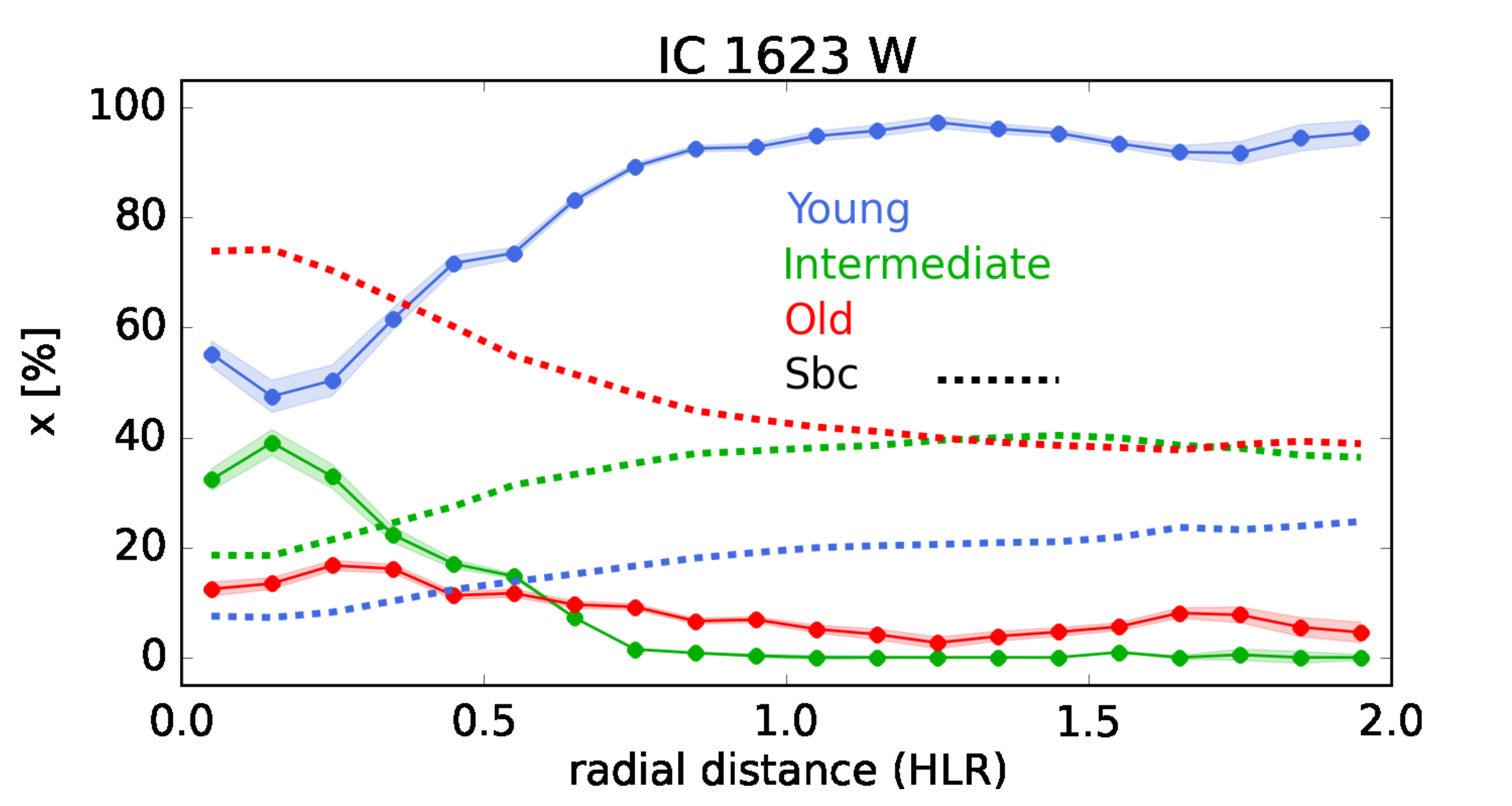}\\
\includegraphics[width=0.56\textwidth]{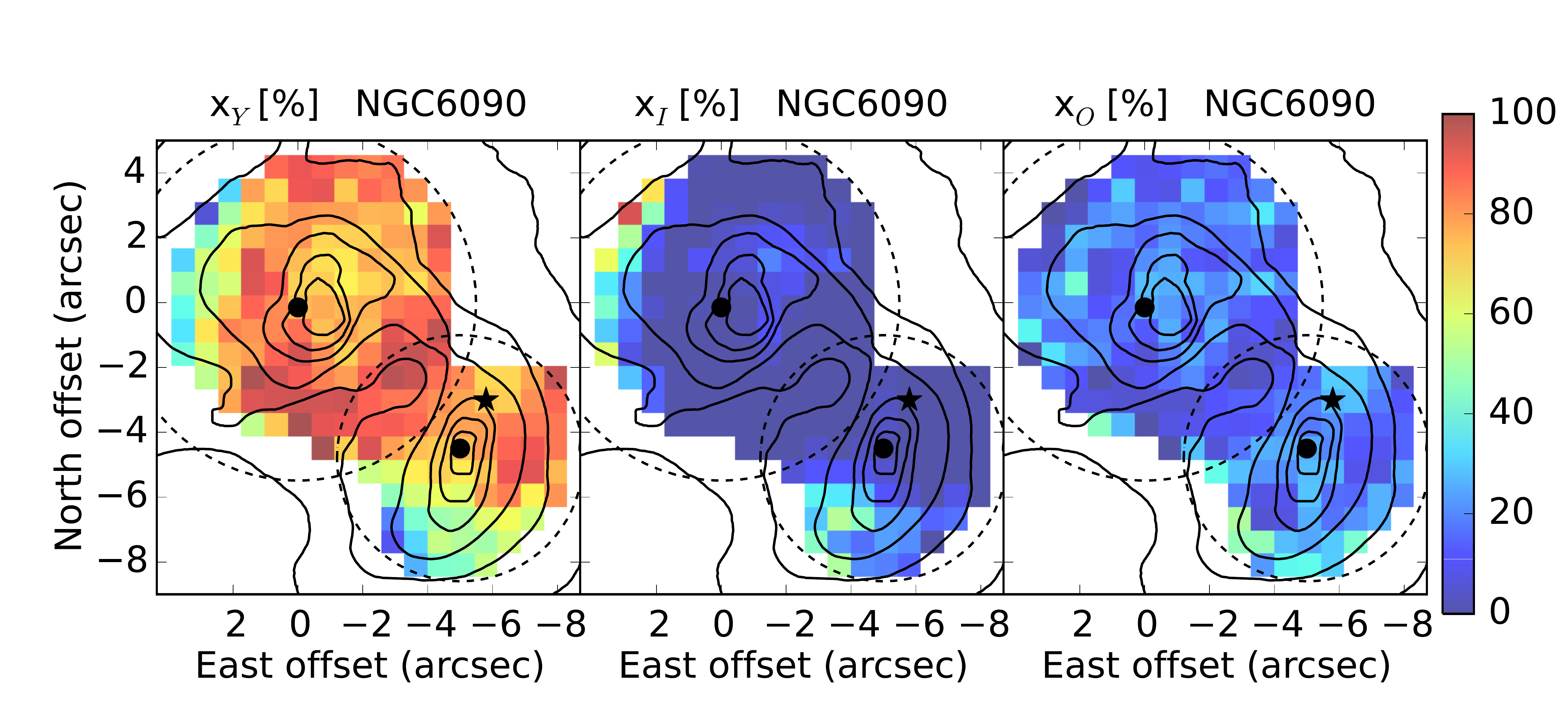} 
\includegraphics[width=0.43\textwidth]{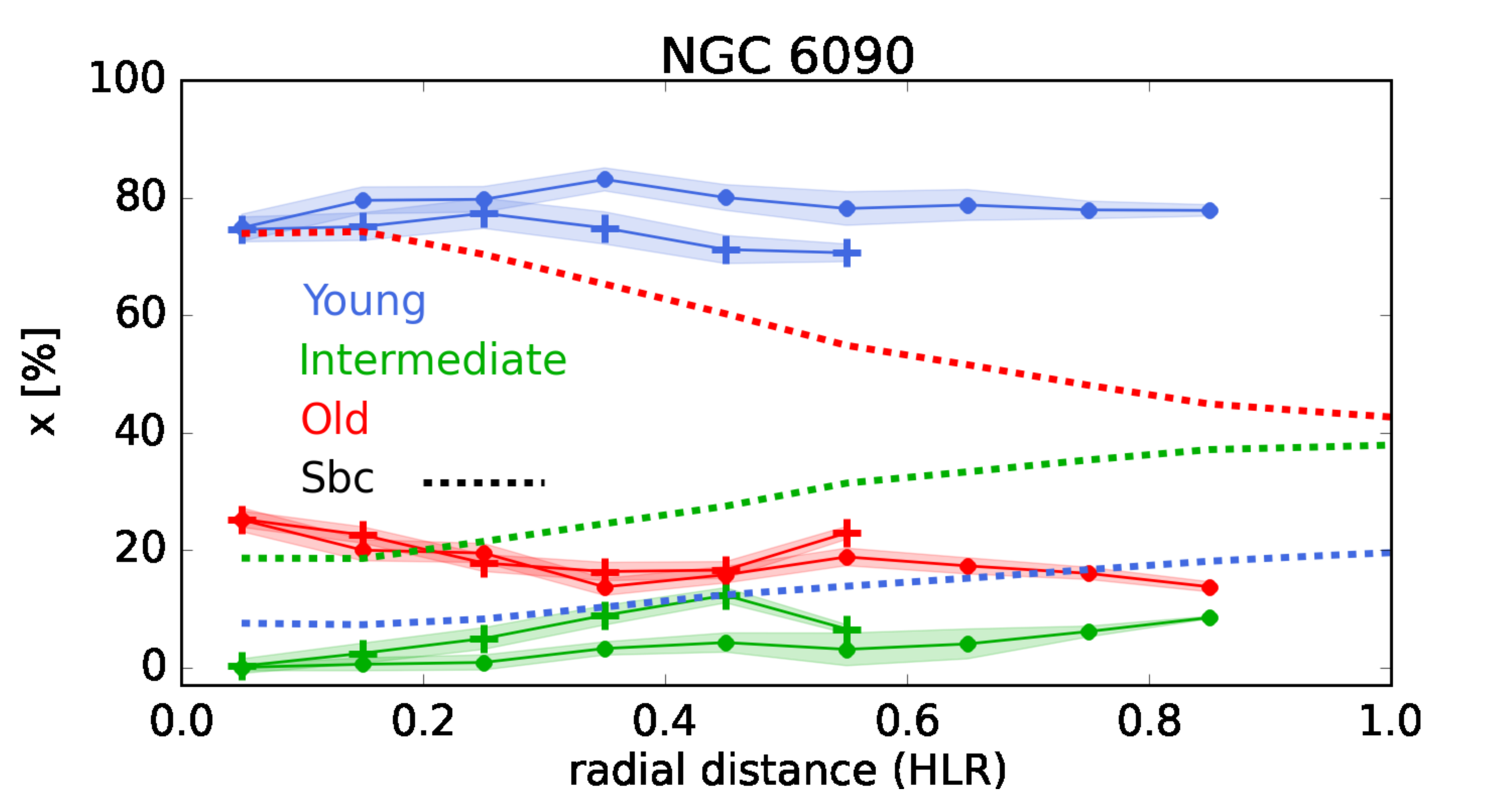}\\
\caption{Left panels: Contribution in percentage of 
(a) young t $<$ 140 Myr , (b) intermediate (140 Myr $<$ t $<$ 1.4 Gyr), and 
(c) old  t $>$ 1.4 Gyr stellar populations to the observed light at 5635 $\rm \AA$.
Right panels: Radial profiles of the light contributions from 
(a) young x$_{Y}$ (blue), (b) intermediate x$_{I}$ (green), 
and (c) old populations x$_{O}$ (red), with the radial distance 
in HLR. For comparison, the dotted lines with the same colours are the average profiles 
for the Sbc galaxies in CALIFA.
Top panels: IC 1623 W. Bottom panels: NGC 6090. The NGC 6090 NE radial 
profile has circle symbols, while the NGC 6090 SW 
profile has plus symbols.
We find that both early-stage merger systems are clearly
dominated by young stellar populations 
across the full observed extents of the systems.}
\label{Fig_8}  
\end{center}
\end{figure*} 

\begin{figure*}
\begin{center}
\includegraphics[width=0.56\textwidth]{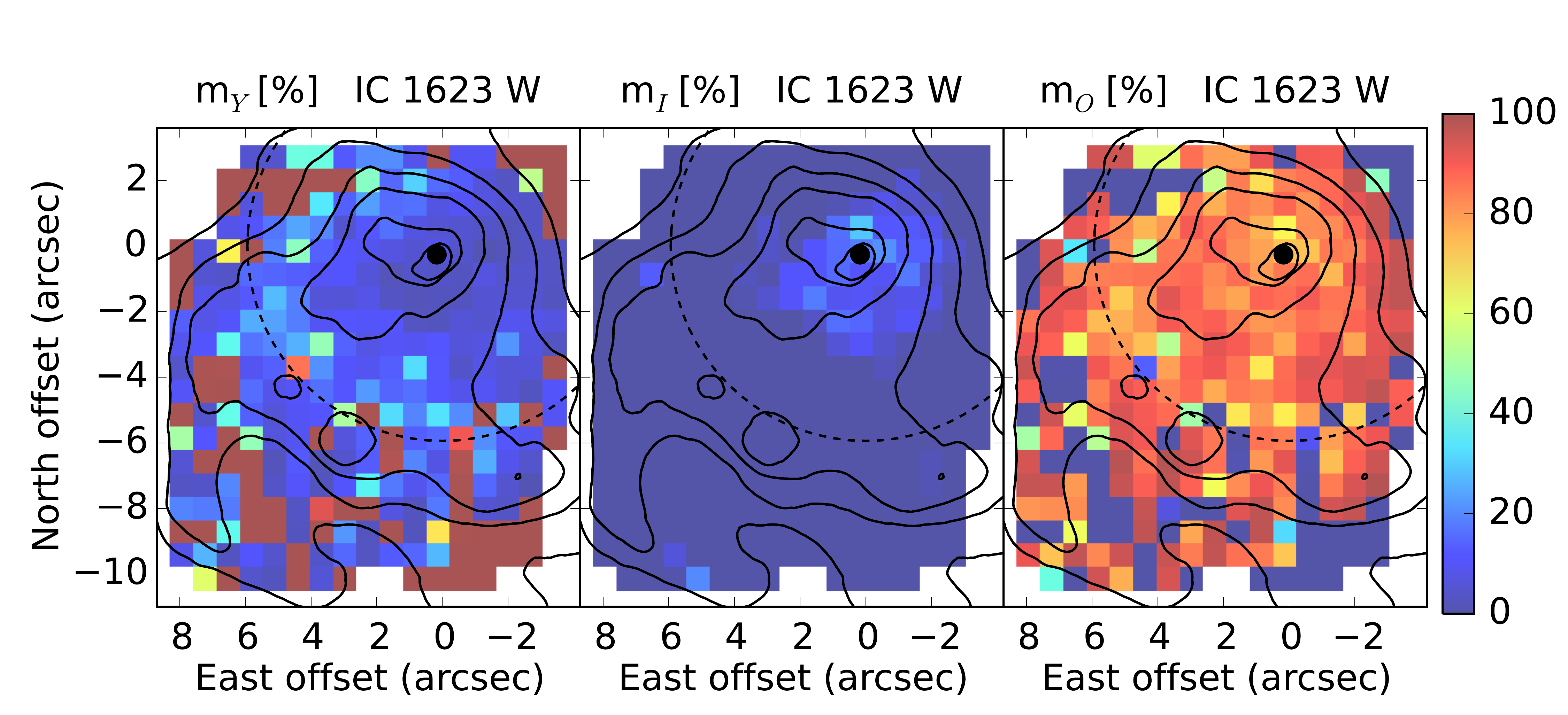} 
\includegraphics[width=0.43\textwidth]{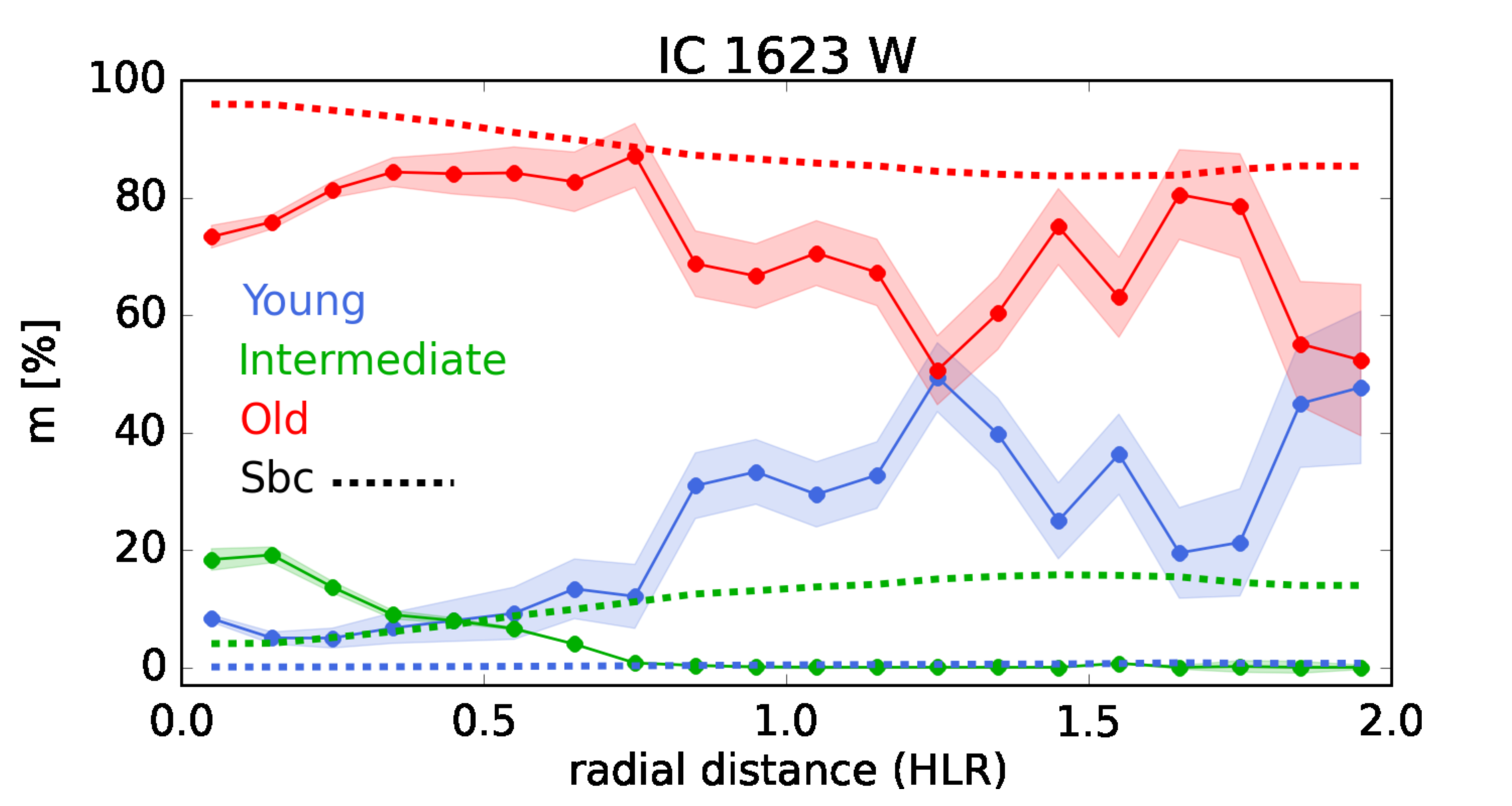} \\
\includegraphics[width=0.56\textwidth]{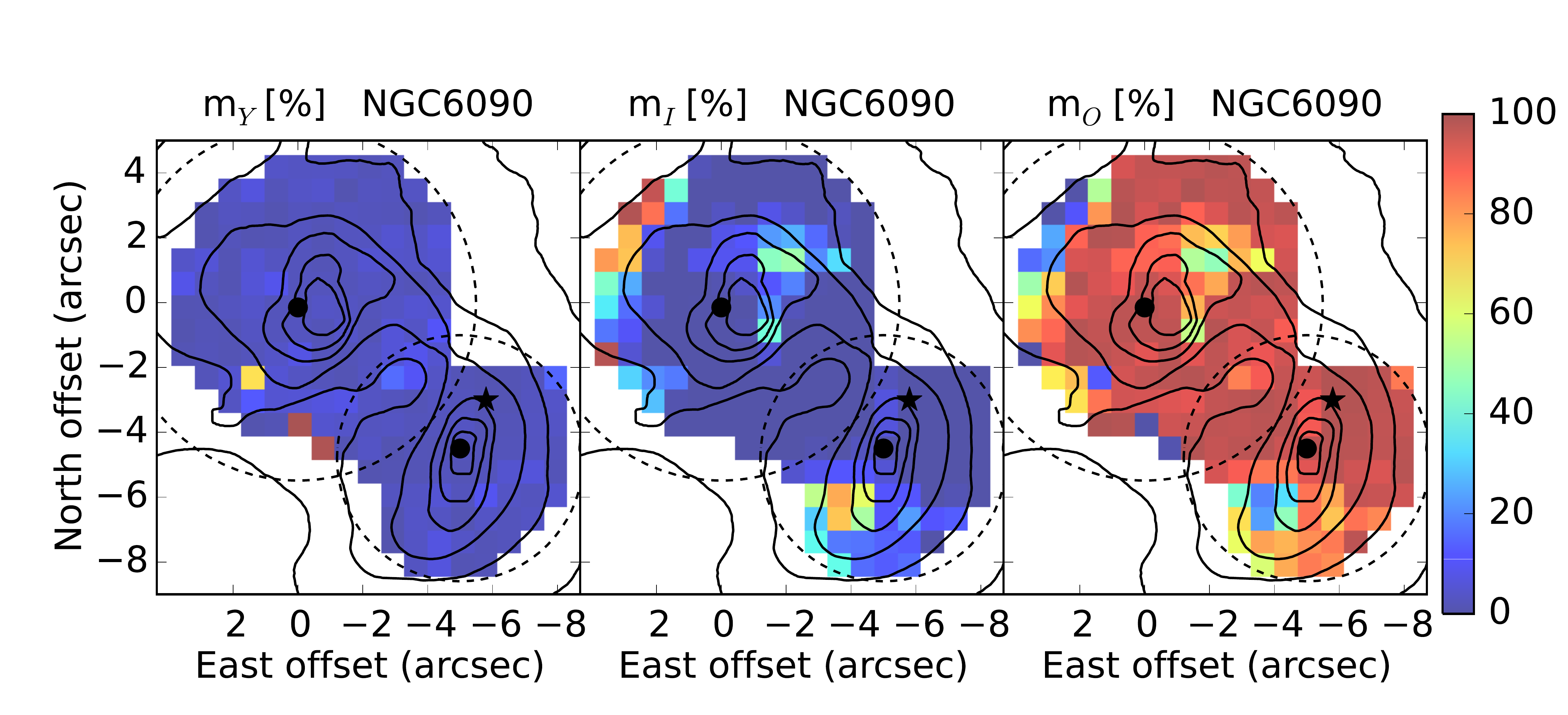} 
\includegraphics[width=0.43\textwidth]{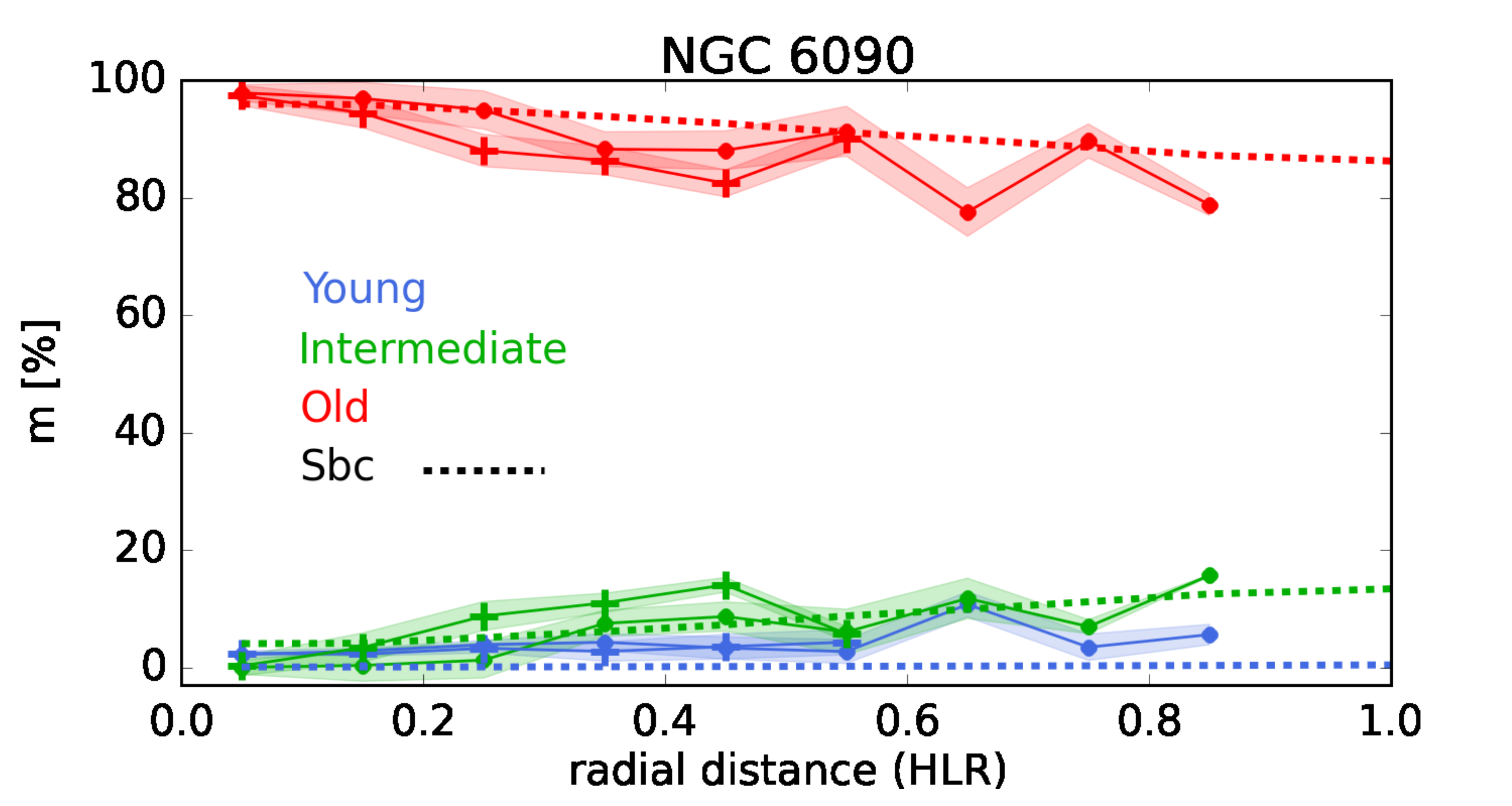} \\
\caption{As Figure 8 but for contributions to the stellar mass.}
\label{Fig_9}  
\end{center}
\end{figure*}

\subsection{Metallicity}\label{4.6}
The mean metallicity has been obtained in a similar way to the 
mean light weighted ages, but weighing by the mass:
\begin{equation}
\langle \log \ Z_{\star} \rangle_{M} = \sum_{t,Z} \mu_{t,Z} \log Z 
\end{equation}
where $\mu_{t,Z}$ is the mass fraction in the base 
population with age $t$ and metallicity $Z$. This is the same definition 
adopted in GD2015. 
The mass-weigthed metallicity maps 
are shown in Figure 7. 
The scale range in both maps is the same, and spans  the 
full range of the base (from $-0.7$ to $+0.2$ in log-solar units).

As can be seen in Figure 7, the 
scatter is large, maybe due to the weakness of the metallic 
absorption lines in the spectra. Moreover, our 
results could be affected by the age-metallicity degeneracy, 
although the spectral synthesis has been proven to be better 
than the index method to break it \citep{sanchez-blazquez2011}.

Focusing on the radial averages, it seems that 
IC 1623 W, NGC 6090 NE, and SW, have global 
mean metallicities around $\sim$ 0.6 Z$_{\odot}$ 
(but ranging from 0.4 to 0.8 Z$_{\odot}$).  
We find that beyond 0.7 HLR in IC 1623 W the 
metallicities are similar to those of Sbc galaxies, but in the central 
regions (in the inner 0.5 HLR) the metallicities are smaller, more 
similar to Sc galaxies. Something similar happens in NGC 6090 NE, 
where a decrease in the metallicity below 0.4 HLR is found.
The metallicity radial profiles 
in the inner 1 HLR of these three galaxies 
have flat or even slightly positive gradients, 
contrary to what happens in 
spirals earlier than Sc, 
which show a negative gradient (GD2015). However, 
we cannot claim a dilution or interpret this further, 
given the large dispersion of up to 0.25 dex in the metallicity 
of the control samples. 

\subsection{Contribution of young, intermediate and old populations}\label{4.7}
Population synthesis studies in the past found that 
a useful way to summarize the SFH is to condense the age 
distribution encoded in the population vector into age 
ranges. This strategy originated in methods  
based on fitting equivalent spectral indices \citep{bica1988,cidfernandes2003}, 
but it was also applied to 
full spectral fits \citep{gonzalezdelgado2004}.
We follow this strategy by dividing into: young stellar 
populations (YSP,  $t \le 0.14$ Gyr), intermediate age 
stellar populations (ISP, $0.14 < t \le 1.4$  Gyr), and 
old stellar populations (OSP, $t > 1.4$ Gyr).
We select these three age bins to correspond to the lifetimes 
of populations with distinctly different optical 
line and continuum features. 

Figure 8 presents the maps (left) and radial profiles (right) 
of the light fractions due 
to YSP, ISP, and OSP (x$_{Y}$, x$_{I}$, and x$_{O}$).
IC 1623 W is shown in the top panels, 
and NGC 6090 in the bottom panels.
We find that both early-stage merger systems are 
dominated by young stellar populations over their 
full observed extents. 

In \textbf{IC 1623 W} we find that the contribution of the YSP 
to light is around 90$\%$ everywhere, except
in the centre, where it is  around 50$\%$. 
In contrast, in Sbc galaxies the YSP 
contribution is much lower, $\sim$ 10$\%$ in the centre, 
and 20$\%$ in the disc.
In IC 1623 W the ISP has a significant contribution only in 
the inner 0.3 HLR, with contributions of 30--40$\%$, higher 
but still similar to  Sbc galaxies, $\sim 20\%$.
The contribution of OSP is everywhere 
below $20 \%$ in IC 1623 W, while in Sbc galaxies it ranges from 
$70\%$ in the nucleus to $40\%$ at 2 HLR.

In \textbf{NGC 6090} we find that almost all spaxels have 
YSP contribution above 
$70\%$ (much higher than the 10--$20\%$ in Sbc galaxies), 
except for some spaxels located at the easternmost end of NGC 6090 NE and 
at the south edge of NGC 6090 SW, where the contribution of 
ISP is also significant.
However, the average contribution of ISP is small,
below 10$\%$ for both progenitors, smaller 
than the 20--40$\%$ typical of Sbc galaxies. 
The OSP contributes around $20\%$ of the light of
both progenitors over their full extents, 
much smaller than its contribution in Sbc galaxies.

Analogously, in Fig.\  9 we present the maps (left) and radial profiles (right) 
of the mass contribution by percentage of the  YSP, ISP, and 
OSP (m$_{Y}$, m$_{I}$, and m$_{O}$). 
For both IC 1623 W (top panels) and NGC 6090 (bottom) most of 
the mass is in old stellar populations.

In \textbf{IC 1623 W} this is true everywhere, except for
some spaxels in the outer regions which are
completely dominated by young components.
The radial profiles of  the OSP and YSP mass contributions 
evolve inversely to each other.
In a non-uniform way, the OSP contribution decreases
from $80\%$ in the nucleus to $40\%$ at 2 HLR.
Conversely, the mass in YSP grows from $5\%$
in the nucleus to $60\%$ at 2 HLR. 
In terms of the mass in ISP, they range 
from $15\%$ in the nucleus to almost 0 beyond $0.7$ HLR.

The same happens with \textbf{NGC 6090}, with the OSP 
dominating the mass everywhere except in the 
easternmost end of NGC 6090 NE, in some 
spaxels to the north west of NGC 6090 
NE nucleus, and to the south of NGC 6090 SW, 
where the contributions of ISPs are also 
large (50--$80\%$, and up
to $90\%$ in some spaxels).
From the radial profile, we see that for
both progenitors the OSP is responsible for more than 
$80\%$ of the mass, 
and this is true over the full observed extent of
the system.
The fractions of stellar mass in YSP and ISP 
are comparatively low.

We have also computed the total masses in YSP, ISP, and OSP. 
From the total stellar mass in \textbf{IC 1623 W} 
($3.9 \times10^{10}$ M$_{\odot}$), 
the mass in young, intermediate age, and old 
populations are $1.6 \times 10^{9}$ ($4\%$), 
$6.4 \times 10 ^{8}$ ($1 \%$), and 
$3.7 \times 10 ^{10}$ M$_{\odot}$ ($95\%$), respectively.
In \textbf{NGC 6090} 
($6.8 \times 10^{10}$ M$_{\odot}$), the masses in YSP, ISP, and OSP are
$9.5 \times 10 ^{8}$ ($1\%$), 
$5.1 \times 10 ^{9}$ ($8\%$), and 
$6.2 \times 10 ^{10}$ M$_{\odot}$ ($91\%$), respectively.

In addition, we have compared the mass in clusters, as derived 
in Section \ref{3}, with the mass in YSP derived using the 
model base CB, which was also used in the 
photometric study. 
On the one hand, we found that the mass in star clusters 
represents only  3$\%$ and 4$\%$ of 
the total stellar mass of IC 1623 W and NGC 6090, respectively.
On the other hand, in IC 1623 W, the mass in clusters and the mass 
in YSP from the spectroscopy are comparable to each other within 
a factor of $\sim$ 1.3, in agreement with the idea 
that the vast majority of stars forms in clusters rather 
than in isolation \citep{chandar2015}. 
However, in NGC 6090, the mass in star clusters 
is a factor $\sim$ 4 higher than the mass in SSPs with ages $<$ 300 Myr 
from the spectroscopy. This is within the error expected from 
the uncertainties in the exact ages of the clusters. From the 
photometry we found an upper limit for the ages of $\le$ 300 Myr, 
while from the spectroscopy we found that the young components are 
in fact younger than 30 Myr. The ratio of the mass-to-light ratios 
for these populations at Z$_{\odot}$ is approximately 
$\frac{(M/L)_{300 Myr}}{(M/L)_{30 Myr}} \sim$ 3.6. Therefore,  
the age uncertainty in the photometric results could lead to mass 
uncertainties of up to a factor $\sim 4$. 

\section{Ionized gas emission}\label{5}
The {\sc starlight} best-fittings were subtracted from the observed 
spectra to obtain 3D cubes with the pure nebular emission 
line spectra. 
These emission spectra contain information about the ionized 
gas distribution and properties. 
Using the ratio of H$\alpha$ to H$\beta$ intensities, we have 
determined the ionized gas dust attenuation (Section \ref{5.3}), which 
is necessary to compute the total deredened H$\alpha$ luminosity.
The H$\alpha$ luminosity is a commonly used SFR tracer, 
but can be affected by shock emission. 
Following \cite{rich2011}, we have analyze the emission line 
ratios (Section \ref{5.4}) and velocity dispersions (Section \ref{5.5}) 
to confirm which spatial regions 
are indeed dominated by star formation.
 
\begin{figure*}
\begin{center}
\includegraphics[width=0.45\textwidth]{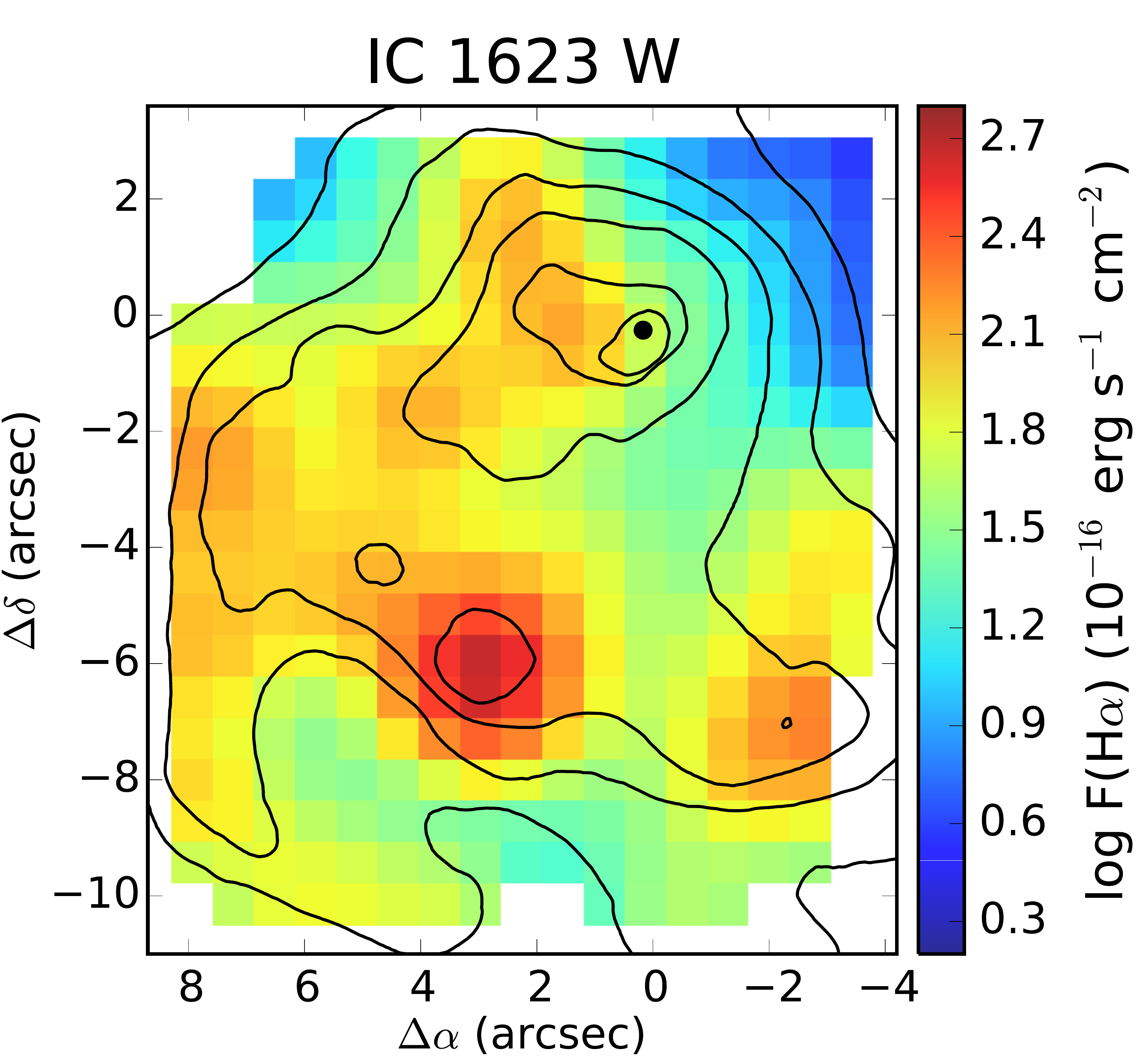} 
\includegraphics[width=0.45\textwidth]{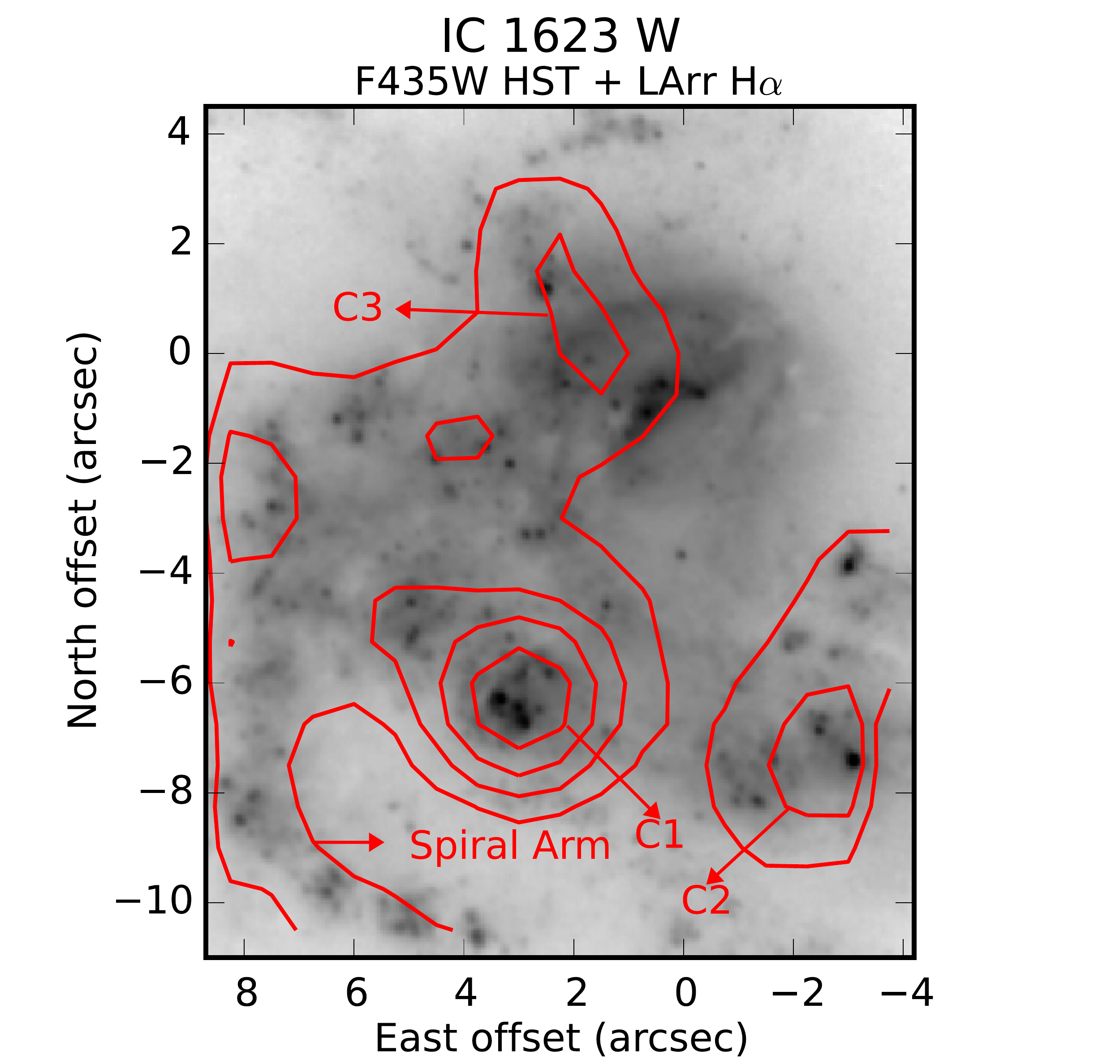}\\
\caption{Left: H$\alpha$ emission line flux (left panel) 
in units of 10$^{-16}$  erg s$^{-1}$ cm$^{-2}$ plotted on a logaritmic scale to enhance the contrast, ranging 
from 0.2 to 2.8. The smoothed HST F435W continuum image is superimposed 
as contours. Right: IC 1623 W HST F435W continuum image 
with LArr H$\alpha$ line emission superimposed as red contours. 
The distorted spiral arm and SSCs aggregates are labelled 
in the Figure.}
\label{Fig_10}  
\end{center}
\end{figure*}

\begin{figure*}
\begin{center}
\includegraphics[width=0.45\textwidth]{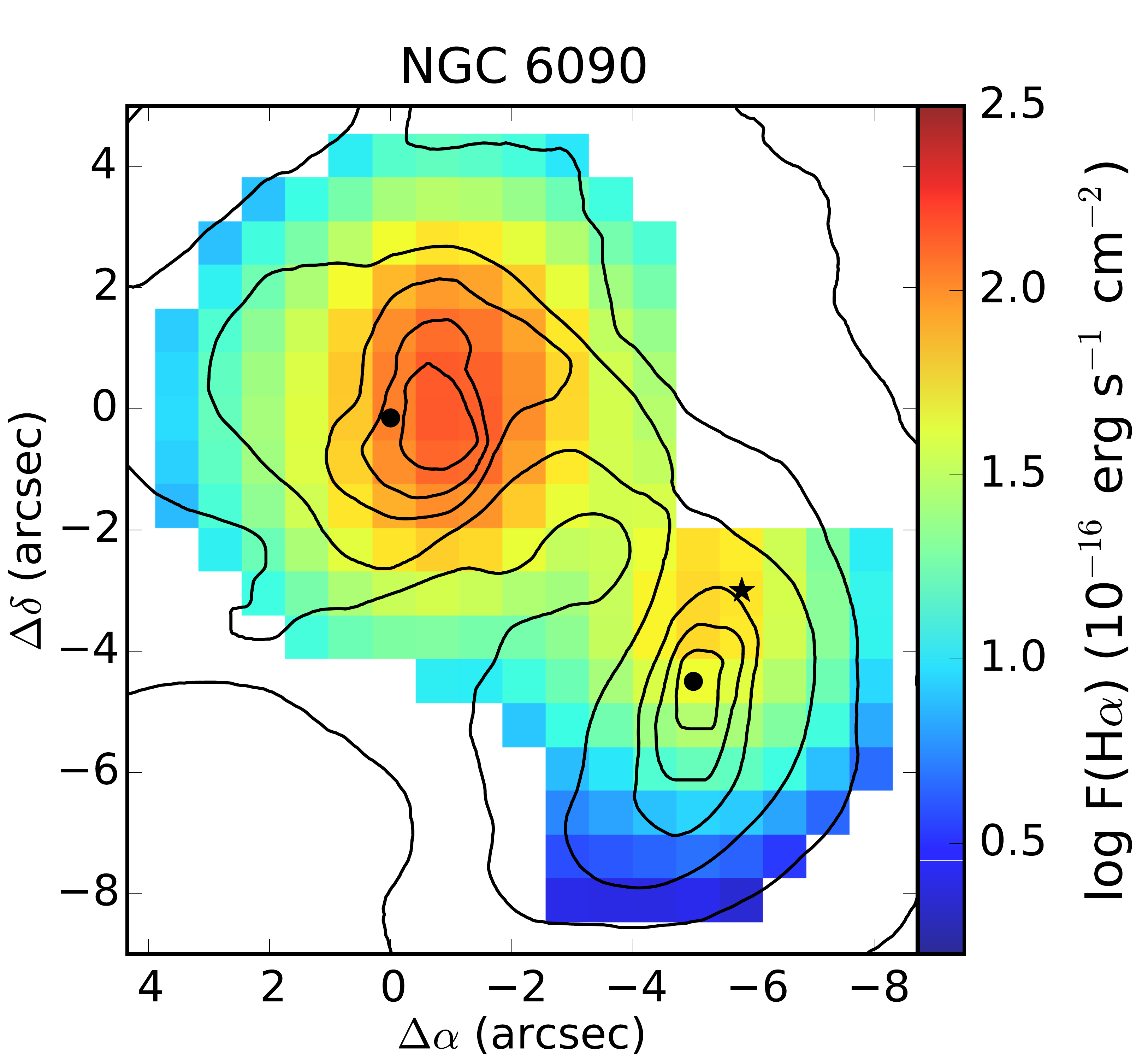} 
\includegraphics[width=0.45\textwidth]{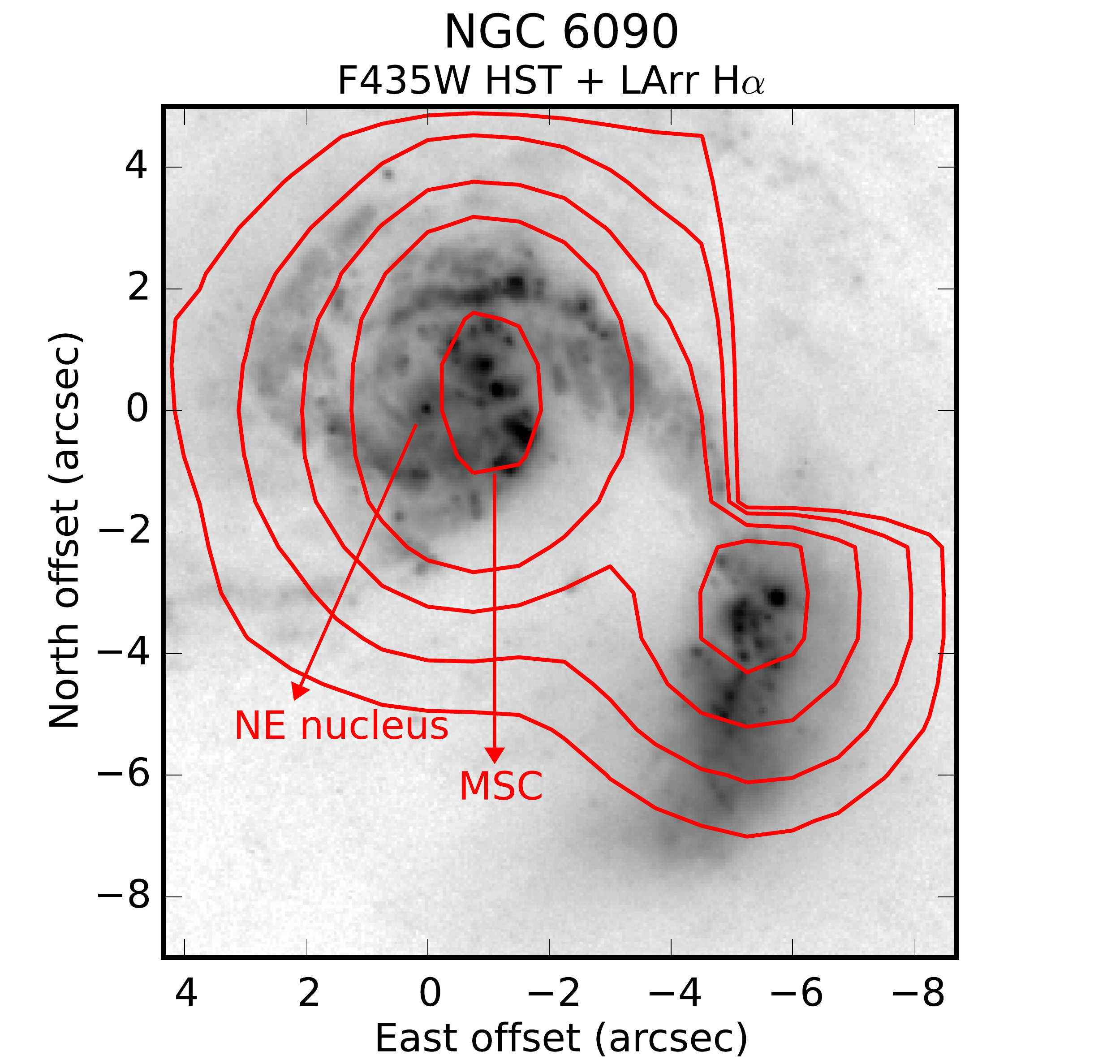}\\
\caption{As Fig. 10 except for NGC 6090. In the right panel, the nucleus of the 
NE galaxy and the region of multiple star clusters (MSC) are labelled.}
\label{Fig_11}  
\end{center}
\end{figure*}

\begin{figure}
\begin{center}
\includegraphics[width=0.45\textwidth]{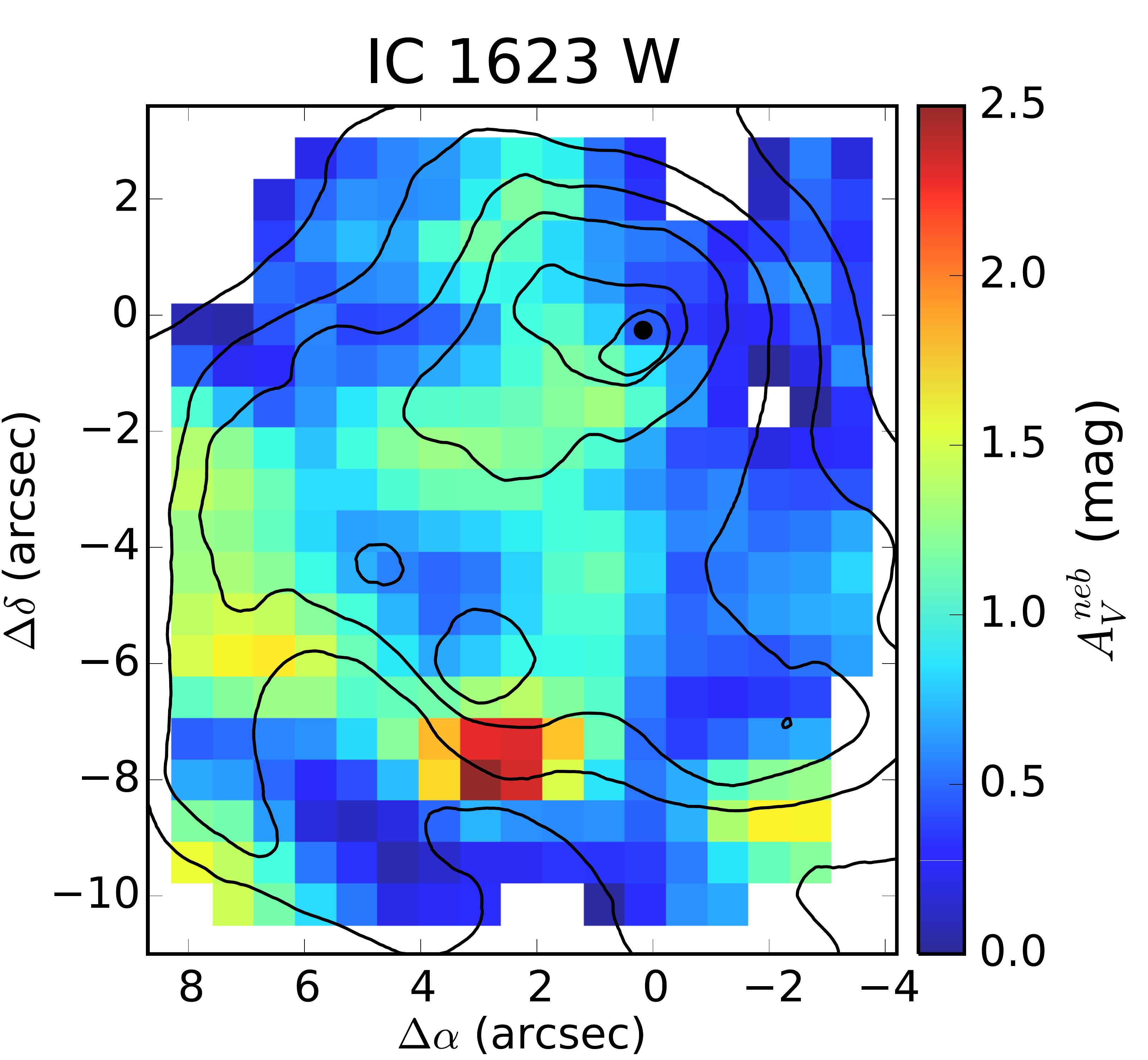} 
\includegraphics[width=0.45\textwidth]{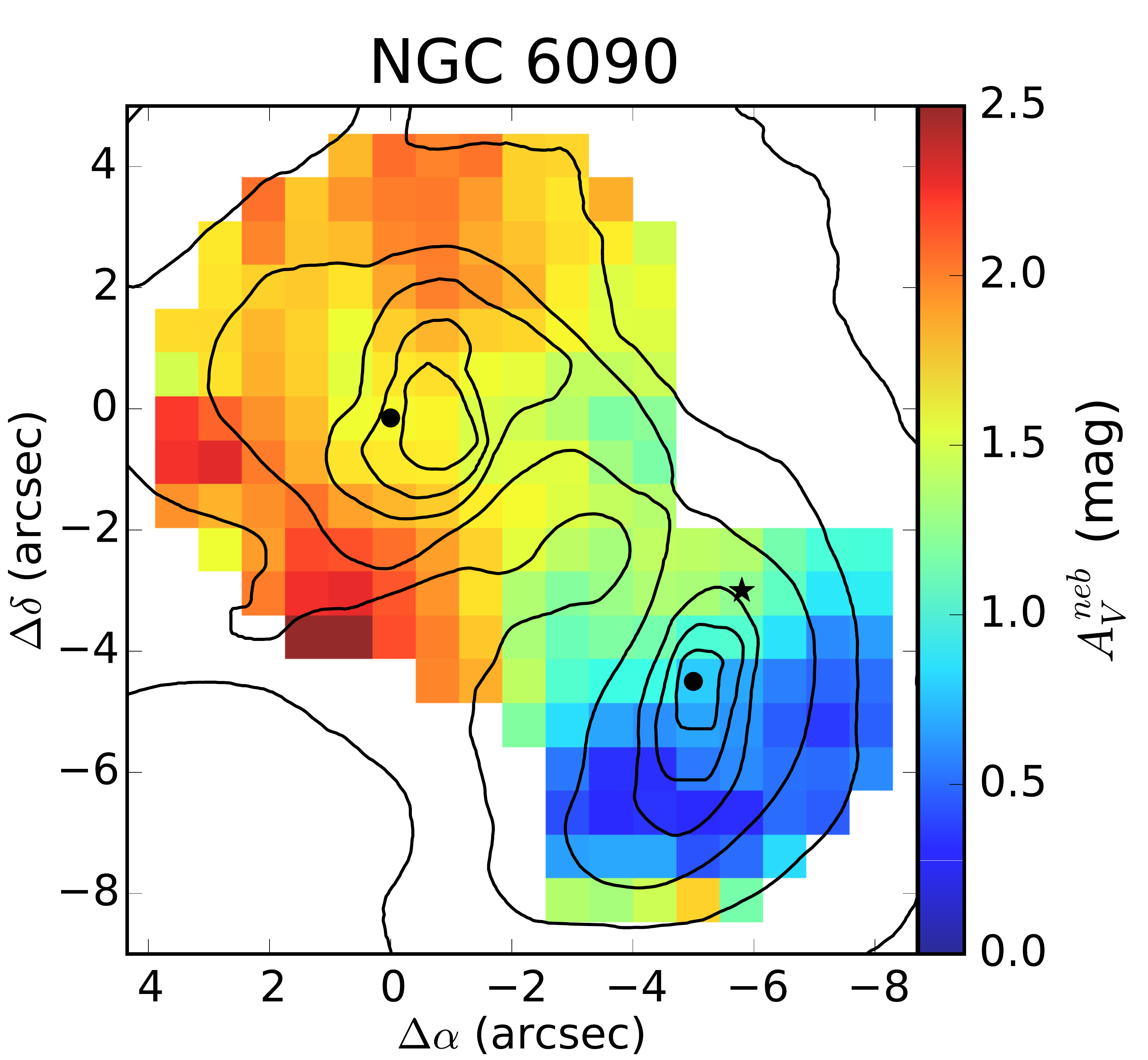} 
\caption{Line attenuation at 5500 $\rm \AA$ (left panels), calculated from the 
H$\alpha$/H$\beta$ emission line ratio assuming a Calzetti-like dust attenuation 
curve for IC 1623 W (top) and NGC 6090 (bottom).}
\label{Fig_12}  
\end{center}
\end{figure}

\begin{figure*}
\begin{center}
\includegraphics[width=1.0\textwidth]{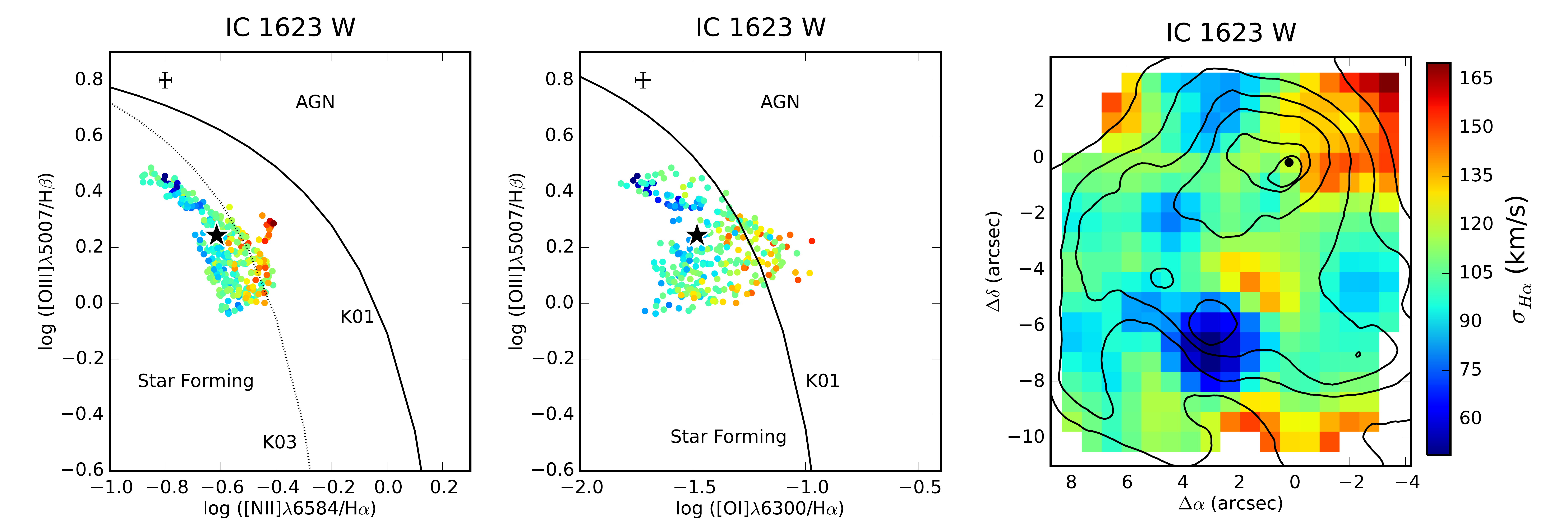}
\includegraphics[width=1.0\textwidth]{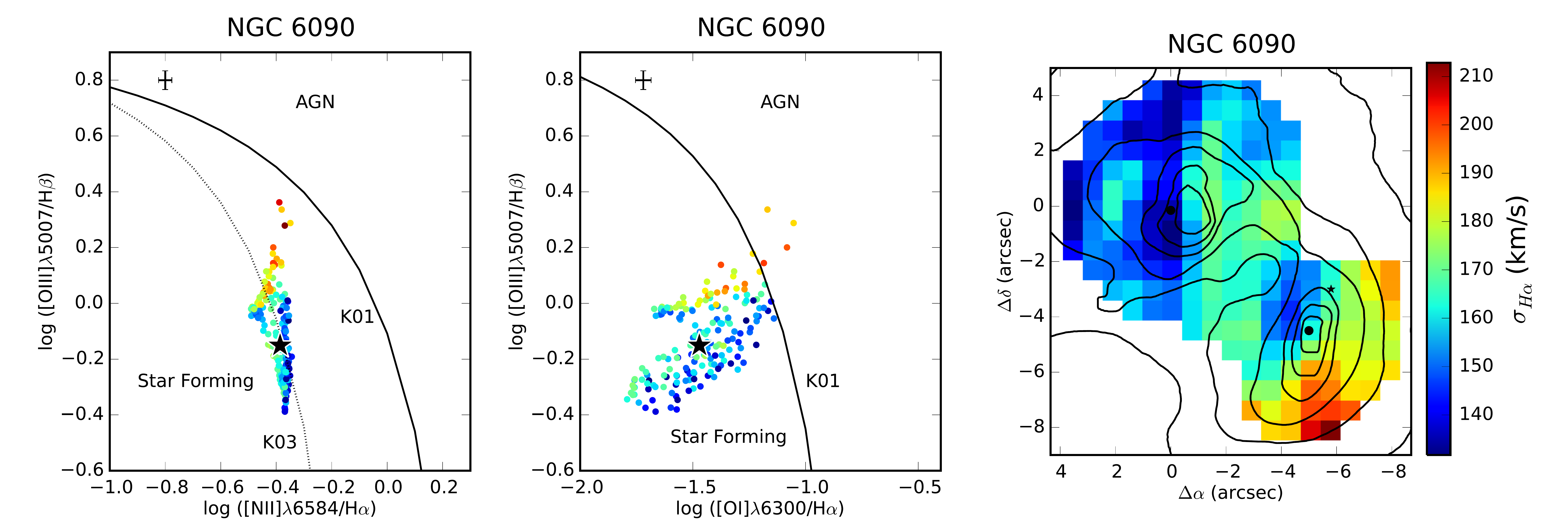}
\caption{[NII]$\lambda$6584/H$\alpha$ vs.\ 
[OIII]$\lambda$5007/H$\beta$ (left panels), and 
[OI]$\lambda$6300/H$\alpha$ vs.\
[OIII]$\lambda$5007/H$\beta$ (middle panels) line ratio diagnostic 
diagrams, and H$\alpha$-based velocity dispersion maps (right panels) for IC 1623 W 
(top) and NGC 6090 (bottom). Spaxels 
are colour coded according to the velocity dispersion. 
The ratios measured in the integrated spectra are shown with black stars.
Solid lines in both diagnostic diagrams mark 
the ``maximum starburst lines" obtained by \protect\cite{kewley2001} (K01) 
by means of photoionization models of synthetic starbursts, and the 
dotted line in the left panels marks the pure SDSS star-forming galaxies 
frontier semi-empirically 
drawn by \protect\cite{kauffmann2003} (K03).
The average error bars are shown in the top left corners.
The velocity dispersions in the right panels
have been corrected for the instrumental 
width ($\sigma_{inst}$ $\sim$ 140 km s$^{-1}$).}
\label{Fig_13}  
\end{center}
\end{figure*} 

\begin{figure}
\begin{center}
\includegraphics[width=0.47\textwidth]{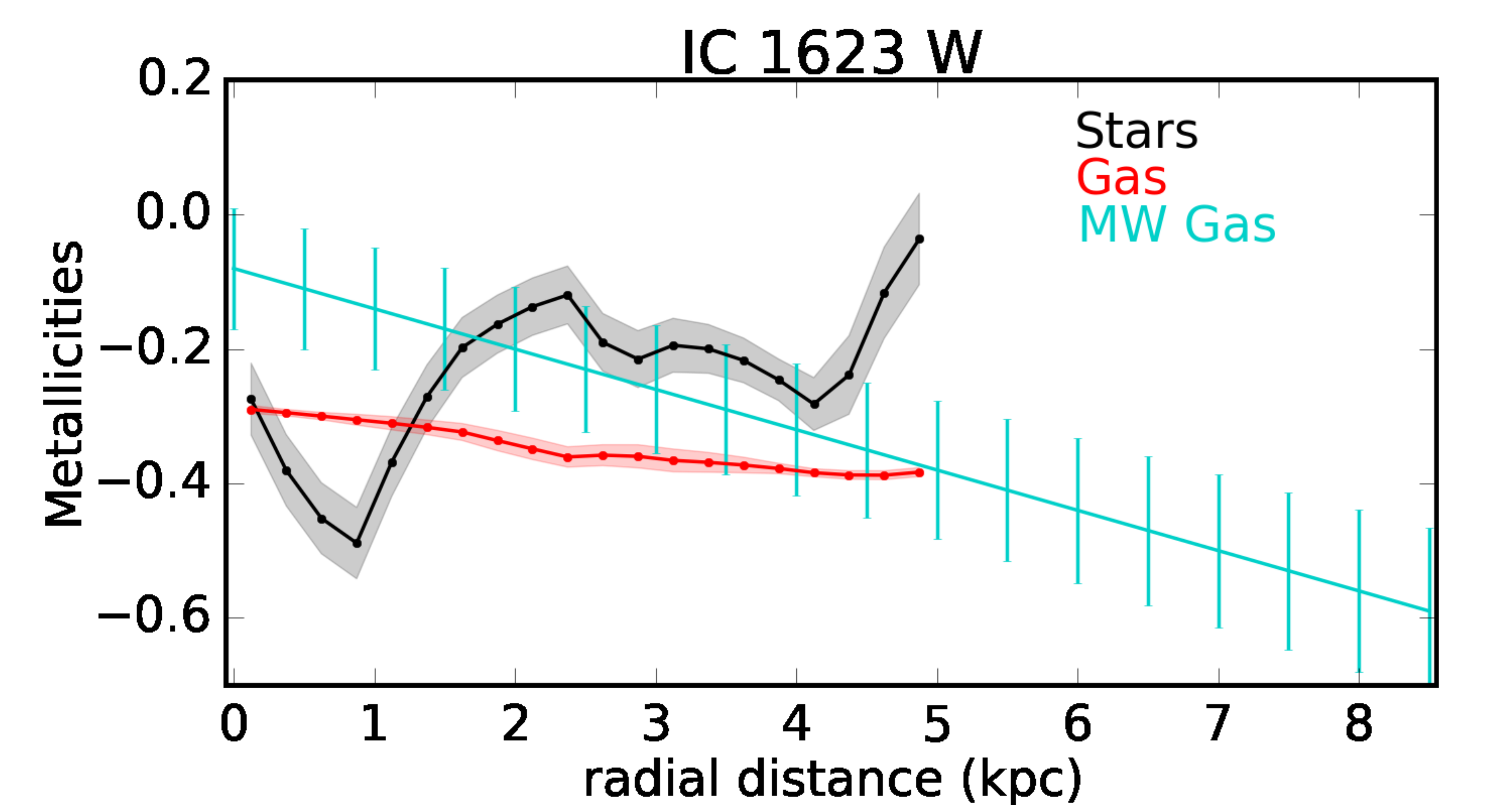} 
\includegraphics[width=0.47\textwidth]{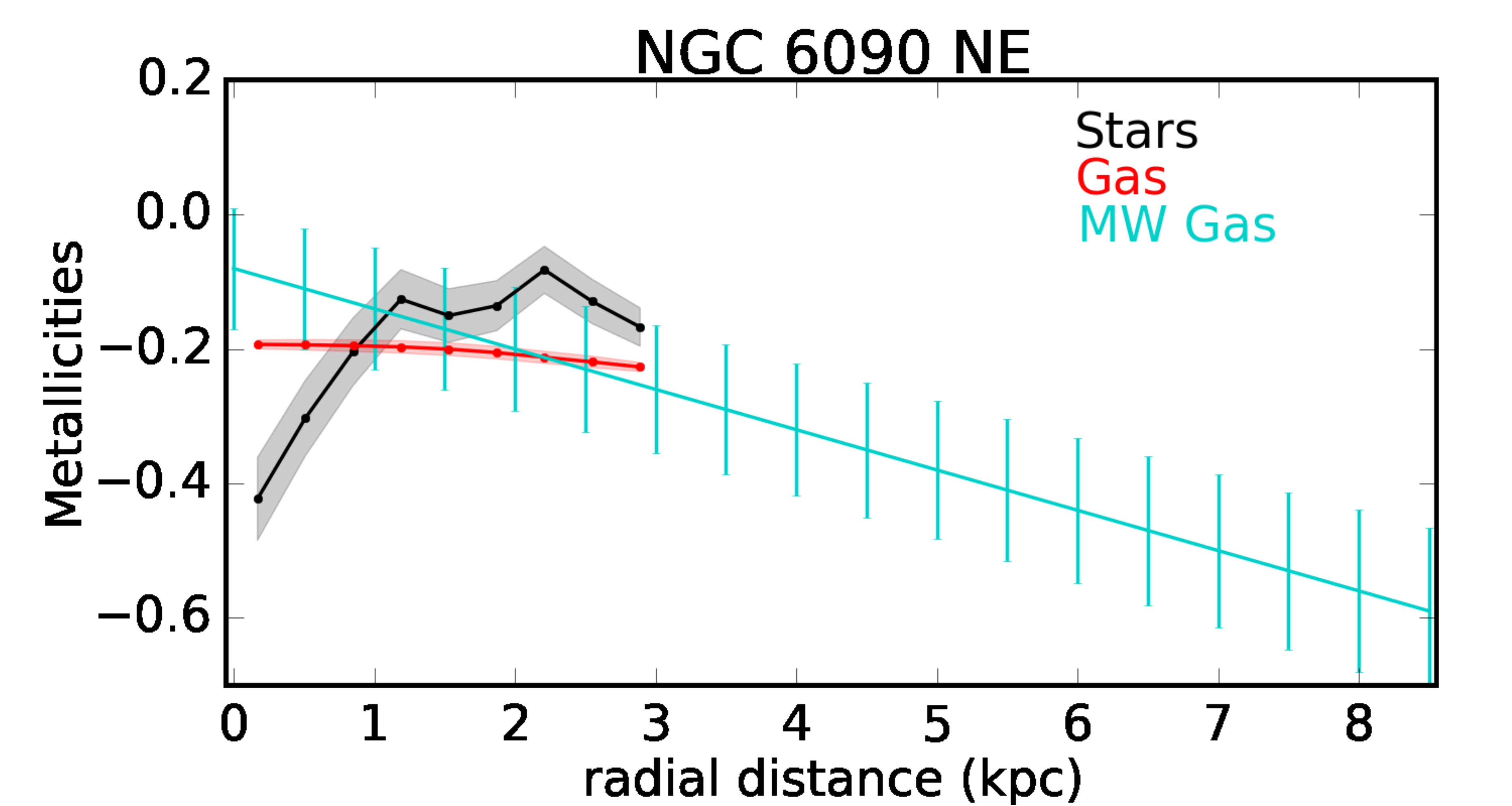} 
\includegraphics[width=0.47\textwidth]{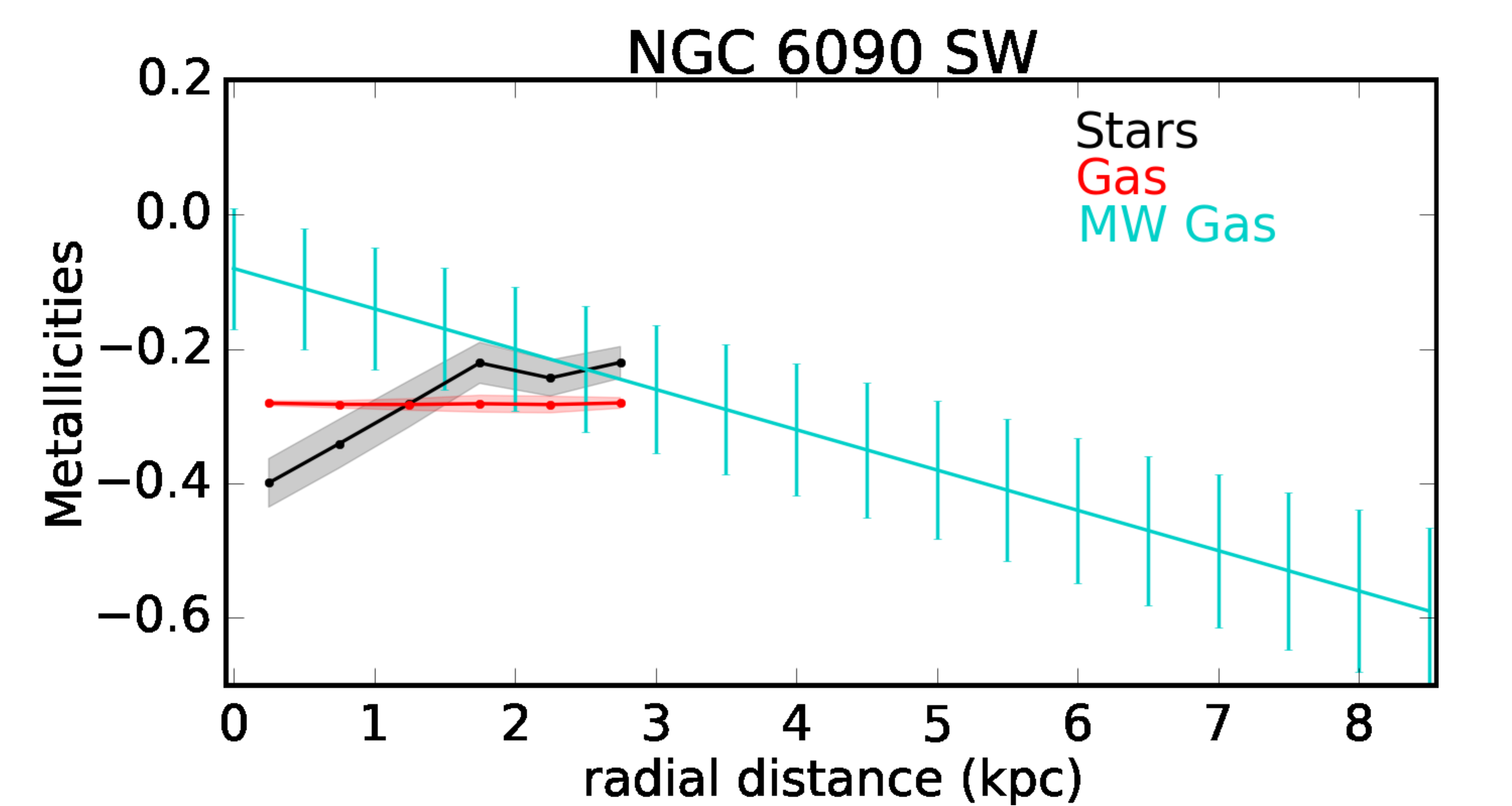}  
\caption{Oxygen metallicity profiles for our early-stage mergers in red, compared
to the stellar metallicities in black, and the Milky Way oxygen metallicity in blue, the latter 
assuming the \protect\cite{afflerbach1997} calibration. For comparison, 
all values are in log-solar units, and the spatial scale is in kpc.}
\label{Fig_16}  
\end{center}
\end{figure}
\subsection{Methodology}\label{5.1}
We have measured the intensities of some of the most prominent 
emission lines  -- [OII]$\lambda$3727, H$\beta$, 
[OIII]$\lambda$5007, [OI]$\lambda$6300, H$\alpha$, 
[NII]$\lambda$6584, and [SII]$\lambda\lambda$6716,6731 --
when present in the spectra.

First, a local continuum was estimated 
using narrow windows located on either side of 
each line.
This local continuum was subtracted to account for 
small mismatches between the {\sc starlight} fit 
and the actual galaxy continuum. 
The emission line intensities in each spectrum were then obtained 
by fitting Gaussians. 
Nearby, blended lines (e.g. [NII] and H$\alpha$) were fitted 
simultaneously using multiple Gaussian profiles. 
In this way, emission-line intensity maps were created for 
each individual line.

The statistical uncertainty in the measurement of the 
line fluxes was calculated from the product of the
standard deviation of the adjacent continuum, $\sigma_{c}$, 
and the FWHM of the emission line. 
In order to avoid spurious weak line measurements, 
we have only considered spaxels where the lines have 
SNR $>$ 3.

\subsection{Ionized gas morphology}\label{5.2}
\textbf{IC 1623 W}. Figure 10 shows the observed 
H$\alpha$ (left panel) emission flux for IC 1623 W 
on a logarithmic scale to enhance the contrast. 
As this maps shows, the ionized gas extends across the 
whole FoV.  
The brightest regions are those that correspond to 
aggregates of SSCs and star forming regions along 
the distorted spiral arm.
To get a better idea of the overall morphology of 
ionized gas in IC 1623 W, 
in the right panel of Figure 10 we have superimposed as red contours 
the LArr H$\alpha$ emission over the HST F435W continuum 
image (greyscale).
We can see that H$\alpha$ emission is concentrated in the 
distorted spiral arm and SSCs aggregates (which we have labelled 
C1 and C2 in the image).
Furthermore, in the nuclear region we detect an elongated 
structure, C3, which appears to leave or enter the nucleus. 
Due to its morphology, at first we thought it might be a 
gas inflow/outflow.   
Yet we see in the HST image that this emission also a 
stellar counterpart with several SSCs along it. 
Moreover, we will see in Section \ref{5.4} that the gas ionization 
is due to star formation in C3. 

\textbf{NGC 6090}. Analogously, in Figure 11 we show the 
observed H$\alpha$ (left panel) map for NGC 6090. 
The ionized gas distribution is extended across 
both progenitors in a more uniform and less 
clumpy way than in IC 1623 W. 
The NE progenitor is more luminous than the SW one. 
In the right panel of Figure 11 we show the HST F435W continuum image
of NGC 6090 with the H$\alpha$ emission superimposed 
as red contours. In the NE progenitor, the 
H$\alpha$ peak is offset to the west of the nucleus, 
spatially coincident with the MSC region.
For the SW progenitor, H$\alpha$ peaks in 
the northern half where most of the star forming 
regions accumulate.

\subsection{Ionized gas dust attenuation}\label{5.3}
The measured ratio of H$\alpha$ to H$\beta$ intensities 
provides an estimate of the nebular dust 
attenuation, $A_{V}^{neb}$, along the line of sight.
We assume an intrinsic ratio of I(H$\alpha$)/I(H$\beta$)=2.86, 
valid for case B recombination with T = 10000 K and electron 
density 10$^{2}$ cm$^{-3}$ \citep{osterbrock2006}, which are 
typical of star forming regions.
We have used the \cite{calzetti2000} reddening function,
as it has been shown to work better for starburst galaxies. 

In Figure 12 we show the  $A_{V}^{neb}$ maps for IC 1623 W (top) 
and NGC 6090 (bottom).

In \textbf{IC 1623 W} the mean nebular extinction is $\sim$ 0.8 mag. 
Following the trend found for the stellar extinction, 
the regions most affected by extinction are located 
near areas where HII regions accumulate. 
Near C1, $A_{V}^{neb}$ reaches its maximum value of 
around 2.5 mag. 
The same happens around C2, with values up to 1.6 mag. 
Also, values between 1.3 and 1.6 mag are found
in the distorted spiral arm, and between 
1.0 and 1.3 mag to the south-east of the nucleus 
and along the elongated structure C3. 
Everywhere else, the extinction is below 1 mag. 

In \textbf{NGC 6090}, the average extinction of 
the NE progenitor is 1.8 mag. $A_{V}^{neb}$ ranges from 
1.5 mag in the nucleus of the galaxy, to
values around 2 mag in the north, and 
up to 2.5 mag in a few spaxels in the
south. For the SW progenitor we found an average
value around 0.8 mag, but it also varies
across the galaxy from 0.5 mag in the south 
to 1 mag in the north.
We found an average difference 
of $\sim$ 1 mag between the two progenitors of the system, as also 
happens with the stellar extinction (see upper right 
panel of Figure 5).
In the overlapping bridge region  $A_{V}^{neb}
\sim1.3$--1.4 mag.

There is a good spatial correlation between
the extended H$\alpha$ emission and the regions where 
the YSP are contributing with $\ge$ 70$\%$ to the total light. 
As we will see in the next Section, the ionization  of the warm gas
in these early-stage merger LIRGs is 
indeed due to star formation, as this correlation suggests.

Moreover, we have compared the ratio between the ionized gas 
extinction to the stellar extinction, using only the spaxels where 
we are able to determine the ionized gas extinction, that is, 
where the SNR of both H$\alpha$ and H$\beta$ are above 3.
More than 95$\%$ of the spaxels have values of the ratio between 
0 and 7. Values above 7 are considered unreliable, as they 
correspond to spaxels with $A_{V}^{stars}$ below the 
uncertainty with which we can determine it 
(of $\sim$ 0.1 mag, \citealt{cidfernandes2013}).
For IC 1623 we found  
$\langle \frac{A_{V}^{neb}}{A_{V}^{stars}} \rangle = 2.2 \pm 1.4$ 
(median and standard deviation), 
while for NGC 6090 
$\langle \frac{A_{V}^{neb}}{A_{V}^{stars}} \rangle = 1.4 \pm 0.8$.
Analysing UV and optical spectra of 39 starburst galaxies, 
\cite{calzetti1994} found 
$\langle \frac{A_{V}^{neb}}{A_{V}^{stars}} \rangle$ = 2.0 $\pm$ 0.4, which 
is compatible with the results from our early-stage merger LIRGs. 
The fact that in starbursts and interacting galaxies 
$A_{V}^{neb}$ is larger than $A_{V}^{stars}$ is 
attributed to the fact that emission line sources 
are preferentially located within dusty birth clouds,
whereas the stellar light mainly samples the diffuse dust 
in the ISM. \cite{kreckel2013} also find a similar result 
analyzing IFS data from 8 nearby galaxies, but noting a 
more complex behaviour of the nebular to stellar extinction ratio, 
which is sensitive to the $H\alpha$ flux.  

\subsection{Ionisation mechanisms: Star-formation and shocks} \label{5.4}
Diagnostic diagrams comparing emission line flux ratios are the most 
widely used method for identifying the nature of the dominant ionisation mechanisms in 
galaxies. Fig. 13 presents two such diagrams: 
[NII]$\lambda$6584/H$\alpha$ vs.\  [OIII]$\lambda$5007/H$\beta$ 
(left panel, henceforth the ``BPT diagram'', after \citealt{baldwin1981}), 
and [OI]$\lambda$6300/H$\alpha$ vs.\  [OIII]$\lambda$5007/H$\beta$ 
(middle panel, following \citealt{veilleux&osterbrock1987}) for both 
IC 1623 (top) and NGC 6090 (bottom).
Panels on the right show the H$\alpha$ velocity 
dispersion ($\sigma_{H\alpha}$) maps.

Solid lines in both diagnostic diagrams mark 
the ``maximum starburst lines" obtained by \cite{kewley2001} by means 
of photoionization models of synthetic starbursts. These lines are often 
interpreted as delimiting the region occupied by ``pure AGN'', even though 
this is not what they were designed for 
(see discussion on \citealt{cidfernandes2011}). The dotted line in the left 
panels marks the frontier semi-empirically drawn by \cite{kauffmann2003} 
to separate SDSS star-forming galaxies from those hosting an AGN. 
Systems between these two lines are often called ``composite", a nomenclature 
that betrays the traditional interpretation of such systems as a mixture 
of star-forming and AGN.

None of the spaxels in either IC 1623 or NGC 6090 lies above the K01 line 
in the BPT diagram (left panels) and only a few do so in the [OI]/H$\alpha$ 
one (middle), indicating that AGN, even if present, contribute little to 
the emission lines. In fact, most of the points lie below the K03 line, 
confirming that photoionization by young stars is the main mechanism of 
line production in these sources. 
Black stars in these panels mark the line ratios of the spatially 
integrated spectra, and their position in the diagnostic diagrams 
further confirms the dominance of star formation as the main ionization source. 

Interestingly, however, in both cases a significant number of spaxels occupy 
the intermediate zone between K03 and K01 in
BPT-coordinates. Moreover, the colour-coding makes it evident to the eye that 
the points which depart the most from the star forming sequence are those 
with high $\sigma_{H\alpha}$ (yellow-red colours). In IC 1623 these large gas 
velocity dispersion regions are mostly located towards the NW of 
the $\sigma_{H\alpha}$ map (top right panel), while in 
NGC 6090 (bottom right) they are concentrated in the 
South of NGC 6090 SW. 

These are signs of tidally induced shocks, 
associated with the interaction process, contributing to 
the ionization of the gas \citep{colina2005}. 
Extended shock ionisation has been previously 
reported in local U/LIRGs, taking advantage of IFS studies 
\citep{monreal-ibero2006,rich2011,rich2014}.
In all these cases, shock ionisation 
exhibits characteristics of extended LINER-like emission 
with broadened line profiles. 
In the case of our IFS data for IC 1623 W and NGC 6090, 
we find broadened line profiles falling in the 
so-called ``composite'' region.
The ``composite'' systems are conventionally associated 
with a superposition of normal HII 
regions and the narrow line region of AGN. 
However, the location of these shocked regions in BPT 
coordinates, together with the spatial coherence of the 
line-ratios and kinematical data revealed in 
our IFS, allow us to circumvent these limitations and correctly identify 
the gas ionization mechanism in our early-stage merger systems.

It is further useful to note that the misidentification of shocked regions 
with a mixture of star-formation and AGN ionization is not the only 
caveat in a naive reading of our diagnostic diagrams. In the case of NGC 6090 
some low $\sigma_{H\alpha}$ spaxels populate the  zone between 
the K03 and K01 demarkation lines (blue points in the bottom left panel 
of Fig. 13). Their off-nuclear locations and closeness to 
the K03 line argue in favour of them being pure HII regions, 
and not ``star-forming $+$ AGN composite'' systems, which their nominal locations 
in the BPT diagram would indicate. This caveat has been noted before 
by \cite{sanchez2015}, who find that some $\sim 14\%$ of the bona fide 
HII regions in the CALIFA sample lie in the (ill-named) ``composite'' region.

\subsection{Ionized gas kinematics: velocity dispersion}\label{5.5}
We trace the gas kinematics of NGC 6090 and IC 1623 W using the results of 
Gaussian fits to their emission lines.
To obtain the final velocity dispersion, we have subtracted the 
instrumental resolution ($\sigma_{inst}$) in quadrature, 
where $\sigma_{inst}$ at H$\alpha$ is $\sim$ 140 km s$^{-1}$.

In the right panel of Figure 13 we show the ionized gas 
velocity dispersion maps for IC 1623 W (top) 
and NGC 6090 (bottom).

In \textbf{IC 1623 W} the velocity dispersion is low in the
star forming regions and in the distorted spiral arms, 
with $\sigma_{H\alpha} <$ 120 km s$^{-1}$.
However, $\sigma_{H\alpha} >$ 120 km s$^{-1}$ in several 
outermost spaxels, with a region to the NW of 
the nucleus where 
$\sigma_{H\alpha}$ reaches up to 170 km s$^{-1}$.

In \textbf{NGC 6090} there exists a gradual increase 
in the velocity dispersion from $\sim 140$--150 km s$^{-1}$ 
to the east side of NE progenitor, to 170 km s$^{-1}$ 
in the bridge between progenitors, and up to 
190 km s$^{-1}$ to the SW of the SW progenitor. 

We interpret the high velocity dispersion regions in 
these early-stage mergers as a consequence 
of shocks producing a broadening of the emission lines.
Moreover, in Section \ref{5.4} we saw that there 
exists a correspondence between these regions and ``composite'' 
excitation conditions in the BPT diagram.  
IFS data for IC 1623 were previously analyzed 
by \cite{rich2011}, who found widespread shocks (LINER-like ionization) in the 
outskirts of the system, beyond the region traced by our data. 
In the region in common (the core of IC 1623 W) 
our results are in total agreement with them, finding similar values of 
the lines ratios [NII]/H$\alpha$ $\sim -$0.9 -- $-0.4$, 
[OI]/H$\alpha$ $\sim -$1.7 -- $-1.0$, consistent with star-formation 
ionization. While \cite{rich2011} are able to further decouple the 
contribution arising from shocks and star formation, we cannot follow 
a similar approach due 
to our lower spectral resolution (140 km s$^{-1}$ vs. 40 km s$^{-1}$ in H$\alpha$).

\subsection{Gas metallicities}\label{5.6}
We estimate the nebular metallicity of the star forming regions using 
the O3N2 $\equiv \log \frac{ [OIII]\lambda5007/H\beta }{ [NII]/H\alpha}$ 
index, as calibrated by  \cite{marino2013}. To exclude shocked regions 
from this analysis in an approximate way we remove spaxels 
with $\sigma_{H\alpha} > 130$ km s$^{-1}$ in IC 1623 W 
and $\sigma_{H\alpha} > 185$ km s$^{-1}$ in NGC 6090, 
corresponding to the points marked in yellow--red in Fig 13. 

The resulting values of 12 + log (O/H) range 
from  8.20 to 8.40 in IC 1623 W, and from 8.40 to 8.52 in NGC 6090. 
For a solar value of  12 + log(O/H) = 8.69 \citep{asplund2009}
these values correspond to $\sim 0.4$ and 0.6 solar (on average) 
for IC 1623 W and NGC 6090, respectively.

Also, in Figure 14 we compare the nebular oxygen metallicity 
in red, with the oxygen 
abundance of the Milky Way (MW) in blue, 
obtained using the calibration by \cite{afflerbach1997}, 
and with the stellar metallicities in black. 
The early-stage mergers have 
lower central oxygen metallicities in comparison with the MW, and 
flatter profiles. 
The nebular and stellar metallicities are similar, both in IC 1623 W and 
NGC 6090, but the stellar metallicities show slightly 
more positive gradients, while the nebular profiles are flat. 
However, a larger sample and more spatial coverage would be necessary to 
assess these results in a statistically significant way.
In Section \ref{6.4} our results are discussed in the context of 
the literature.

\section{Discussion}\label{6}
In this section we discuss our main findings in the context of 
the merger-driven evolution of galaxies. 
The observational characterization of star formation 
in mergers at different stages is necessary to test the validity 
and put contrains on merger simulations.
Merging LIRGs, such as the ones presented in this paper, 
are unique laboratories to study this in detail.
First, we calculate and compare the 
current Star Formation Rate (SFR) of IC 1623 W 
and NGC 6090, using the spectral synthesis, H$_{\alpha}$, 
and MIR data.

\subsection{Multiwavelength star formation rates}\label{6.1}
We explore two ways to estimate 
SFRs from our IFS data.
First, we compute the SFR from the dereddened 
H$\alpha$ luminosity considering that shocks are 
not important, that is, assuming that the gas is photoionized 
by massive stars only. 
We use the conversion to star formation rate provided 
by E. Lacerda  (private communication), calculated for the 
average metallicity of our galaxies 
$\sim$ 0.7 Z$_{\odot}$ and for \cite{bruzual&charlot2003} models. 
The calibration with a Salpeter IMF is very similar to the standard 
calibration of \cite{kennicutt1998a}, calculated for solar metallicity 
models.

Secondly, we use our stellar continuum decomposition 
to estimate the SFR by summing 
all the stellar mass formed since a lookback time $t_{SF}$ and 
performing a mass-over-time average, following equation 11 
in \cite{cidfernandes2013}, that gives the mean SFR 
since $t=t_{SF}$. One can tune $t_{SF}$ to reach
different depths in the past.
For comparison purposes we estimate it 
for $t_{SF} = 30$ Myr \citep{gonzalezdelgado2016,asari2007}, 
which is of the same order of the lifetime of the ionizing 
populations also emitting in H$\alpha$, that is O and 
early B stars.

Additionally, we have used a hybrid calibration, 
including the Spitzer MIPS 24 $\mu$m flux to check whether there is any 
star formation completely obscured from view in the optical,
which is not accounted for by the Balmer 
decrement attenuation correction. We have followed 
the calibration given by \cite{kennicutt2009}.

The three different approaches 
lead to very similar SFR results, with an average 
and dispersion of 23 $\pm$ 4 M$_{\odot}$ yr$^{-1}$ for IC 1623 W, and 
32 $\pm$ 4 M$_{\odot}$ yr$^{-1}$ for NGC 6090, 
for Salpeter IMF, and a factor 1.6--1.7 lower with Chabrier 
IMF models. 
The consistency between the hybrid calibration including MIR
and the optical ones, indicates that, for IC 1623 W and NGC 6090, 
there is no substantial star formation which is not 
accounted for by the optical data.

\subsection{Merger-induced star formation in mergers}\label{6.2}
\subsubsection{Global enhancement of star formation}\label{6.2.1}
In both IC 1623 W and NGC 6090 we find that the SFR 
is increased with respect to the so called
main sequence star forming galaxies (MSSF, \citealt{renzini2015}). 
Using the \cite{gonzalezdelgado2016} calibration for the MSSF in the 
same mass range as our early-stage mergers ($\sim$ Sbc), 
we obtain values of $\sim$ 3.0--4.4 M$_{\odot}$ yr$^{-1}$, 
a factor 6--9 lower than the SFRs of our two early-stage mergers. 
This increase is larger than the predictions from classical merger 
simulations by \cite{dimatteo2008}, who 
found that boosts in the SFR greater than 5 are rare, but still 
possible in about 15$\%$ of major galaxy interactions and mergers. 
We suggest that U/LIRG population could belong to this 15$\%$ in the 
sense that star formation in them is ``caught in the act''. 

As the SFR depends on the total amount of mass, 
we have also calculated the average specific star formation 
rate, sSFR = SFR / M$_{\star}$, which is sSFR $\sim$ 0.60 Gyr$^{-1}$ 
for IC 1623 W, and 0.47 Gyr$^{-1}$ for NGC 6090, 
these are significantly larger, 
a factor 5--6, than the sSFR of the Sbc/Sc control galaxies 
measured at 1 HLR ($\sim$ 0.1 Gyr$^{-1}$). 
\footnote{This comparison can be made because 
we have found that the galaxy-wide average stellar population 
properties are well represented by their values measured 
at 1 HLR \cite{gonzalezdelgado2015conf}.}
Therefore, the star formation 
increase in our two early-stage mergers is still 
significant even after normalization 
by the total stellar mass. This is expected given that 
the control samples have been selected in the same mass 
range as early-stage mergers. 
Moreover, this sSFR can be interpreted as the inverse of a time 
scale for the star formation ($\tau$). 
We find $\tau$ is much shorter in our 
early-stage mergers ($\sim$ 2.0 Gyr) than in 
Sbc/Sc galaxies of similar mass ($\sim$ 10 Gyr). That is, 
the same stellar mass has been built much faster 
in the early-stage mergers 
than in the control samples.

\subsubsection{IC 1623}\label{6.2.2}
From the IFS and star cluster photometry we find that merger-induced 
star formation in IC 1623 is both widespread (out to 1.5 HLR), and recent, 
with an average light weighted age of $\sim$50 Myr in IC 1623 W. 
The centre of IC 1623 W ($<$0.5 HLR) is older, $\sim$300 Myr, 
than the surrounding area, $\sim$30 Myr (Fig. 6), due to the presence 
of a significant contribution of intermediate-age stellar 
populations (Fig. 8) probably induced in a previous stage 
of the merger. However, the light is everywhere dominated 
by young stellar populations, with contributions in the range 50--90\% 
across IC 1623 W compared with 10--20\% in the Sbc/Sc control galaxies. 
The extinction in this progenitor is low $\sim$0.2 mag, and has a flat 
radial profile in comparison with Sbc/Sc control spirals (Fig. 5). 
However, we note that IC 1623 E is affected by much 
higher levels of extinction (up to 2--6 mag) according to the star 
cluster photometry (Fig. 3 and A2). IC 1623 W has a mean 
stellar metallicity of $\sim$0.6 Z$_{\odot}$, with a flat or 
slightly positive radial profile (Fig. 7). The nebular 
metallicity profile is also flatter than in non-interacting 
control spiral galaxies and consistent with the existence of 
gas inflows of metal poor gas to the centre of the system, 
as explained below in Section \ref{6.4}. 

\subsubsection{NGC 6090}\label{6.2.3}
As for IC 1623, the merger-induced star formation in 
NGC 6090 is widespread, as indicated by the star clusters 
detected in the HST images (Fig. 3), and recent, 
with an average light weighted age from the IFS 
of $\sim$80 Myr, with both progenitors, NGC 6090 NE and 
NGC 6090 SW having similar ages (Fig. 6).  
As in IC 1623 W, the age profiles are significantly flatter 
than in Sbc/Sc galaxies, in agreement with a general rejuvenation 
of the progenitor galaxies due to merger-induced star formation.
The light is dominated by young stellar populations in both 
progenitors, with contributions $\sim$70$\%$ (Fig. 8).
There exist a difference of $\sim$ 1 mag between the 
stellar extinctions of the two progenitors, with an average 
of 1.3 mag in NGC 6090 NE, 
and 0.5 mag in NGC 6090 SW (Fig. 5), and whose 
radial profiles are flat in comparison 
with Sbc/Sc control spirals. As in IC 1623 W, NGC 6090 
has a mean stellar metallicity of $\sim$0.6 Z$_{\odot}$, 
with a flat or slightly positive radial profile (Fig. 7), 
and a flat nebular metallicity profile, also in agreement 
with the gas inflow scenario.

\subsubsection{Evolutionary scheme}\label{6.2.4}
Based on our results we propose an evolutionary scheme in which 
the gas components of the progenitors react promptly
to the merger process, probably during the first 
contact (stage II in \citealt{veilleux2002} classification), 
where the discs overlap but 
strong bars and tidal tails have not yet formed. During this stage 
the gas starts inflowing to the dynamical centre of the system, 
and can trigger some star formation. The intermediate age 
stellar population in IC 1623 W nucleus ($\sim$300 Myr) could be a consequence 
of this, and we note that the star formation triggered 
during the first contact stage is still not extended, 
but concentrated in the centre. 
The early-stage merger (III in \citealt{veilleux2002}) starts 
after the first pericenter passage 
(in our systems, in the last $<$ 50--100 Myr), where most 
of the extended star formation is due to the fragmentation of gas clouds 
produced by 
the increase of the supersonic turbulence of the ISM, 
as predicted from high resolution models in the literature 
\citep{teyssier2010,hopkins2013,renaud2015}. 
Therefore, at this stage, the star formation is mainly 
in the form of star clusters. We found 
that the mass in the star clusters  is similar to that of
the 140 Myr age stellar populations derived from the IFS data 
(within factors of 1.3 and $\sim$ 4 for IC 1623 W and 
NGC 6090, respectively.)
Spatially extended star formation has also been previously 
observed in other early-stage merger systems 
\citep{wang2004,elmegreen2006,hattori2004,garcia-marin2009,arribas2012}.
The central nebular metallicities are already 
diluted at the early merger stage, at least in some cases. 

\subsection{Ionization mechanism}\label{6.3}
In both IC 1623 W and NGC 6090 the dominant ionization mechanism is 
photoionization by young massive stars. 
For a few spaxels we cannot discard composite 
photoionization by hot stars and shock ionisation, 
given also the high velocity dispersion in the north west 
of IC 1623 W and to the south of NGC 6090 SW.  
This result is in agreement with previous results from 
the literature \citep{rich2011,rich2014}.

\subsection{Nebular metallicity gradients}\label{6.4}
Finally, in the literature there exists some controversy about the evolution 
of the gas metallicity in mergers, with some studies finding 
lower central metallicity in mergers in comparison to control galaxies, 
and a flattening of the metallicity profiles, possibly related with 
gas inflows \citep{kewley2010,rich2012,guo2016}, and others finding a 
similar central metallicity in mergers and control samples, with a decrease 
compared to the control sample objects at the large radii. This suggests that 
other processes 
such as stellar feedback can contribute 
to the metal enrichment in interacting 
galaxies \citep{barrera-ballesteros2015}. 

Nebular metallicities estimated with the O3N2 method 
are $\sim 0.4$ and $0.6$ solar in IC 1623 W and 
NGC 6090, respectively. In Section \ref{5.6} we showed the oxygen 
metallicity profiles of our early-stage mergers.
In particular, for IC 1623 W we found a gradient of -0.020 $\pm$ 0.013 dex/kpc, 
which is in agreement with the gradient determined by \cite{rich2012} 
using \cite{kewley&dopita2002} calibration (-0.016 dex/kpc). 
\cite{rich2012} also compute a set of merger simulations to trace the oxygen 
metallicity variation. For close pairs (stage 2) they found a metallicity gradient 
of -0.025 $\pm$ 0.015 dex/kpc, consistent with that found by us in IC 1623 W.
In NGC 6090 we find even flatter profiles of -0.012 $\pm$ 0.013 dex/kpc in 
the NE progenitor, and 0.000 $\pm$ 0.014 dex/kpc in the SW progenitor, 
but still within the observational range of close pairs in the \cite{rich2012} study.  
All of them are flatter than the MW gradient 
(-0.060 $\pm$ 0.035 dex/kpc), and the control sample of isolated 
star forming spirals of \cite{rupke2010b} (-0.041 $\pm$ 0.009 dex/kpc), 
in agreement with the gas inflow scenario as the main cause of 
the dilution.

\section{Summary and conclusions}
Though Integral Field Spectroscopy PMAS LArr data and multiwavelength HST imaging, 
and with the {\sc starlight} full spectral fitting code, in this paper we characterize 
the star cluster properties, stellar populations, ionized gas properties (ionization mechanism 
and kinematics), and star formation rates in the early-stage merger 
LIRGs IC 1623 W and NGC 6090. Early-stage merger results have been compared with 
control Sbc and Sc galaxies from CALIFA (GD2015) in the same mass range.
Our main results are:

\begin{itemize}
\item[-] From the IFS and star cluster photometry we find that merger-induced 
star formation in these two LIRGs is widespread (out to 1.5 HLR) and recent, 
as revealed by the young light-weighted ages (50--80 Myr vs. $\sim$1 Gyr in Sbc/Sc control galaxies) 
and the high contributions to light of young stellar populations (50--90$\%$ vs. 10--20$\%$ in Sbc/Sc
galaxies). These results are in agreement with high resolution merger simulations in the literature.

\item[-] The A$_V$ profile in the inner 1 HLR of these two early-stage mergers 
is flat in comparison with the negative gradient in Sbc/Sc control spirals. 

\item[-] The age profiles are significantly flatter than in Sbc/Sc galaxies, 
in agreement with a general rejuvenation of the progenitor galaxies due 
to merger-induced star formation.

\item[-] The average stellar mass weighted metallicities at 1 HLR are 
$\sim$ 0.6 Z$_{\odot}$, similar to Sbc galaxies. 
Although with a large scatter, the stellar metallicity profiles are comparatively 
flat or slightly positive as compared to Sbc/Sc galaxies. 

\item[-] Nebular abundances estimated with the O3N2 method are 
$\sim 0.4$ and $0.6$ solar in IC 1623 W and 
NGC 6090, respectively.

\item[-] The nebular abundance profiles are flatter 
(-0.020 $\pm$ 0.013 dex/kpc gradient in 
IC 1623 W, -0.012 $\pm$ 0.013 dex/kpc in NGC 6090 NE, and 
0.000 $\pm$ 0.014 dex/kpc in NGC 6090 SW) than the ones in the Milky Way 
and in a sample of isolated star forming spirals, in agreement with the 
scenario where gas inflows are responsible of the dilution.

\item[-] From the [NII]$\lambda$6583/H$\alpha$ 
vs. [OIII]$\lambda$5007/H$\beta$, and [OI]$\lambda$6300/H$\alpha$ 
vs. [OIII]$\lambda$5007/H$\beta$ diagnostic diagrams we find that both 
IC 1623 W and NGC 6090 are dominated by star formation in the 
region mapped in this work.

\item[-] In some regions we can not discard a mix of photoionization 
by hot stars and by shocks, given the high velocity dispersion to the south of 
NGC 6090 SW, and to the north west of IC 1623 W, consistent with previous 
results from the literature.

\item[-] We obtained several estimations of the current SFR 
using the spectral synthesis (SFR(30 Myr)), the dereddened H$\alpha$ luminosity 
(SFR(H$\alpha$)), and a hybrid calibration 
using the observed H$\alpha$ luminosity and MIR 24 $\mu m$ luminosity. 
All of them lead to very similar values of the SFR, with an 
average of 23 $\pm$ 4 M$_{\odot}$ yr$^{-1}$ for IC 1623 W, and 
32 $\pm$ 4 M$_{\odot}$ yr$^{-1}$ for NGC 6090. 
From the good consistency between the different estimations we conclude that 
for IC 1623 W and NGC 6090 there is no substantial star formation 
which is not accounted for by the optical data.

\item[-] The average SFR in our early-stage mergers is 
enhanced by factors of 6 to 9 with respect to main-sequence Sbc star 
forming galaxies. This is slightly above the predictions from classical 
merger simulations, which indicate that SFR enhancement factors greater 
than 5 are rare, but still possible in about 15$\%$ of major galaxy 
interactions and mergers. We suggest that U/LIRGs could belong 
to this 15$\%$ in the sense that star formation in them 
is caught in the act.

\item[-] The sSFR in IC 1623 W ($\sim$ 0.60 Gyr$^{-1}$) and 
NGC 6090 ($\sim$ 0.47 Gyr$^{-1}$) is also enhanced by factors of 5 to 6 
with respect to Sbc/Sc galaxies ($\sim$ 0.1 Gyr$^{-1}$).

\item[-] Interpreting the sSFR as the inverse 
of a time scale for the SF, our results indicate that, at their 
present SFR, early-stage mergers build their stellar 
mass much faster ($\sim$ 2 Gyr)
than Sbc/Sc control galaxies ($\sim$ 10 Gyr).

\item[-] Our results are consistent with 
an evolutionary scheme where some star formation can be 
initially triggered in the central regions during the first 
contact stage, followed by extended star formation after the 
first pericenter passage (early stage mergers). 
\end{itemize}

\section*{Acknowledgements}
CALIFA is the first legacy survey carried out at Calar Alto. The CALIFA 
collaboration would like to thank the IAA-CSIC and MPIA-MPG as major partners 
of the observatory, and CAHA itself, for the unique access to telescope 
time and support in manpower and infrastructures. We also thank 
the CAHA staff for the dedication to this project. Support from the Spanish 
Ministerio de Econom\'{\i}a y Competitividad, through projects AYA2014-57490-P, 
AYA2010-15081, and Junta de Andaluc\'{\i}a FQ1580.
ALdA, EADL and RCF thanks the hospitality of the 
IAA and the support of CAPES and CNPq. RGD acknowledges the support of 
CNPq (Brazil) through Programa Ciência sem Fronteiras (401452/2012-3). 








\appendix

\section{Star cluster photometry}
\subsection{Method}\label{A.1} 
In Section \ref{3} we have summarized the stellar population 
properties of the super star clusters 
(SSCs) in IC 1623 W and NGC 6090 using the 
HST images from FUV to NIR. 
Their characteristics are summarized 
in Table A.1 and A.2, respectively. 
\begin{table*}
 \caption{Summary of HST data for IC 1623}
 \label{tab:natbib}
 \begin{tabular}{ccccccccc}
  \hline
\hline

	&	    & Detector/  & Plate Scale          & Observation &t$_{exp}$  & Proposal ID,  \\ 
Filter & Instrument & Camera     & (arcsec pixel$^{-1}$)   & Date        & (s)     & PI  \\ 
 
\hline
F25SRF2 & STIS	     & FUV-MAMA  &  0.025                &2000-12-25 & 2371.0 & 8201, G. Meurer \\
F25QTZ  & STIS	     & NUV-MAMA  &  0.025                &2000-12-25 &  750.0 & 8201, G. Meurer  \\
F435W   & ACS	     & WFC       &  0.05                 &2006-07-12 & 1260.0 & 10592, A. Evans \\
F814W   & ACS	     & WFC       &  0.05                 &2006-07-12 &  720.0 & 10592, A. Evans \\
F110W   & NICMOS     & NIC2      &  0.075                &1998-08-03 & 223.8  & 7219, N. Scoville \\
F160W   & NICMOS     & NIC2      &  0.075                &1998-08-03 & 223.8  & 7219, N. Scoville \\
\hline
 \end{tabular}
\end{table*}
\begin{table*}
 \caption{Summary of HST data for NGC 6090}
 \label{tab:natbib}
 \begin{tabular}{ccccccccc}
  \hline
\hline

       &	   & Detector/         & Plate Scale	      & Observation &t$_{exp}$  & Proposal ID,  \\ 
Filter & Instrument & Camera   & (arcsec pixel$^{-1}$)   & Date	     & (s)     & PI  \\ 
 
\hline
F140LP  & ACS	    & SBC      &  0.025 	       &2008-08-09 & 2648.0  & 11110, S. McCandliss \\
F330W	& ACS	    & HRC      &  0.025 	       &2005-11-12 &  800.0  & 10575, G.Ostlin  \\
F435W	& ACS	    & WFC      &  0.05  	       &2005-09-18 & 1380.0  & 10592, A. Evans \\
F814W	& ACS	    & WFC      &  0.05  	       &2005-09-18 &  800.0  & 10592, A. Evans \\
F110W	& NICMOS     & NIC2	&  0.075		 &1997-11-10 & 383.6   & 7219, N. Scoville \\
F160W	& NICMOS     & NIC2	&  0.075		 &2003-12-01 & 599.4   & 9726, R. Maiolino \\
\hline
 \end{tabular}
\end{table*}

Here we describe in detail the methodology we have followed 
to perform the star cluster photometry.
The clusters were detected in a combination of the 
HST optical images in F435W and F814W filters. 
We registered and resampled 
all the HST images to a common pixel scale, 
using GEOMAP/GEOTRAN IRAF tasks. 
Prior to the detection, we had to remove large-scale 
background variations in this optical image in two steps. 
First, it was smoothed with a $9 \times 9$ box median filter, 
and then, the smoothed image was subtracted 
from the original. 
The input object list for photometry was obtained 
by running DAOFIND task in DAOPHOT package on this 
background-subtracted image.
This procedure is similar to that 
followed by \cite{larsen&brodie2000}.
As an additional selection criterion, only clusters 
with  S/N $\geq$ 5 in optical filters were considered for
subsequent photometry.
The detected clusters were shown in Figure 3. 

Photometry was obtained using the phot task in APPHOT. 
It gives the fluxes of the clusters in all the filters, by specifying
the aperture radius and the position and width of the sky ring for background
determination. 
We have followed the recommendations of the reference guide to IRAF/APPHOT 
package (Lindsey E. Davis), selecting an aperture radius  $\sim$ FWHM$_{PSF}$, 
and $\sim$ 4 $\times$ FWHM$_{PSF}$ for the sky ring inner radius and width.
The aperture photometry of sources requires the correction for the light loss
due to the finite size of the aperture used, that is called the aperture 
correction.
The best way to determine aperture corrections is from real 
point sources (stars) on the images.
However, isolated  bright stars to obtain a meaningful 
correction were only present in optical images, but nor in UV neither in NIR.
Thus, aperture corrections were determined from analytically
generated point sources using Tiny Tim software \citep{krist2011}.

We have converted
these fluxes into STMAG magnitudes, using equation (6) in 
\cite{sirianni2005}.
The errors have been calculated through error propagation:
\begin{equation}
\label{eq:ErrF}
\Delta F = \sqrt{F+n_{aper}\times (sky +\sigma ^{2}_{RO}) \times (1+ \frac{n_{aper}}{n_{sky}})}
\end{equation}
where F is the star cluster flux within the aperture, n$_{aper}$ is the 
number of pixels in the aperture, sky is the sky
background value per pixel, measured in the sky annulus around the cluster,
n$_{sky}$ is the number
of pixels in the background annulus, and
$\sigma _{RO}$ is the readout noise of the CCD.
Errors in the corresponding magnitudes were computed from eq.\ \ref{eq:ErrF} using $\Delta m = 2.5 \log e \times \Delta F/F$.


\subsection{Cluster ages}\label{A.2} 
To estimate cluster properties, 
we have compared the colours and magnitudes of IC 1623 and NGC 6090 
clusters with those of Charlot \& Bruzual SSPs models 
(2007, unpublished) in a range
of ages between 1 Myr to 13 Gyr.
These models are similar to those 
in \cite{bruzual&charlot2003}, 
but replacing STELIB \citep{leborgne2003} by a combination of 
the MILES \citep{sanchez-blazquez2006} and $\textsc{granada}$ 
\citep{martins2005} libraries. 
Chabrier IMF and Padova 1994 evolutionary tracks were used.
Model colours have been computed using STSDAS.SYNPHOT software. 
Using the \cite{calzetti2000}  law we have computed also the colours of
the models reddened by up to A$_V = 6$ mag.
Despite recent
improvements in modelling tecniques, it is very
difficult from photometry to estimate metallicity on individual clusters.
From the spectroscopy we found that the average value for 
the metallicity in these systems is $\sim$ 0.6 Z$_{\odot}$ 
(see Section \ref{4.6}). 
Taking this into account, we have used solar metallicity SSP 
models (1 Z$_{\odot}$) in this analysis. However, we have 
checked that 0.4 Z$_{\odot}$ models predict similar results.

\begin{figure}  
\begin{center}
\includegraphics[width=0.45\textwidth]{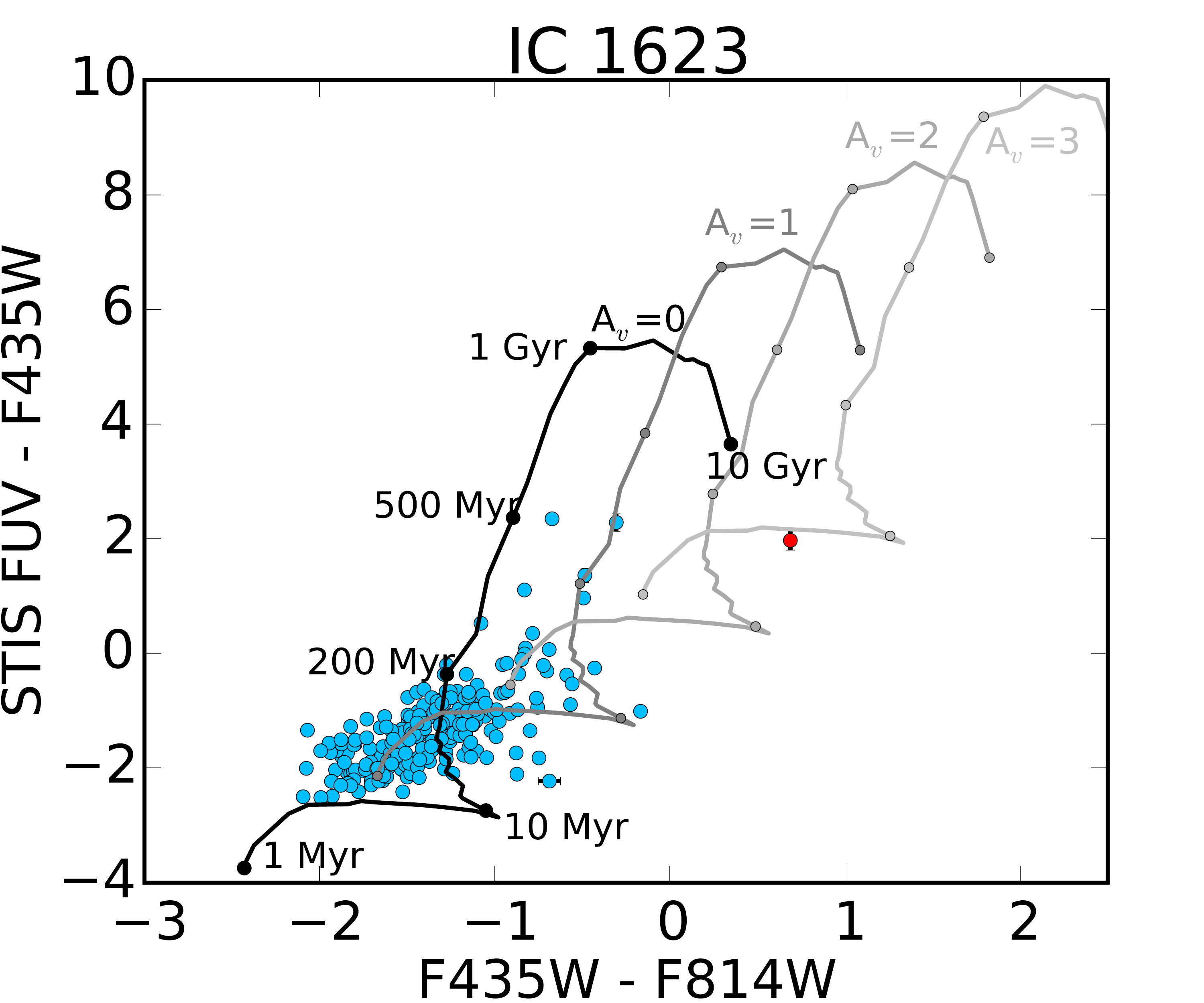} 
\includegraphics[width=0.45\textwidth]{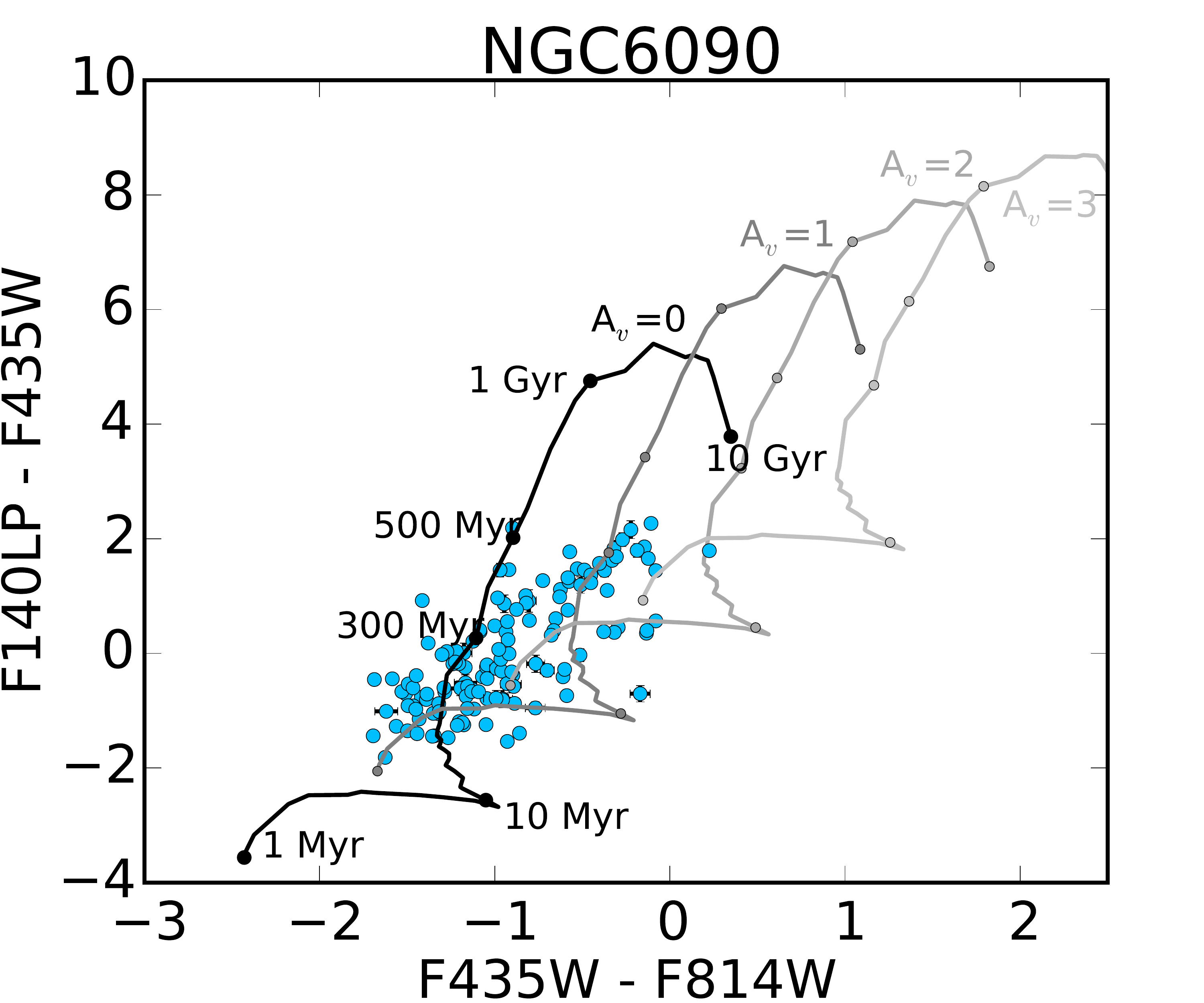} 
\caption{FUV - F435W vs F435W - F814W diagram for IC 1623 clusters (top panel) 
and NGC 6090 clusters (bottom panel).
The colour coding is the same as in Figure 3, 
but only those clusters with SNR $>$ 5 in FUV, F435W, and F814W filters 
are included in the plots. The black solid line is the path described by SSP 
from 1 Myr to 13 Gyr, Z$_{\odot}$, and A$_{V}$=0 mag. The greyscale 
lines are the paths for the same models reddened by 1 to 3 mag, 
with the lighter shades tracing the more extincted models.
For both systems most clusters
are consistent with ages younger than 200--300 Myr, 
assuming very low to almost no foreground extinction.
If affected by some extinction 0.5 - 1.0 mag, they 
could be even younger than 10 Myr.}
\label{Fig_A1}  
\end{center}
\end{figure} 

In Figure A1 we show the FUV $-$ F435W vs F435W $-$ F814W diagram 
for IC 1623 (top panel) 
and NGC 6090 (bottom panel), which is the best to break 
the age-A$_V$ degeneracy.
The colour coding is the same as in Figure 3,
but only those clusters with SNR $\geq$ 5 
in FUV, F435W, and F814W filters are included in the plots.
The black solid line is the path described by SSP models
from 1 Myr to 13 Gyr, Z$_{\odot}$, and A$_{V}$ = 0 mag. 
The greyscale lines are the paths 
for the same models reddened by 1 to 3 mag, 
with the lighter shades tracing the more reddened models.

Comparing the position of the clusters in colour-colour diagrams
with respect to the model positions we can estimate 
the range of cluster ages. The main results are summarized 
in Section \ref{3}.

\begin{figure}
\begin{center}
\includegraphics[width=0.45\textwidth]{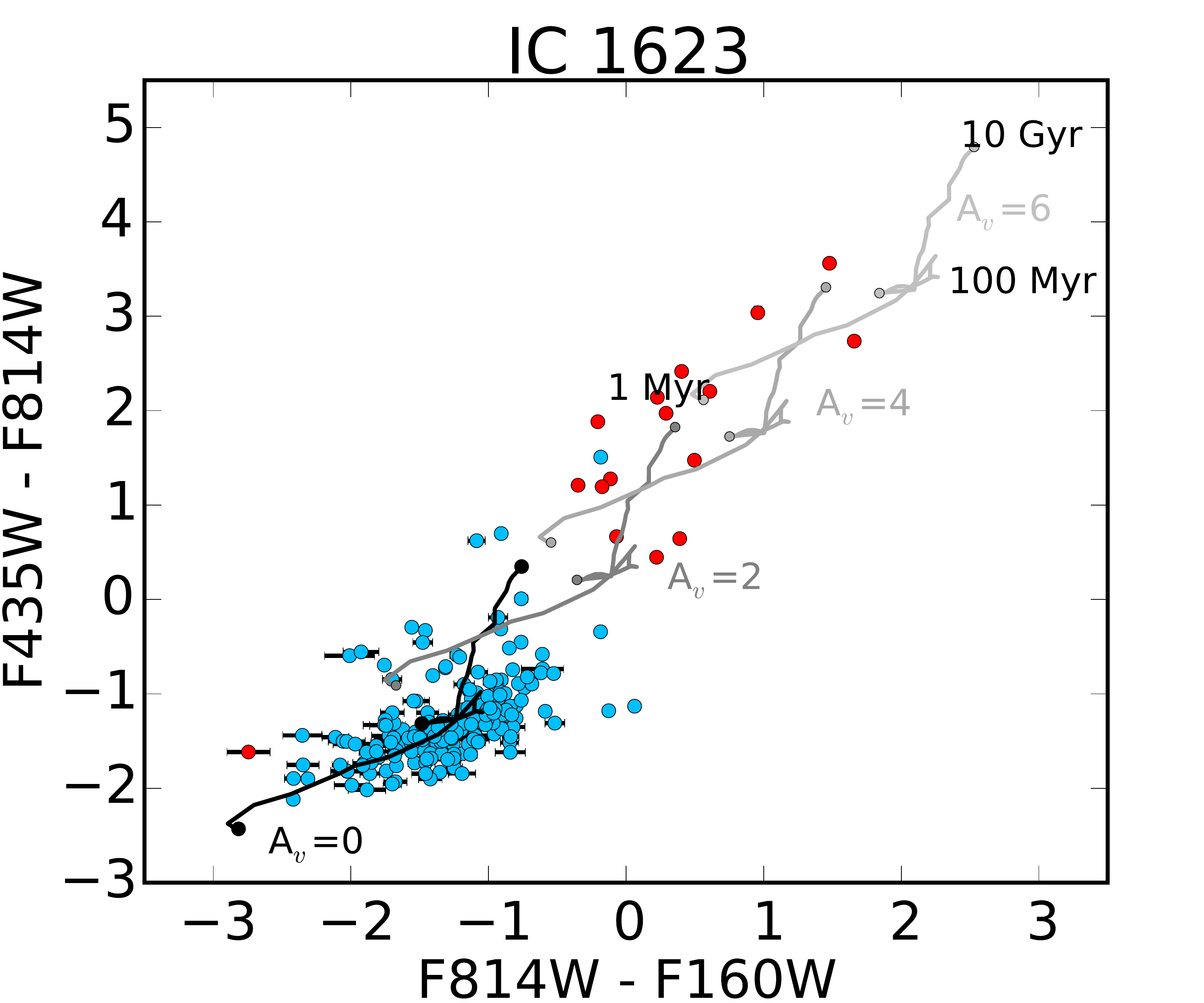} 
\includegraphics[width=0.45\textwidth]{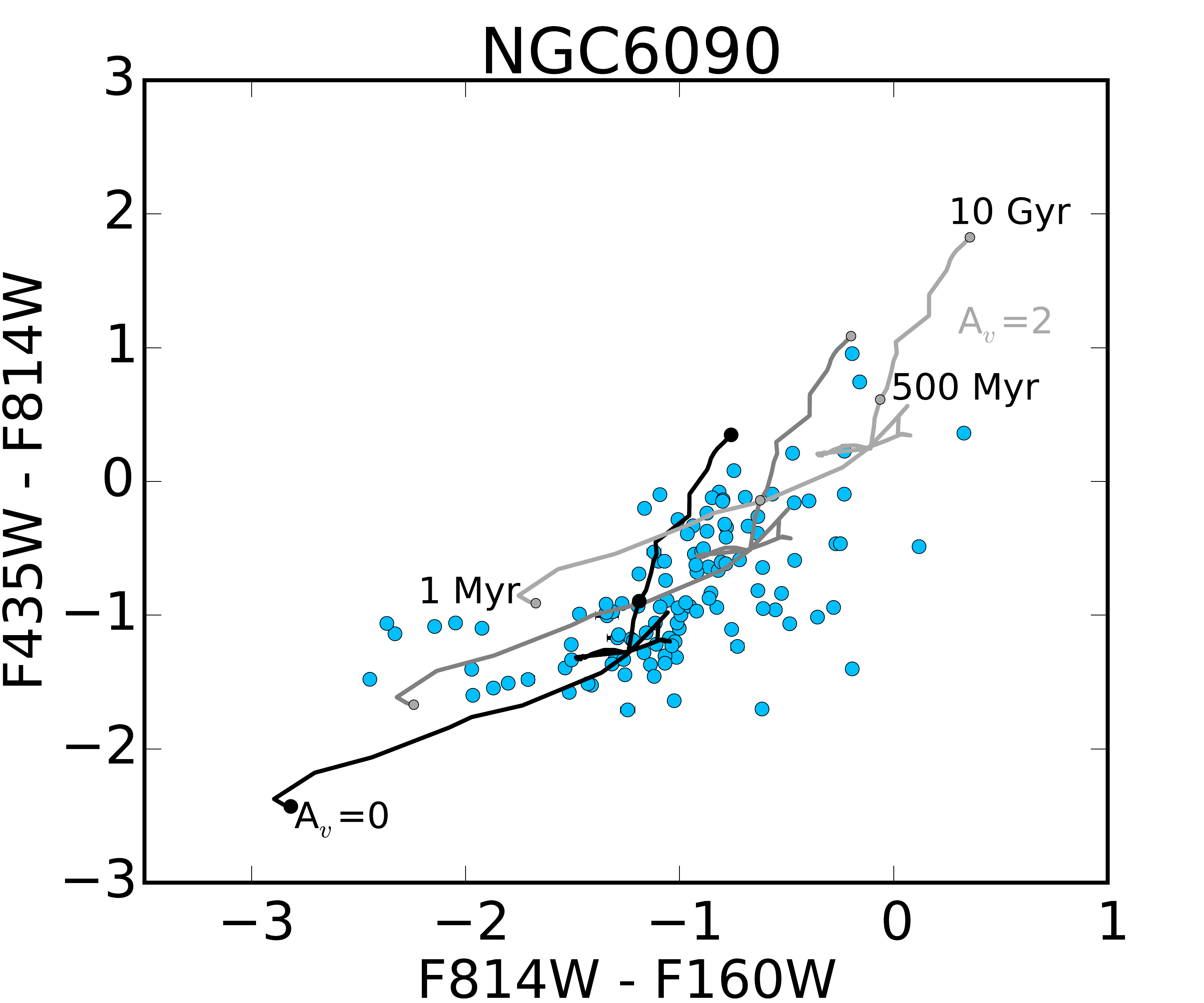} 
\caption{F435W - F814W vs F814W - F160W diagram for IC 1623 clusters (top panel) 
and NGC 6090 (bottom panel). 
Only clusters with SNR $>$ 5 in F435W, F814W, and F160W filters are shown.
The black solid line is the path described by SSPs 
from 1 Myr to 13 Gyr, Z$_{\odot}$, and A$_{V}$=0 mag. The greyscale lines 
are the paths for the same models but reddened by 1, 2, 4 and 6 mag.}
\label{Fig_A2}  
\end{center}
\end{figure} 

\begin{figure}  
\begin{center}
\includegraphics[width=0.45\textwidth]{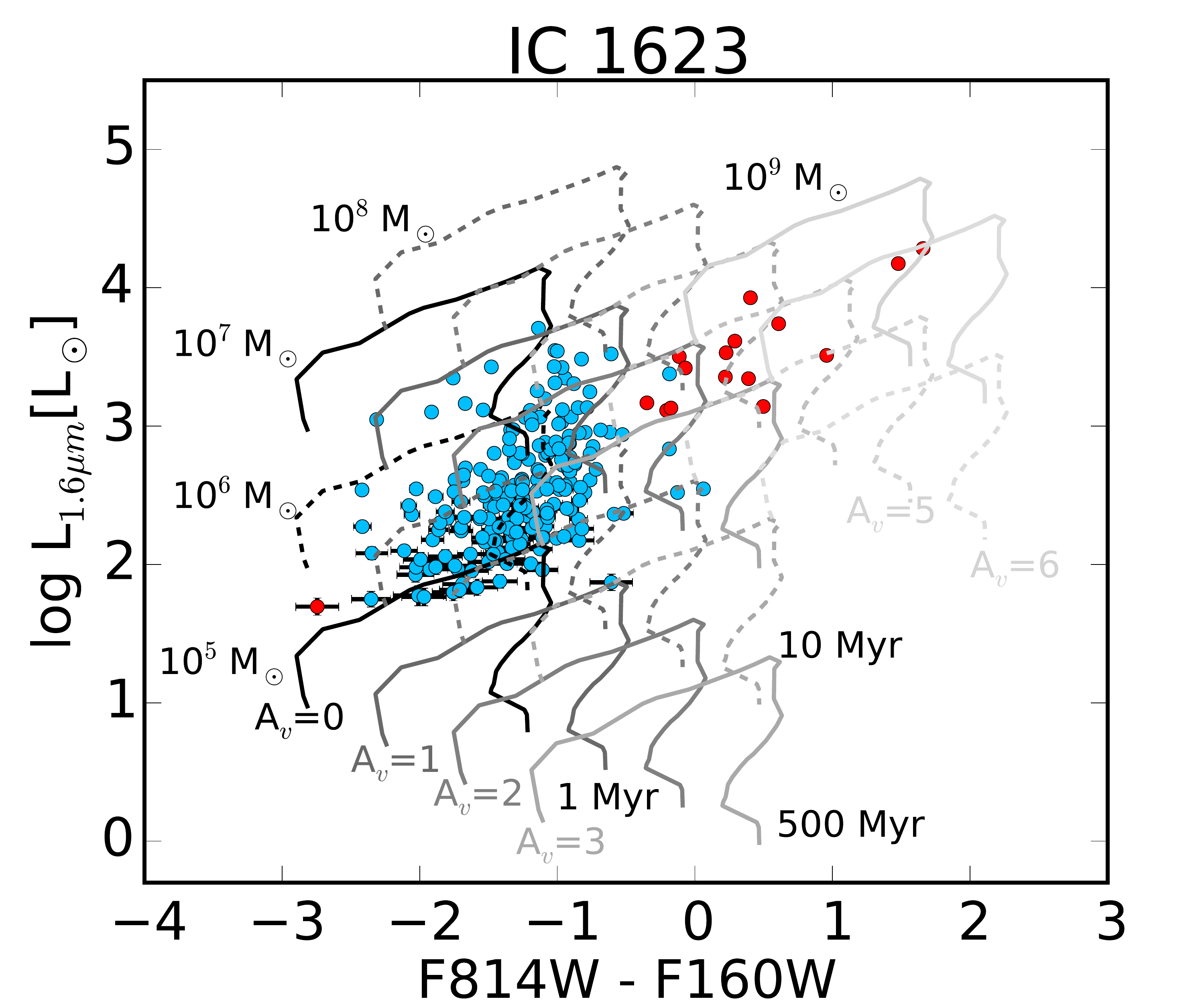} 
\includegraphics[width=0.45\textwidth]{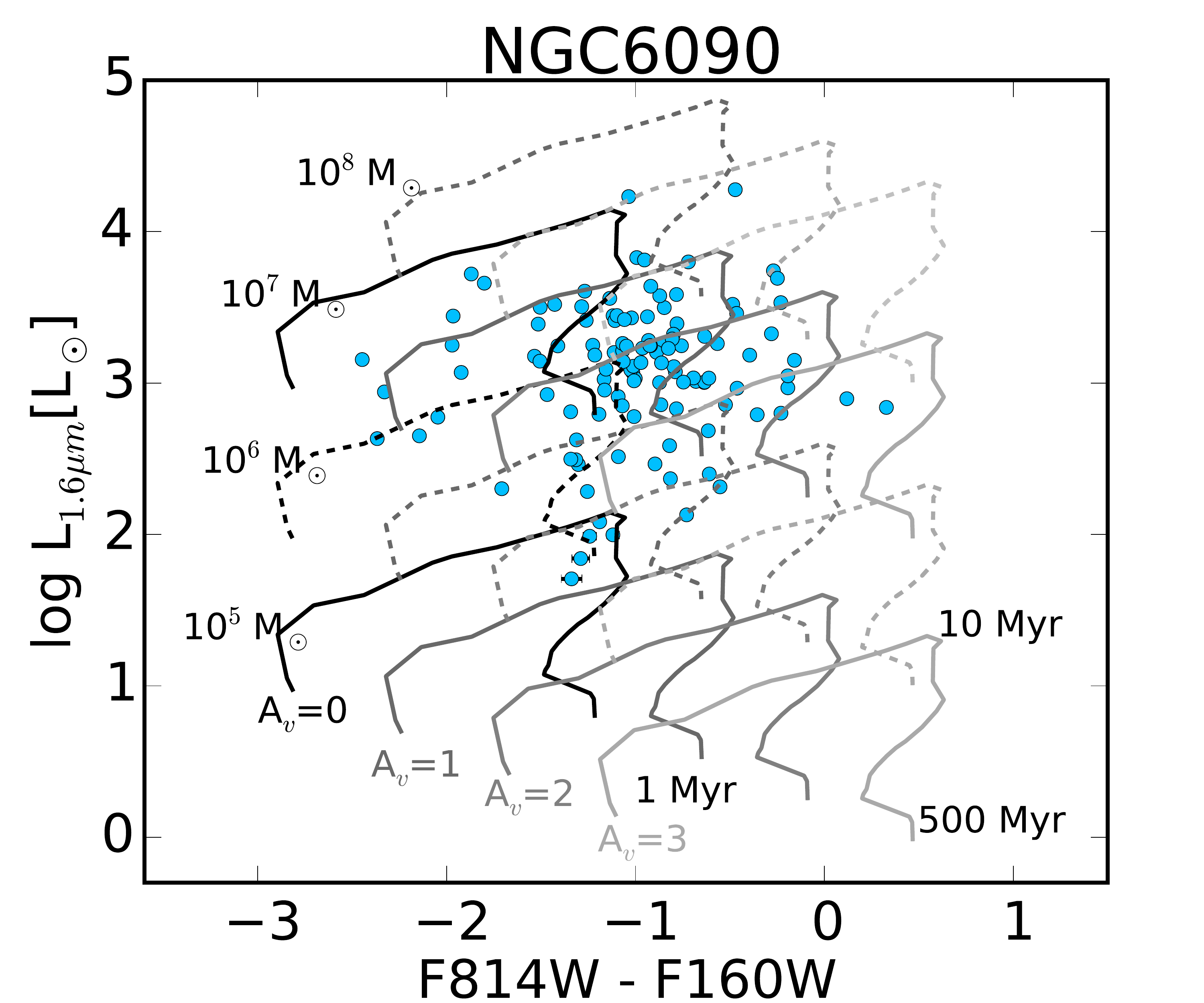} 
\caption{Absolute NIR 1.6$\mu m$ magnitude vs F814W - F160W diagram 
for IC1623 clusters (top panel) and NGC 6090 clusters (bottom panel).
Only clusters with SNR $>$ 5 in F435W, F814W and F160W filters are shown.
The colours of the clusters are the same as in Figure 3.
The paths from models of 10$^{5}$, 10$^{6}$, 10$^{7}$, 10$^{8}$ and 
10$^{9}$ M$_{\odot}$ have been shown as solid and dashed lines.
The models have been cut between 1 - 500 Myr, as we known 
from previous diagrams that the clusters are not older 
than 500 Myr in these two early-stage merger LIRGs. 
The greyscale coding represents the variation in 
stellar extinction as in previous figures.}
\label{Fig_A3}  
\end{center}
\end{figure}

\subsection{Cluster extinction}\label{A.3}
Figure A2 shows the F435W $-$ F814W vs.\
F814W $-$ F160W diagram for IC 1623 (top panel) and 
NGC 6090 (bottom panel). This diagram allows to 
estimate the range of extiction affecting the clusters
which are obscured in the FUV. 
Only clusters with SNR $\geq$ 5 
in F435W, F814W, and F160W filters are shown.
Again, the black solid line is the path described 
by SSPs from 1 Myr to 13 Gyr, Z$_{\odot}$, and 
A$_{V}$ = 0 mag. The greyscale lines are the paths 
for the same models reddened by up to 6 mag, with the 
lighter shades tracing the more extincted models.

We note that, while in NGC 6090 the difference in 
the values of extinction between the two progenitors 
is relatively small, $\sim$ 1 mag, for IC 1623 the contrast 
between both progenitors is much higher, possibly 
up to 4--6 mag.

\subsection{Cluster mass}\label{A.4}
We found that NICMOS 1.6 $\mu m$ band is the most sentitive 
to mass variations. 
We  estimate the mass range of the star clusters by 
comparing their 1.6 $\mu m$ absolute magnitude with the 
absolute magnitude of SSPs models of masses ranging from 10$^{5}$ 
to 10$^{9}$ M$_{\odot}$.
The results are shown in Figure A3 for IC 1623 (top panel) 
and NGC 6090 (bottom panel).
This is a NIR 1.6$\mu m$ absolute magnitude vs 
F814W - F160W diagram.
The paths from models of 10$^{5}$, 10$^{6}$, 10$^{7}$, 10$^{8}$, and 
10$^{9}$ M$_{\odot}$ are shown as solid and dashed lines.
The models have been cut between 1 - 500 Myr, as we know from the 
previous section that these are the older ages our clusters can have.

We have performed a rough estimate of the total mass 
in clusters in these systems by assigning a certain 
average mass for the clusters in each luminosity 
L$_{1.6 \mu m}$ range. In particular, we have 
separared the clusters in four possible 
luminosity ranges: from 
0.9 $<$ log L$_{1.6 \mu m}$ $<$ 1.5 
we assign a mass of $\sim$ 10$^{5}$ M$_{\odot}$, 
if 1.5 $<$ log L$_{1.6 \mu m}$ $<$ 2.65 
then $\sim$ 10$^{6}$ M$_{\odot}$, 
if 2.65 $<$ log L$_{1.6 \mu m}$ $<$ 3.5 
then $\sim$ 10$^{7}$ M$_{\odot}$, and 
above log L$_{1.6 \mu m}$ $>$ 3.5 we assign
$\sim$ 10$^{8}$ M$_{\odot}$. 
In the case of IC 1623 W we have 
0, 143, 62 and 1 clusters in each range, leading 
to a clusters mass 
M$_{IC 1623 W} ^{clus}$ $\sim$ 9 $\times$ 10$^{8}$ M$_{\odot}$.
In NGC 6090 we have 0, 19, 79, 19 clusters in each 
luminosity range, leading to 
M$_{NGC 6090} ^{clus}$ $\sim$ 2 $\times$ 10$^{9}$ M$_{\odot}$.

\subsection{Cluster detection limits}\label{A.5}
A major question to be addressed with star cluster photometry 
is to know if older/more reddened clusters could have been detected.
As the star cluster analysis was a complementaty goal, 
we did not make a proper completeness 
study as in \cite{davies2016}, where synthetic clusters of 
different ages/optical depths are introduced in 
the cluster detection image to quantify which percetange 
of them could be recovered.
However, we followed a more simplistic approach to estimate 
which clusters could be, in principle, detected, and which not.
Using the SSP models of 10 Myr, 100 Myr, 200 Myr, 300 Myr, 500 Myr, 
and 1 Gyr, and the transmission curves from F435W and F814W filters, 
we compute the 435 and 814 magnitudes for clusters of different masses 
($10^5 M_{\odot}$, $10^6 M_{\odot}$, $10^7 M_{\odot}$, 
and $10^8 M_{\odot}$) and affected by extinctions of 
0 mag, 1 mag, 2 mag, 4 mag, 6 mag and 10 mag. 
These magnitudes are compared with the 435 and 814 limiting magnitudes 
expected for point sources (most clusters are not resolved) with 
SNR=5, obtained from Figure 10.2 and 10.32 of the ACS Instrument Handbook, 
given the exposure times of the HST observations used in 
this paper ($\sim$ $10^{3}$ s). 

The less massive the cluster the more difficult to 
detect it as it ages/is more reddened. 
In IC 1623, clusters of $10^5 M_{\odot}$ up to 300 Myr can be detected 
for no extinction, for $A_{V}$ = 1 mag, clusters older than 100 Myr 
cannot be detected, and for $A_{V}$ = 2--4 mag, only clusters of 
10 Myr can be detected. In NGC 6090, whose distance is larger 
than IC 1623 one, only $10^5 M_{\odot}$ clusters up to 100 Myr 
can be detected for no extinction, while for $A_{V}$ = 1--2 mag, 
only 10 Myr clusters can be detected.

From our observations, we found 
that most clusters in IC 1623 and NGC 6090, 
have masses between $10^{6}-10^{7} M_{\odot}$. In principle, 
$10^6 M_{\odot}$ clusters up to 1 Gyr (in IC 1623) and 500 Myr (NGC 6090) 
could be detected up to extinction levels of $A_{V} = 2$ mag. 
For $A_{V} = 4$ mag, only clusters up to 300 Myr (IC 1623) and 
100 Myr (NGC 6090) can be detected, and for $A_{V} = 6$ mag, 
only 10 Myr clusters can be detected. Above that extinction level, 
we cannot detect clusters anymore.

The age/extinction limits reach further for 
the more massive clusters. 
For example, $10^{7}-10^{8} M_{\odot}$ clusters in IC 1623 E can 
be detected up to ages of 1 Gyr, and $A_{V} = 6$ mag.

Because point sources are much harder to detect on top of the 
diffuse/changing galaxy background than when isolated 
(which is what the HST manual assumes), we acknowledge 
that our detection limits are in fact upper limits, 
as they consider cluster detection under optimal conditions.
More realistically, the detection limits become more restrictive 
depending on the galaxy background.

From this analysis, we conclude that star clusters 
photometry only allows to trace the recent star formation history 
of these mergers, $<$500 Myr--1 Gyr (at best, for no extinction), 
but not in previous epochs. 
So if older star formation occurred, complementary methods 
will be necessary to detect it, as stellar population synthesis.  
Therefore, star cluster studies are preferably indicated 
in early-stage mergers where most of the merger-induced 
star formation is recent, 
or in more advanced mergers if we are only interested 
in the youngest populations formed.

\section{Alternative radial profiles for NGC 6090 SW}\label{B.1}
As commented in Section \ref{2.4}, the exact position of NGC 6090 SW 
nucleus is not clear. In the radial profiles shown in previous 
sections we considered as the nucleus the luminosity peak in the IFS 
continuum, marked as a black dot in the maps. 
However, there exists the possibility that the 
real nucleus is the bright knot located to the 
North West of this progenitor, which is indicated with 
a km s$^{-1}$bol in the maps. 
As a sanity check we have repeated the radial 
profiles of the stellar population properties 
taking the knot position as origin.
The only property for which the radial 
profile changes is the metallicity. 
The trend is the same, metallicity grows with 
distance from the core, but the gradient is steeper 
from the knot than from our nucleus. From the knot 
it goes from $\langle \log \ Z_{\star}/Z_\odot \rangle_{M} = -0.6$ to $-0.15$,
while for our nucleus it goes from $-0.4$ to $-0.25$.
For the other properties the profiles are very similar 
and the differences are not significant.


\bsp	
\label{lastpage}
\end{document}